\documentclass[journal]{IEEEtran}
\usepackage{cite}
\usepackage{amsmath,amssymb,amsfonts}

\usepackage{graphicx,color}
\usepackage{textcomp}
\usepackage{graphicx}

\usepackage{graphicx}
\usepackage{amsmath}
\usepackage{amssymb}
\usepackage{booktabs}

\usepackage{graphicx}
\usepackage{algorithm}
\usepackage{algpseudocode}
\usepackage{float}
\usepackage{stfloats}
\usepackage{caption}
\usepackage{color} 
\usepackage[margin=0pt,position=bottom,font=footnotesize,labelfont=bf]{caption}
\usepackage[margin=0pt,position=bottom,font=footnotesize,labelfont=bf]{subcaption}
\usepackage[export]{adjustbox}
\usepackage{graphics,cite}
\usepackage{array}
\usepackage{amsmath,amsfonts, enumitem}
\usepackage{siunitx}

\usepackage{flushend}
\usepackage{pict2e}
\usepackage{tikz}
\usepackage{wrapfig}

\usepackage{hyperref}

\begin{document}
	
	\title{Coded Illumination for Improved Lensless Imaging}
	
	\author{Yucheng Zheng, \textit{Student Member, IEEE} and M. Salman Asif \textit{Senior Member, IEEE}%
		\thanks{This work was supported in part by NSF Awards 2046293, 2133084 and AFOSR Award FA9550-21-1-0330.}%
		\thanks{Y. Zheng and M. Asif are with the Department of Electrical and Computer Engineering, University of California, Riverside,
			CA, 92521 USA (e-mail: yzhen069@ucr.edu; sasif@ucr.edu).}}
	
	\maketitle
	
	\begin{abstract} 
		Mask-based lensless cameras can be flat, thin, and light-weight, which makes them suitable for novel designs of computational imaging systems with large surface areas and arbitrary shapes. Despite recent progress in lensless cameras, the quality of images recovered from the lensless cameras is often poor due to the ill-conditioning of the underlying measurement system. In this paper, we propose to use coded illumination to improve the quality of images reconstructed with lensless cameras. In our imaging model, the scene/object is illuminated by multiple coded illumination patterns as the lensless camera records sensor measurements. We designed and tested a number of illumination patterns and observed that shifting dots (and related orthogonal) patterns provide the best overall performance. We propose a fast and low-complexity recovery algorithm that exploits the separability and block-diagonal structure in our system. We present simulation results and hardware experiment results to demonstrate that our proposed method can significantly improve the reconstruction quality. 
	\end{abstract}
	

	\IEEEpeerreviewmaketitle

	%%%%%%%%%%%%%%%%%%%%%%%%%%  body  %%%%%%%%%%%%%%%%%%%%%%%%%%
	\section{Introduction}
	
	\IEEEPARstart{L}{ensless} cameras provide novel designs for extreme imaging conditions that require small, thin form factor, large field-of-view, or large-area sensors \cite{asif2017flatcam,boominathan2016lensless,antipa2018diffusercam,boominathan2022recent}. Compared to conventional lens-based cameras, lensless cameras can be flat, thin, light-weight, and potentially flexible because the physical constraints imposed by a lens are relaxed. FlatCam is an  example of a lensless camera \cite{asif2017flatcam}, which belongs to a broader class of coded-aperture cameras that replace the lens with a coded mask \cite{cannon1980coded_aperture,fenimore1978ura}. The image formed on the sensor with a coded mask is a linear combination of multiple shifted versions of the scene. To recover the scene image from the sensor measurements, we need to solve a linear inverse problem. The quality of the recovered image depends on the conditioning of the linear system; especially in the absence of any prior knowledge about the scene.  
	
	\begin{figure*}[ht]
		\centering
		\includegraphics[width=0.8\linewidth]{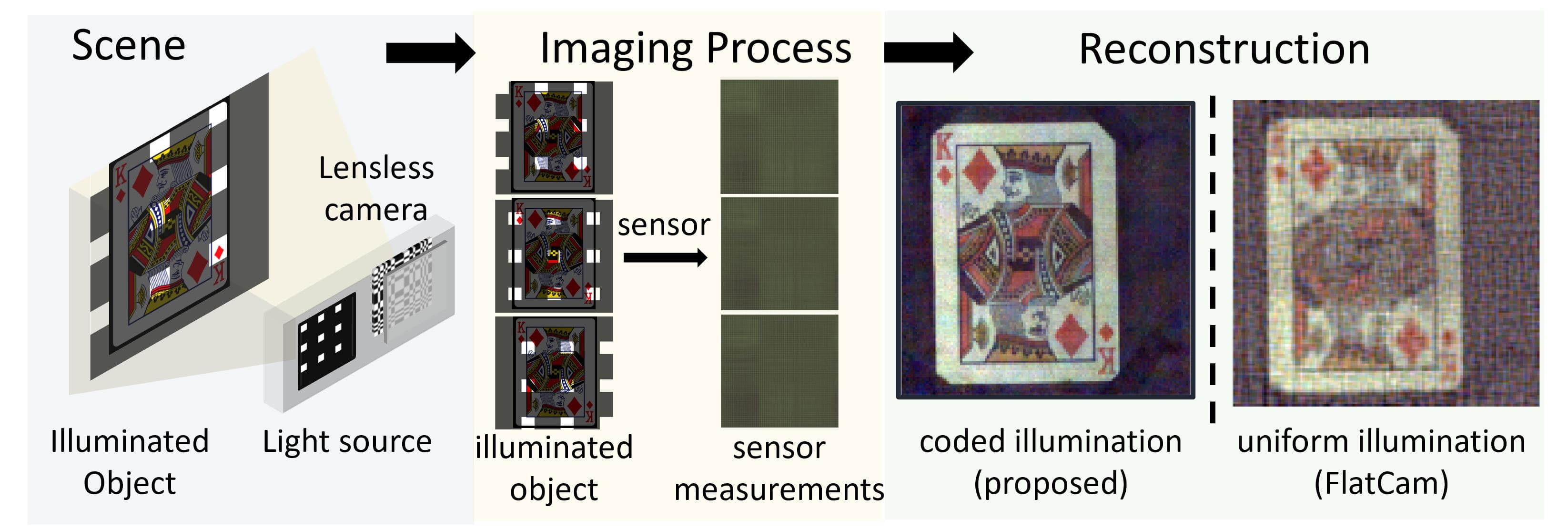}
		\caption{An overview of our proposed method. We project a sequence of binary illumination patterns onto the object and capture the sensor measurements corresponding to each illumination pattern. The reconstruction result using multiple coded illumination patterns significantly outperforms the conventional method where the scene is illuminated by uniform illumination. }
		\label{fig:intro}
	\end{figure*}

	In this paper, we propose a new method that combines coded illumination with mask-based lensless cameras (such as FlatCam) to improve the quality of recovered images, illustrated in Fig.~\ref{fig:intro}. A 2D scene is illuminated with multiple coded patterns during data acquisition. We capture sensor measurements for each illumination pattern and use a fast recovery algorithm to reconstruct the scene using all the measurements.
	The main contributions of this paper are as follows
	\begin{itemize}[leftmargin=4mm,topsep=0pt,itemsep=0ex,partopsep=1ex,parsep=1pt]
		\item We propose a new framework to combine coded illumination with lensless imaging.
		\item We propose a fast and low-complexity algorithm that exploits separability of the mask and illumination patterns and avoids storing all the measurements or creating large system matrices during reconstruction.
		\item We design shifting dots and  orthogonal patterns that provide best overall performance in terms of quality and computational complexity. 
		\item We present simulation results to show that our method can improve the image reconstruction quality under different system conditions. 
		\item We present real experiments using a prototype we built to evaluate the performance of our imaging system and algorithm in the real environments.
	\end{itemize}
	
	Our main objective is to demonstrate that we can significantly improve the conditioning of a lensless imaging systems and the quality of the reconstruction by using coded illumination. 
	Our experiments show that the image quality improves in almost all cases as we increase the number of illumination patterns. Our method also has an inherent trade-off between imaging speed and the quality of the recovered images.
	In this paper, we mainly implement and discuss the design of FlatCam \cite{asif2017flatcam} with separable masks and illumination patterns. Nevertheless, the ideas presented here can be used to improve the imaging performance of other 
	lensless systems, such as DiffuserCam \cite{antipa2018diffusercam}, fresnel zone aperture \cite{shimano2018fresnel}, or phlatcam \cite{boominathan2020phlatcam}.

	The proposed framework can be used to build large-area,  high-resolution, and compact cameras that can potentially be used for under-the-display fingerprint and vein imaging. Such cameras can use the display panels for coded illumination \cite{yang2021designing, yin2020underdisplay,yokota2021flexible}. Lensless microscopes  \cite{adams2017rice_depth, antipa2018diffusercam,Chakrova2015illumination} can also benefit by combining coded illumination with data capture. Another potential application for such lensless cameras is in different wearable devices. In particular, our inspiration came from a project in which we seek to embed lensless sensors on a soft robot that will be used to assist infants in rehabilitation. The goal is to track infant arm motion using lensless sensors on a wearable sleeve while they reach objects. In all these cases, the objects in the scene can potentially be illuminated with a built-in display, a mini projector, or an add-on light source combined with a shifting mask,.

	\section{Related Work}
	
	The mask-based lensless camera such as FlatCam \cite{asif2017flatcam} is an extended version of pinhole cameras. Although a pinhole camera is able to image the scene directly on its sensor, it often suffers from noise \cite{yedidia2018analysis_aperture}. Coded aperture-based cameras alleviate this problem by using multiple pinholes placed in a designed pattern \cite{asif2017flatcam, fenimore1978ura, busboom1998ura, cannon1980coded_aperture, boominathan2016lensless}. In contrast to conventional lens-based cameras that capture images of the scene directly, coded mask-based cameras capture linear measurements of the scene and perform reconstruction by solving a linear inverse problem. Coded aperture-based cameras can also recover the depth information of a scene \cite{levin2007_mask_on_lens,asif2017greedy_depth, antipa2018diffusercam, adams2017rice_depth,zheng2019joint,hua2020sweep,zheng2021simple}. The main advantage of FlatCam is the thin and flat  form factor, which also makes the system ill-conditioned and affects the quality of reconstructed images.
	
	Signal recovery from ill-conditioned and under-determined systems is a classical and long-standing problem in signal processing. A number of methods have been proposed to tackle these problems over the decades \cite{candes2006compressive,rudin1992tv,golub1996matrix}. 
	An ill-conditioned system is unstable as its solution can change dramatically with tiny perturbations; thus, such a system rarely generates good results in a signal recovery problem. An under-determined system has fewer measurements than the number of unknowns; thus, it admits infinitely many solutions, and one can hardly determine the true solution using only the measurements. 
	A standard approach to deal with ill-conditioned and under-determined systems is to add a signal-dependent regularization term in the recovery problem, which constrains the range of solutions. Popular methods include adding sparse and low-rank priors on the signals \cite{candes2006compressive, rudin1992tv, donoho2006compressive, baraniuk2007compressive,recht2010matrix} and natural-image-like generators prior \cite{rakib2019generator, bora2017generator, hand2018phase}. Another approach is to capture multiple, diverse measurements of the scene that makes the modified imaging system well-condition and the reconstruction more accurate \cite{hua2020sweep,zheng2021simple}.
	Recently, a number of methods have been proposed that use deep networks to reconstruct images from lensless measurements \cite{khan2020flatnet,boominathan2020phlatcam,kristina2019learning,kristina2021untrained}. Some of these methods provide an exceptional improvement over traditional optimization-based methods. Nevertheless, deep learning-based methods in general, and end-to-end methods in particular, provide a huge variation in performance for simulated and real data (mainly because of mismatch in the simulated/actual mask-sensor-projector configuration and scenes). In contrast to deep learning methods, our method seeks to improve the conditioning of the underlying linear system and offer better generalization and robust results for arbitrary scenes without the need for any learning from data.

	Multiplexed illumination analysis is discussed in \cite{schechner2007multiplexlighting}, which shows how overall reconstruction SNR can be improved using coded illumination with Hadamard codes. Plenoptic imaging and noise analysis for reconstruction from multiplexed measurements are discussed in \cite{Wetzstein2012PMultiplexing}. Even though this paper is focused on lensless imaging with coded illumination, the noise analysis we discuss in Sec. \ref{sec:noise} follows similar arguments as  \cite{schechner2007multiplexlighting,Wetzstein2012PMultiplexing} to show the relationship between reconstruction error and overall system conditioning and singular values.
	
	Our proposed approach is an active imaging approach combining coded modulation or structured illumination method with coded aperture imaging
	\cite{gustafsson2008structured, nayar2012diffuse, fofi2004survey}.
	Structured illumination schemes are commonly used for imaging beyond diffraction in microscopy. These schemes use multiple structured illumination patterns to down-modulate high spatial frequencies in a sample into a low-frequency region that can be captured by the microscope \cite{gustafsson2008structured,rainer1999SIM,gustafsson2000surpassing}. Structured light is also widely used in multiplexing scene recovery to improve image quality and SNR \cite{schechner2007multiplexlighting, gu2011multiplex, mitra2014multiplex, ratner2007multiplexed}.  
	Other active imaging approach includes time-of-flight sensors \cite{gokturk2004tof,heide2013tof} that estimate the 3D scene by sending out infrared light and measuring its traveling time in reflection. Coded diffraction imaging is used to recover complex-valued wavefront from Fourier measurements \cite{candes2013phaselift,shechtman2015phase}. In coded diffraction imaging, the signal of interest gets modulated by a sequence of illumination patterns before the K-space measurements were captured \cite{candes2015diffraction, miao2008diffraction, jagatap2020ptychography}. Ptychography is another related method for capturing high-resolution microscopy images by capturing multiple images of the scene using a sequence of coded illumination patterns \cite{rodenburg2008ptychography, lei2014fourierptychography}.
	Dual photography \cite{sen2005dual} and compressive sensing \cite{duarte2008single_pixel,sen2009compressive} schemes also use coded illumination to sample the scenes. A dual photography system can create a high-resolution image of the scene by scanning the entire scene one pixel at a time, but it requires a fast laser projector. Single-pixel camera collects thousands of multiplexed measurements of the scene and solves a regularized optimization problem to reconstruct the image. In our method, we use lensless camera to capture tens of sensor frames with shifting coded illumination patterns and get better reconstruction with improved system conditioning. 
	
	A random illumination patterns-based lensless imaging method with simulations was presented in \cite{zheng2020icassp}. In contrast, we design shifting dots and orthogonal patterns that are  significantly superior to random patterns in terms of quality of reconstruction and computational complexity. We provide detailed simulations and experimental results on real data captured with a custom-built prototype.
	
	%% Methods 
	\section{Methods}
	\subsection{Separable Imaging Model} 
	FlatCam \cite{asif2017flatcam} consists of an amplitude mask placed on top of a bare sensor, and every sensor pixel records a linear combination of the entire scene. Suppose the sensor plane is at the origin of the 3D Cartesian coordinates $(u,v,z)$ and an amplitude mask is placed parallel to the sensor at distance $d$. We can model the  measurement recorded at sensor pixel $(u,v)$ as 
	\begin{equation}
	y(u,v) = \int x(u', v', z) \, \varphi(u', v', u,v, z) \, du' dv' dz, 
	\label{eq:integration}
	\end{equation}
	where $x(u',v',z)$ represents the intensity and $\varphi(u',v',u,v,z)$ represents the sensor response of a point source at $(u',v',z)$. 
	In this paper, we assume that the scene consists of a single plane at a known depth; therefore, we can ignore the depth parameter and represent the sensor measurements as
	\begin{gather}
	y(u,v) = \int  x(u', v') \, \varphi(u', v', u,v)\, du' dv' \notag \\ \Rightarrow ~~\mathbf{y = \Phi x},
	\label{eq:general_case_imaging_equation}
	\end{gather}
	where $\mathbf{x}$ denotes the scene intensity vector, $\mathbf{\Phi}$ denotes the system matrix, and $\mathbf{y}$ denotes the sensor measurement vector. The computational and memory complexity of the general imaging model in \eqref{eq:general_case_imaging_equation} makes it unsuitable for systems with a large number of scene and sensor pixels. We can overcome this challenge in a number of ways; for instance, we can use a separable model as in FlatCam~\cite{asif2017flatcam, adams2017rice_depth} or a convolutional model as in DiffuserCam~\cite{antipa2018diffusercam}. We use a separable system in this paper.

	A separable mask pattern that is aligned with the sensor grid yields a separable imaging system, which can be represented as
	\begin{equation}
	y(u,v) = \int \int x(u', v') \, \varphi_L(u', u)\, du'~  \varphi_R(v',v)\, dv'. \label{eq:separable_cont}
	\end{equation}
	The product of $\varphi_L(u',v)$ and $\varphi_R(v',v)$ represents the separable system response for point sources along $u,v$ axes. 
	Let us assume that $X$ represents an $n\times n$ image of the scene intensities at a fixed plane and $Y$ denotes $m\times m$ sensor measurements, then we can represent the separable system in \eqref{eq:separable_cont} as 
	\begin{equation}
	Y = \Phi_L X \Phi_R^\top, \label{eq:separable_disc}
	\end{equation}
	where $\Phi_L, \Phi_R$ denote the system matrices for $u,v$ axes, respectively. We assume square shapes for the scene and sensor to keep our discussion simple, but the ideas can be extended to arbitrary shapes. 
	
	\subsection{Coded Illumination and Reconstruction} 
	The effect of illumination can be modeled as an element-wise product between the scene and the illumination patterns. In our experiments, we use a laser projector placed next to the lensless camera to illuminate the object. In other applications, such as under-the-display cameras, an LED screen can be used for illumination. 
	To simplify the recovery process, we further assume that the illumination patterns are separable and drawn from columns of $n\times k$ matrices $P_L$ and $P_R$. Let us denote a pattern as $P_{i,j} = p_{Li}p_{Rj}^\top$, where $p_{Li}$ and $p_{Rj}$ are $i$th and $j$th columns of $P_L$ and $P_R$, respectively. 
	We can describe sensor measurements for any given illumination pattern $P_{i,j}$ as
	\begin{equation}\label{eq:illum_forwardModel}
	Y_{i,j} = \Phi_L (P_{i,j} \odot X)\Phi_R^\top + E_{i,j}, 
	\end{equation}
	where $\odot$ represents element-wise multiplication operator and $E_{i,j}$ denotes measurement noise.

	To recover image $X$ from the sensor measurements $Y_{i,j}$, we can solve the following $\ell_2$-regularized least-squares problem:
	\begin{equation}
	\underset{X}{\text{argmin}}~\sum_{i,j}\|Y_{i,j}-\Phi_L(P_{i,j} \odot X)\Phi_R^\top\|_2^2+\lambda\|X\|_2^2,
	\label{eq:tikhonov}
	\end{equation}
	where $\lambda >0 $ is a regularization parameter. An optimal solution of \eqref{eq:tikhonov} must satisfy the following conditions (which can be derived by setting the gradient to zero) with $\mathbf{Q}=\sum_{i,j} (\Phi_L^\top Y_{i,j} \Phi_R) \odot P_{i,j}$: 
	\begin{align}
	\small
	\mathbf{Q} &= \underbrace{(\Phi_L^T\Phi_L \odot P_LP_L^T)}_{\mathbf{A}_L} X\underbrace{(\Phi_R^T\Phi_R \odot P_RP_R^T)}_{\mathbf{A}_R} + \lambda X, \notag \\
	\normalsize 
	\Rightarrow \mathbf{Q} &= \mathbf{A_L} X \mathbf{A_R} + \lambda X, \label{eq:tikhonov_sep}
	\end{align}
	where $\mathbf{A_L}$, and $\mathbf{A_R}$ are $n\times n$ matrices. 
	The solution of \eqref{eq:tikhonov_sep} can be written in closed form using the eigen-decomposition of $\mathbf{A}_L,\mathbf{A}_R$ \cite{asif2017flatcam} as 
	\begin{equation}
	\hat X = \mathbf{V_L[(V_L^\top Q V_R)}./(\mathbf{s_Ls_R}^\top+\lambda\mathbf{11}^\top)]\mathbf{V_R^\top}, 
	\label{eq:closed_form}
	\end{equation}
	where $\mathbf{V_L,V_R}$ denote the eigenvectors and $\mathbf{s_L,s_R}$ denote the eigenvalues of $\mathbf{A_L,A_R}$, respectively, $./$ denotes element-wise division of entries in two matrices, and $\mathbf{1}$ denotes a vector with all ones.

	A na\"ive approach would require storing all the measurements $Y_{i,j}$, which increases the storage complexity of the system proportional to the number of illumination patterns. The procedure described above avoids this cost, as we can recursively update an estimate of all the matrices and vectors needed for image recovery without any additional storage overhead. Thus, the storage cost of our method remains constant regardless of the number of illumination patterns. 
	Every captured sensor measurement requires some processing, so the computational cost per recovered image increases linearly with the number of illumination patterns. 
	
	The required storage space for all the parameters is $\mathcal{O}(n^2)$ because $\mathbf{Q}, \mathbf{A_L}$, and $\mathbf{A_R}$ are $n\times n$ matrices. 
	We only need to compute $\mathbf{A_L},\mathbf{A_R}$ once, each of which costs $O(mn^2+kn^2)$. Eigendecomposition of $n\times n$ matrices is $O(n^3)$. 
	The most expensive step in our method is computing $\mathbf{Q}$, which we can perform by in-place addition of $(\Phi_L^\top Y_{i,j} \Phi_R) \odot P_{i,j}$ as we acquire measurements for all $i,j$. In this manner, we never need to store any of the captured measurements. The complexity of updating $\mathbf{Q}$ is $\mathcal{O}(k^2 (nm^2+ mn^2))$. 
	
	\subsection{Choice of Illumination Patterns}
	
	One of our goals is to select the $n\times k$ illumination pattern matrices $P_L,P_R$ that maximize the quality of reconstruction for fixed $\Phi_L,\Phi_R$. %
	The quality of reconstruction in \eqref{eq:closed_form} directly depends on the conditioning of the $\mathbf{A_L,A_R}$ matrices in \eqref{eq:tikhonov_sep}, which in turn depends on the mask and illumination patterns.  
	One possible approach to improve the conditioning of $\mathbf{A_L,A_R}$ is to make them diagonal or diagonally dominant \cite{golub1996matrix}, which we can achieve by enforcing the same structures in  $P_LP_L^\top,P_RP_R^\top$. 
	
	In principle, we can make $P_LP_L^\top$ diagonal or even identity by using $P_L$ as an identity matrix, which requires $k=n$. This would be equivalent to scanning the entire scene by illuminating one pixel at a time. We can also make $P_LP_L^\top$ identity by selecting $P_L$ as any orthogonal matrix, which also requires $k=n$. 
	In a practical scenario, we can only use a small number of illumination patterns; therefore, $k \ll n$. Below we discuss how we can get diagonally dominant $\mathbf{A_L,A_R}$ using small values of $k$. 
	
\begin{figure*}[t]
		\centering
		\includegraphics[width=0.8\linewidth, keepaspectratio]{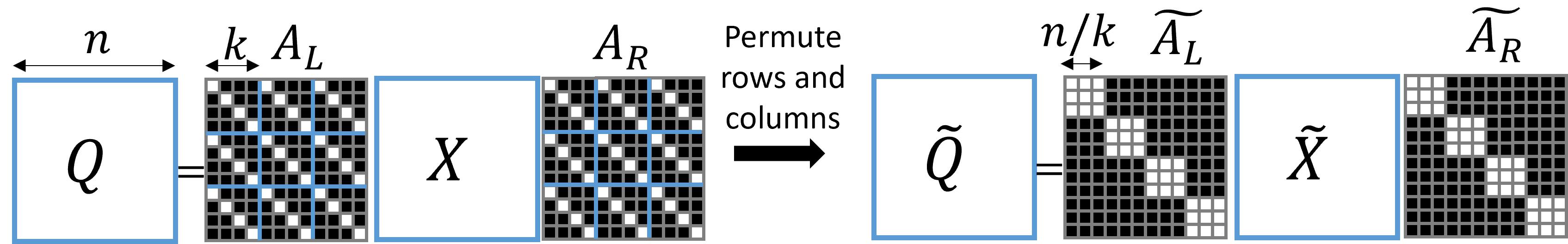}
		\caption{Illustration of system in  \eqref{eq:tikhonov_sep} when the illumination patterns form orthogonal basis over $k\times k$ image patch. (Left) $A_L$ and $A_R$ are $n\times n$ block matrices with diagonal blocks of size $k\times k$. (Right) Permuting rows and columns results in block diagonal matrices with block size $\frac nk \times \frac nk$. The recovery performance of this system depends on the conditioning of each block. We can solve the block diagonal system by recovering $\frac nk \times \frac nk$ patches in $X$ independently, in parallel. }
		\label{fig:Q_permute_overall}
	\end{figure*}

	We propose to use illumination patterns that constitute an orthogonal basis over $k\times k$ blocks and repeat the same patterns across the entire scene. 
	The simplest example of such patterns is a dot pattern in which two adjacent dots are placed $k$ scene pixels apart. We can then shift the dot pattern across horizontal and vertical directions, one pixel at a time,  to capture $k^2$ shifting dots patterns. These shifting dots patterns are separable and orthogonal over every $k\times k$ block. 
	More generally, we can use any sequence of orthogonal separable patterns over $k\times k$ blocks. 
	Let us assume the separable illumination patterns can be drawn from $P_L,P_R$ that are defined as 
	\begin{equation}
	P_L = P_R = 
	\scriptscriptstyle
	\begin{bmatrix}
	\psi_k \\
	\vdots \\
	\psi_k, 
	\end{bmatrix} 
	\displaystyle
	\Rightarrow 
	P_LP_L^\top = P_RP_R^\top=
	\scriptscriptstyle
	\begin{bmatrix}
	I_k & \ldots & I_k \\
	\vdots& \ddots & \vdots \\
	I_k &  \ldots & I_k 
	\end{bmatrix}, 
	\label{eq:shifting_dots}
	\end{equation}
	where $\psi_k$ and $I_k$ denote $k\times k$ orthogonal and identity matrices, respectively. 
	The resulting $P_LP_L^\top, P_RP_R^\top$ matrices (shown above) will be block matrices with $k\times k$ identity blocks, and the  $\mathbf{A_L,A_R}$ matrices (shown in Fig.~\ref{fig:Q_permute_overall}) will be block matrices with $k\times k$ diagonal blocks.

	Recall that $\mathbf{A_L,A_R}$ are system matrices for the linear system we need to solve in \eqref{eq:tikhonov_sep}. We can permute the rows and columns of $\mathbf{Q}$, which is equivalent to permuting rows of $\mathbf{A_L}$  and columns of $\mathbf{A_R}$, without affecting the solution of \eqref{eq:tikhonov_sep}. Let us represent the resulting permuted equations as 
	\begin{equation}
	\widetilde{\mathbf{Q}} = \mathbf{\widetilde{A}_L} X \mathbf{\widetilde{A}_R} + \lambda X.  \label{eq:tikhonov_blkDiag}
	\end{equation}
	We illustrate the permuted system in Fig.~\ref{fig:Q_permute_overall}, where $\mathbf{\widetilde{A}_L}, \mathbf{\widetilde{A}_R}$ represent $n\times n$ block diagonal matrices, with $k$ blocks along the diagonal each of size $\frac n k \times \frac n k$. As we increase the value of $k$, the system matrices $\mathbf{\widetilde{A}_L}, \mathbf{\widetilde{A}_R}$  become diagonally dominant and the overall conditioning of the system improves. 
	
	We can exploit the block diagonal structure of the system matrices to solve the system in \eqref{eq:tikhonov_blkDiag} in a reliable and computationally efficient manner. Note that the separable, block diagonal system can be divided into $k^2$ independent systems, each involving an $\frac n k \times \frac n k$ patch in $\mathbf{X}$. We can solve all these systems in parallel to speed up recovery. The overall complexity of the inversion also reduces from $O(n^3)$ to $O(n^3/k)$. 
	Furthermore, the conditioning of the overall system now depends on the conditioning of each $\frac nk\times \frac nk$ block in $\mathbf{\widetilde{A}_L}, \mathbf{\widetilde{A}_R}$. As long as all the blocks are well-conditioned, we can recover the underlying signal accurately.

	%% Simulations 
	\section{Simulations} 
	\subsection{Simulation Setup}
	\label{sec:simu_setup}
	To validate the performance of the proposed algorithm, we simulate a lensless imaging system where a coded-mask is placed on top of an image sensor. We use a separable maximum length sequence (MLS) mask pattern. The size of each mask feature is 60$\mu$m, and the sensor-mask distance is 2mm. The sensor pitch in the simulation is 11.72$\mu$m and the total number of pixels on the sensors is $512\times512$. We simulate a $128\times128$ planar scene that is 40cm away from the sensor, and the height/width of the scene is 12cm. 
	The simulated sensor noise includes photon noise and read noise\cite{hua2020sweep}, and the noisy sensor measurements can be described as 
	\begin{equation}
	\mathbf{Y}_n = \frac{G}{F}(\text{Poisson}(\frac{F}{G}\mathbf{Y})+N(0,\sigma^2)),
	\end{equation}
	where $\mathbf{Y}$ and $\mathbf{Y}_n$ refers to original and noisy measurements, $F$ stands for the full-well capacity, and $G$ represents the gain value. The variance $\sigma=F\times10^{-R/20}$ and $R$ is the dynamic range.
	We show the reconstruction results on a few example scenes using different illumination patterns; additional results can be found in the supplementary material.

	\subsection{Effect of Illumination on Reconstruction}
	
	We first evaluate the conditioning of different illumination patterns by observing the singular values of the system matrices in \eqref{eq:tikhonov_sep}. The matrices that have flat singular value spectrum provide better recovery performance \cite{asif2017flatcam,antipa2018diffusercam,golub1996matrix}. 
	We tested different types of binary, separable illumination patterns for this experiment. We generate different instances of matrices $P_L,P_R$ and use outer products of all pairs of columns to generate the illumination patterns. We ensure that the union of all the patterns should illuminate all the pixels (i.e., if we add columns of $P_L,P_R$, they should be nonzero everywhere). 
	\textbf{Uniform:} One pattern that illuminates all the pixels simultaneously; $P_L,P_R$ are vectors of all ones. 
	\textbf{Random:} $P_L$ and $P_R$ are $k\times n$ binary random matrices that generate $k^2$ patterns. 
	\textbf{Orthogonal:} We tested two types of orthogonal patterns (shifting dots and repeated Hadamard) that yield identical system matrices in \eqref{eq:tikhonov_blkDiag}. 
	\textbf{Shifting dots:} $P_L,P_R$ are $k\times n$ matrices, each of which consists of $k\times k$ identity matrices stacked on top of each other (as described in \eqref{eq:shifting_dots}). The base illumination pattern consists of dots separated by $k$ pixels along the horizontal and vertical directions. We generate a total of $k^2$ illumination patterns, each of which is a shifted version of the base pattern. The summation of all the patterns will give us a uniform illumination pattern.
	\textbf{ Repeated Hadamard:} As an extension to shifting dots, $P_L,P_R$ are $k\times n$ matrices, each of which consists of the same $k\times k$ orthogonal Hadamard matrix stacked on top of each other. 
	We can use grayscale or color patterns to illuminate the scene. In real experiments, the calibration of the projector and nonlinearity of color/intensity ranges pose additional challenges. 
	
	\begin{figure}
		\begin{subfigure}[b]{0.45\linewidth} 
			\centering
			\includegraphics[width=1\linewidth]{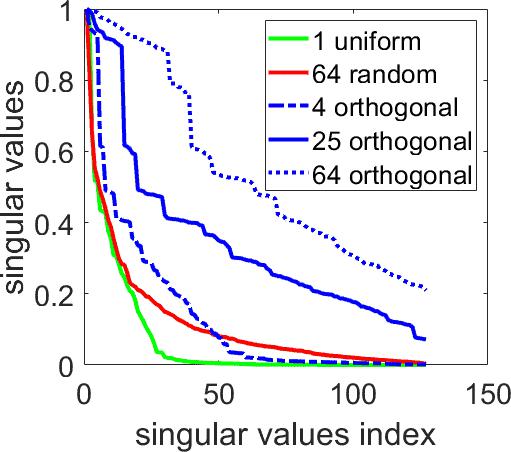}
			\caption{System singular values.}\label{fig:singular_values} 
		\end{subfigure}
		\hfill 
		\begin{subfigure}[b]{0.43\linewidth}
			\includegraphics[width=1\linewidth,keepaspectratio]{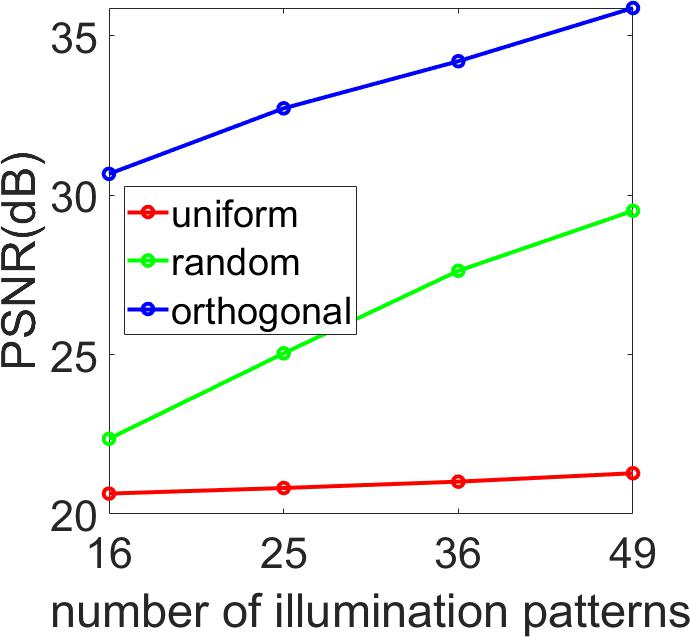}
			\caption{Average PSNR values.  }	\label{fig:simu_illumPattern}
		\end{subfigure}
		\caption{Recovery performance of the imaging system. (a) Singular values of the system matrices with uniform, 64 random, and 4, 25, and 64 orthogonal shifting dots patterns. (b) Average PSNR of 8 test images  reconstructed with  different numbers of illumination patterns. Red dashed line shows results with uniform illumination. Reconstruction quality improves as we increase the number of illumination patterns. The orthogonal patterns outperform random and uniform patterns. }
	\end{figure} 
	
	To evaluate the effect of illumination on the lensless imaging system, we observe the decay of singular values of the system matrices in \eqref{eq:tikhonov_sep} as we increase the number of illumination patterns. Figure~\ref{fig:singular_values} plots the singular values for different illumination patterns. The singular values of the original system matrix with one uniform illumination decay sharply. We tested various patterns and found that the shifting dots or orthogonal patterns provide the best overall conditioning (flat SVD curve) for a given budget of measurements. As we increase the number of illumination patterns, the singular values spectrum becomes flatter, which corresponds to a system that is nearly orthogonal. 
	In principle, we can use a dot projector to create shifting dots patterns, but we need a programmable projector for Hadamard-like patterns. In our experiments, we use a laser projector for illumination.

	We present simulation results for reconstruction of 8 test images using different types and number of illumination patterns in Fig.~\ref{fig:simu_illumPattern}. We observe that orthogonal patterns outperform other patterns in terms of PSNR. Additional simulation results are available in the supplementary material.

	\subsection{Effect of Sensor-to-Mask Distance}
\begin{figure}[t]
		\setlength\tabcolsep{1pt}
		\renewcommand{\arraystretch}{1} % Default value: 1
		\footnotesize
		\begin{subfigure}[b]{0.68\linewidth}
			% \centering
			\begin{tabular}{cccc}
				750$\mu$m &
				2000$\mu$m &
				750$\mu$m &
				2000$\mu$m
				\\
				\rotatebox{90}{\parbox{1.4cm}{\centering  uniform}}
				\includegraphics[width=0.22\linewidth,keepaspectratio]{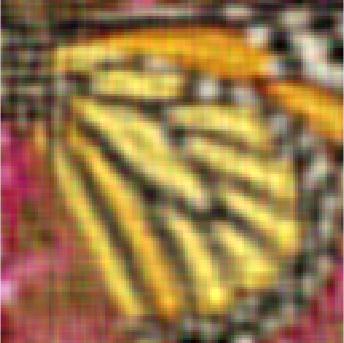} &
				\includegraphics[width=0.22\linewidth,keepaspectratio]{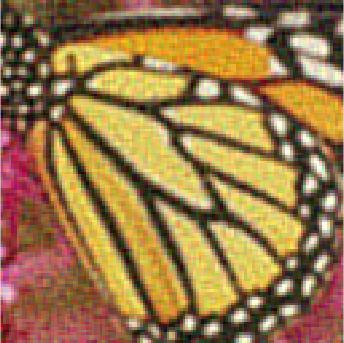} &
				\includegraphics[width=0.22\linewidth,keepaspectratio]{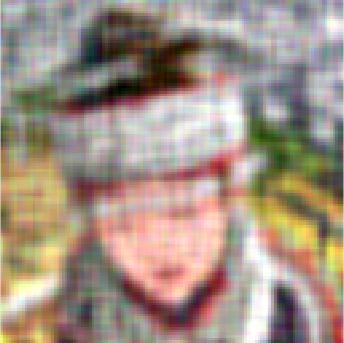} &
				\includegraphics[width=0.22\linewidth,keepaspectratio]{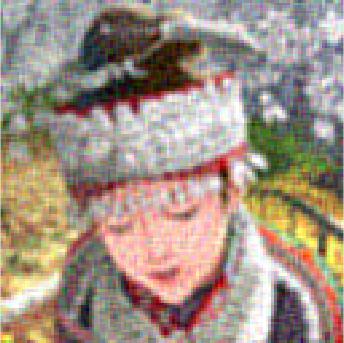} 
				\\
				PSNR: 16.87dB &
				21.76dB &
				18.96dB & 
				22.40dB
				\\
				\rotatebox{90}{\parbox{1.4cm}{\centering  49 shift}}
				\includegraphics[width=0.22\linewidth,keepaspectratio]{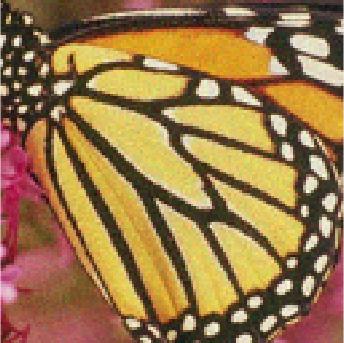} &
				\includegraphics[width=0.22\linewidth,keepaspectratio]{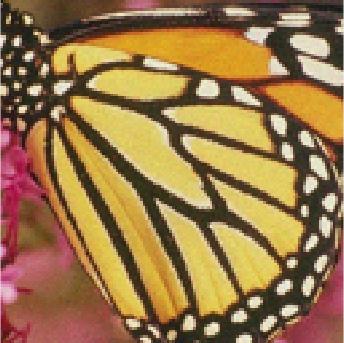}  &
				\includegraphics[width=0.22\linewidth,keepaspectratio]{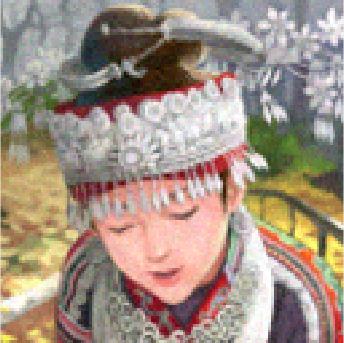} &
				\includegraphics[width=0.22\linewidth,keepaspectratio]{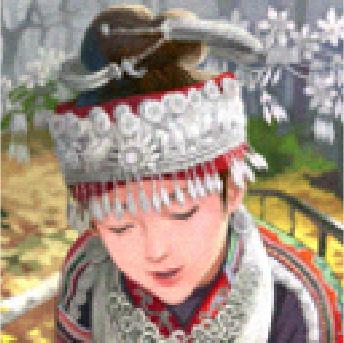} 
				% 		&
				\\
				PSNR: 30.44dB &
				35.33dB &
				29.93dB &
				34.39dB
			\end{tabular}
			\caption{Examples of reconstructed images with sensor-to-mask distance at 750$\mu$m and 2000$\mu$m. }
		\end{subfigure}
		~
		\begin{subfigure}[b]{0.28\linewidth}
			% \centering
			\includegraphics[width=1\linewidth,keepaspectratio]{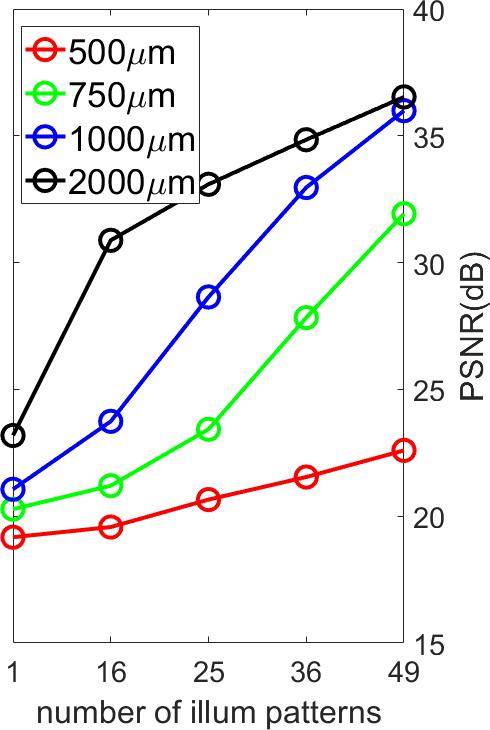}
			\caption{Average PSNR of all test images. }
		\end{subfigure}
		\caption{Simulation results for reconstruction with different sensor-to-mask distances in uniform illumination and shifting dots illumination patterns. The reconstruction quality improves as the sensor-to-mask distance increases. }
		\label{fig:simu_sensor2mask}
	\end{figure}

	The sensor-to-mask distance of a lensless imaging system greatly influences the conditioning of the system and the quality of reconstruction. We present simulation results to evaluate the performance with different sensor-to-mask distances.
	We keep the sensor size fixed at $512\times512$ pixels, and test the sensor-to-mask distance at  $500\mu$m, $750\mu$m, $1000\mu$m, and $2000\mu$m. The reconstruction results for two test images are presented in Fig.~\ref{fig:simu_sensor2mask}, along with the PSNR plot for different numbers of shifting dots patterns. We observe that multiple coded illumination patterns outperform the results with a single uniform pattern at all the sensor-to-mask distances. Even if the sensor-to-mask distance is 750$\mu$m, the results with 25 coded illumination patterns are comparable to the uniform illumination results at 2000$\mu$m. Additional simulation results are available in the supplementary material.
	
	\subsection{Comparison with Multishot Lensless Methods}
\begin{figure}[t]
		\setlength\tabcolsep{1pt}
		\renewcommand{\arraystretch}{1} % Default value: 1
		\footnotesize
		\begin{subfigure}[b]{0.64\linewidth}
			% \centering
			\begin{tabular}{cccc}
				uniform &
				SweepCam\cite{hua2020sweep} &
				ours
				\\
				\includegraphics[width=0.3\linewidth,keepaspectratio]{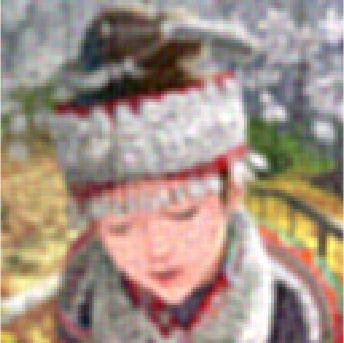} &
				\includegraphics[width=0.3\linewidth,keepaspectratio]{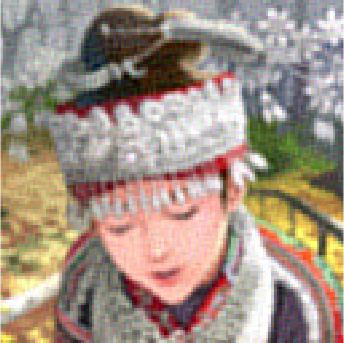} &
				\includegraphics[width=0.3\linewidth,keepaspectratio]{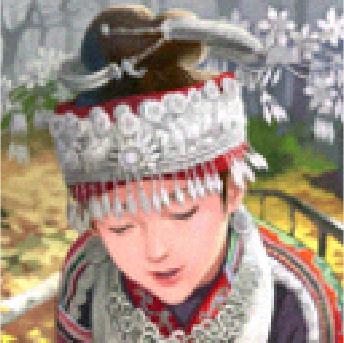} &
				\\
				PSNR: 22.85dB &
				26.83dB & 
				34.01dB
				\\
				\\
				\includegraphics[width=0.3\linewidth,keepaspectratio]{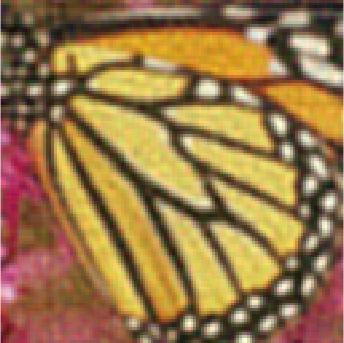} &
				\includegraphics[width=0.3\linewidth,keepaspectratio]{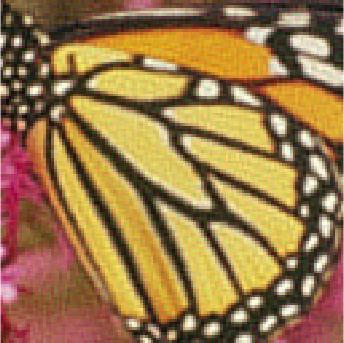}  &
				\includegraphics[width=0.3\linewidth,keepaspectratio]{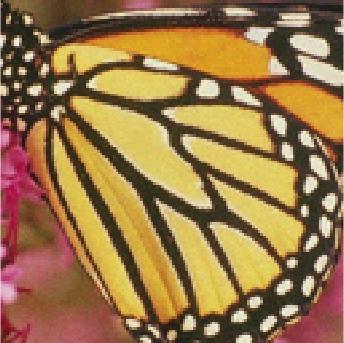} &
				% 		&
				\\
				PSNR: 21.29dB &
				24.88dB &
				35.53dB
			\end{tabular}
			\caption{Example results with uniform illumination, 49 shifting masks, and 49 shifting dots. }
		\end{subfigure}
		\begin{subfigure}[b]{0.34\linewidth}
			% \centering
			\includegraphics[width=1\linewidth,keepaspectratio]{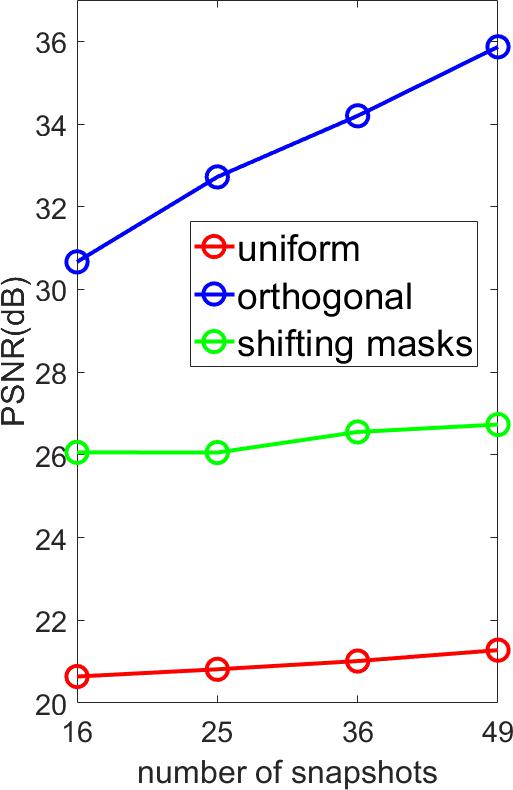}
			\caption{Average PSNR of all test images. }
		\end{subfigure}
		\caption{Simulation results for reconstruction using uniform illumination, shifting masks from SweepCam\cite{hua2020sweep} and shifting dots illumination patterns using our method at different number of measurements instances. Our method outperform uniform the other methods. }
		\label{fig:simu_methods}
	\end{figure}

	Multishot lensless methods are also applied in \cite{hua2020sweep,zheng2021simple}, where multiple frames of lensless measurements are captured with a shifting or programmable mask on top of the image sensor.
	We present the simulation results comparing with a shifting mask-based multishot lensless imaging system~\cite{hua2020sweep} in Fig.~\ref{fig:simu_methods}. We keep the sensor size fixed at $512\times512$ and sensor-to-mask distance at $2000\mu$m. The mask feature size is $60\mu$m in all cases. We simulate the imaging process for SweepCam~\cite{hua2020sweep} with the mask at multiple shifting positions between $-15$ and $+15$ mask feature pixels (i.e., $-900\mu$m to $+900\mu$m physical shift). The number of captured frames for SweepCam~\cite{hua2020sweep} are the same as our method with coded illumination patterns. The simulation results in Fig.~\ref{fig:simu_methods} show that our method with the same number of coded illumination patterns provides significantly better results compared to SweepCam.

	\section{Experiments}
	\subsection{Experiment Setup}
	\label{sec:exp_setup}
\begin{figure}[thb]
		\centering
		\includegraphics[width=1\linewidth]{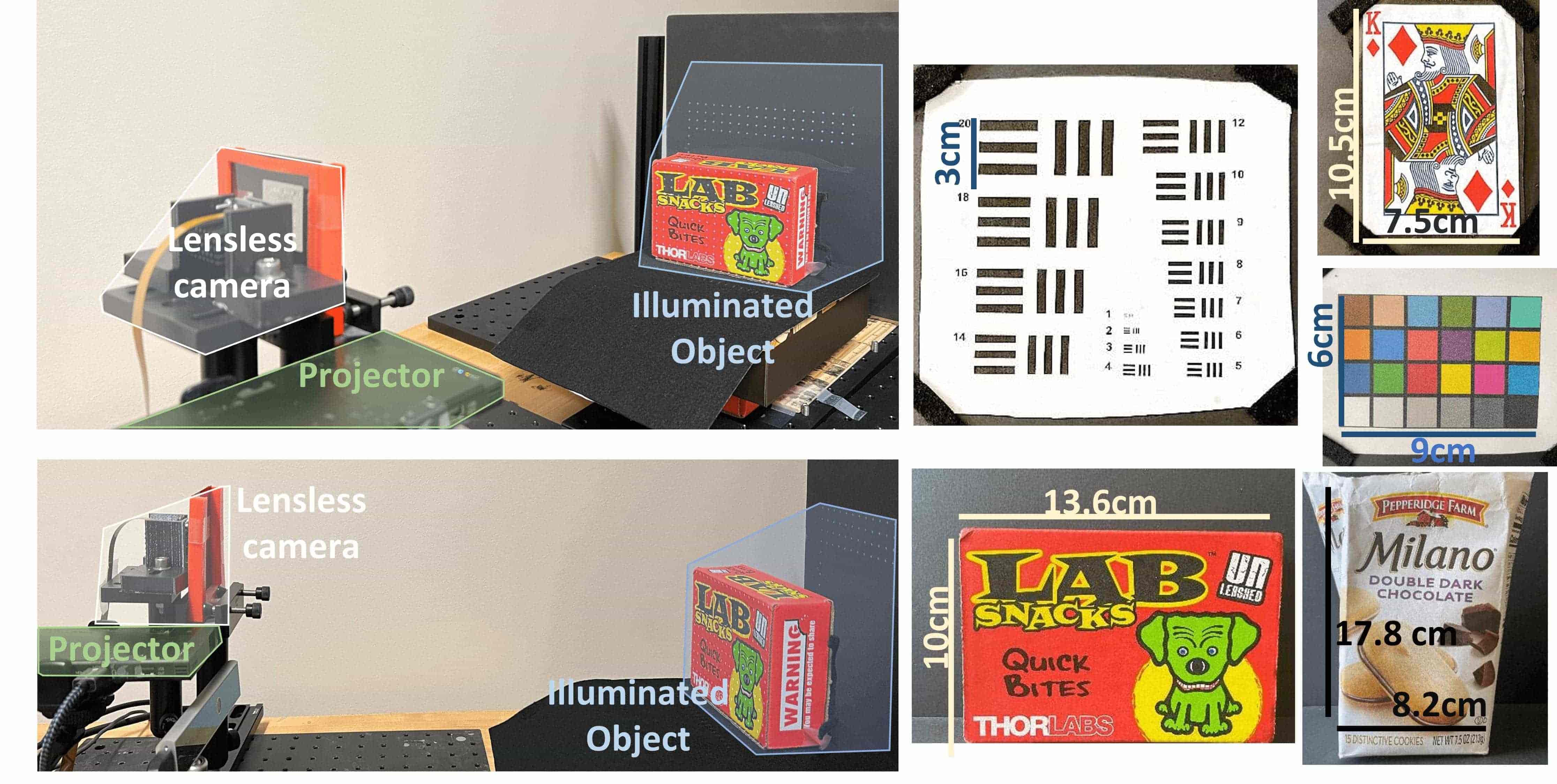}
		\caption{The experiment setup and five test scenes (annotated and scaled to proportional size). The projector is placed right next to the lensless camera. The target scenes/objects are 40cm away from the camera. }
		\label{fig:exp_hardware}
	\end{figure}

	We build a prototype with a lensless camera and a Sony MP-CL1 laser projector, shown in Fig.~\ref{fig:exp_hardware}. The lensless camera prototype consists of an image sensor with a coded mask on top of it. The mask has a separable MLS pattern described in \cite{asif2017flatcam}. The mask has $511\times 511$ square features, each of length/width 60$\mu$m. The sensor-mask distance is 2mm. We use a Sony IMX249 sensor that has $1920\times1200$ pixels with 5.86$\mu$m pixel pitch. We bin $2\times 2$ sensor pixels and record $512\times 512$ measurements from the center of the sensor. The effective sensor pitch is 11.72$\mu$m and the effective sensor area is nearly $6\times6$ mm. The target objects are 40cm away from the camera, and the projector illuminates $12\times12$ cm area on the scene plane. 
	Finally, we reconstruct $128\times128$ pixels in the illuminated area, which results in the effective sampling interval of 120mm/128 = 0.93mm per pixel in the reconstructed images.

	In our experiment, the pixel grid of the scene, illumination patterns $P_L$, $P_R$, the system matrices $\Phi_L$, $\Phi_R$ must be correctly aligned; otherwise, we will get artifacts in the reconstruction. To avoid any grid mismatch, we use the same projector to calibrate the system matrices and generate the illumination patterns in our experiments. 
	{We estimate the system matrices $\Phi_L,\Phi_R$ following the Hadamard pattern-based calibration procedure in \cite{asif2017flatcam}. Let us denote the Hadamard patterns as an $n\times n$ orthogonal matrix $\mathbf{H} = [\mathbf{h}_1, \ldots, \mathbf{h}_n]$, where each $\mathbf{h}_i$ represents a Hadamard pattern of length $n$. To calibrate the system for an $n\times n$ pixel grid, we project $n$ horizontal and $n$ vertical (rank-one) Hadamard patterns on a flat surface in the scene and record their response on the sensor. Every horizontal pattern can be represented as an $n\times n$ rank-one matrix $X_i = \mathbf{h}_i \mathbf{1}^\top$, where $\mathbf{1}$ denotes a vector of all ones. The corresponding sensor response can be represented as a rank-one matrix $Y_i = \Phi_L (\mathbf{h}_i \mathbf{1}^\top)  \Phi_R^\top \equiv \mathbf{u}_k \mathbf{v}^\top$, where $\mathbf{u}_i = \Phi_L \mathbf{h}_i$ and $\mathbf{v} = \Phi_R \mathbf{1}$. We can concatenate all the $\mathbf{u}_i$ as columns in a matrix as $\mathbf{U} = [\mathbf{u}_1 ~\cdots ~ \mathbf{u}_n] = \Phi_L \mathbf{H}$ and estimate $\Phi_L = \mathbf{UH^\top}$. We can repeat the same procedure with vertical Hadamard patterns $\mathbf{1h}_i^\top$ and estimate $\Phi_R$.}
	Finally, we conduct the experiments with coded illumination while the position and angle of the projector are fixed. This ensures that the pixel grids of the projector and the transfer matrices are identical.
	
	\newcommand{\figwidth}{0.24\linewidth}
	\begin{figure}[t]
		\setlength\tabcolsep{1pt}
		\centering
		\footnotesize
		\begin{tabular}{cccc}
			Uniform &
			16 Shifting dots &
			25 Shifting dots &
			49 Shifting dots 
			\\
			\includegraphics[width=\figwidth,keepaspectratio]{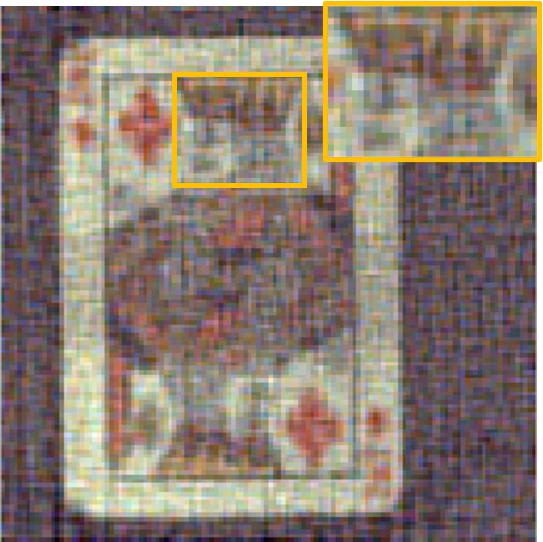} &
			\includegraphics[width=\figwidth,keepaspectratio]{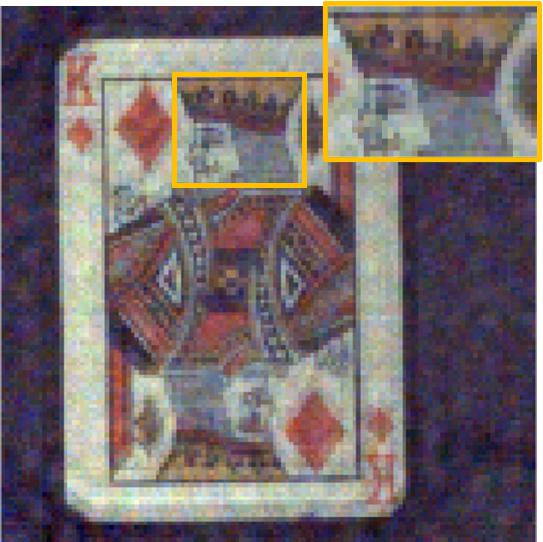} &
			\includegraphics[width=\figwidth,keepaspectratio]{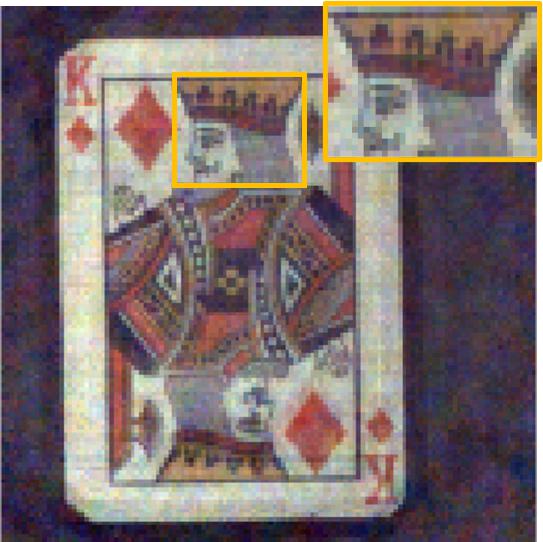} &
			\includegraphics[width=\figwidth,keepaspectratio]{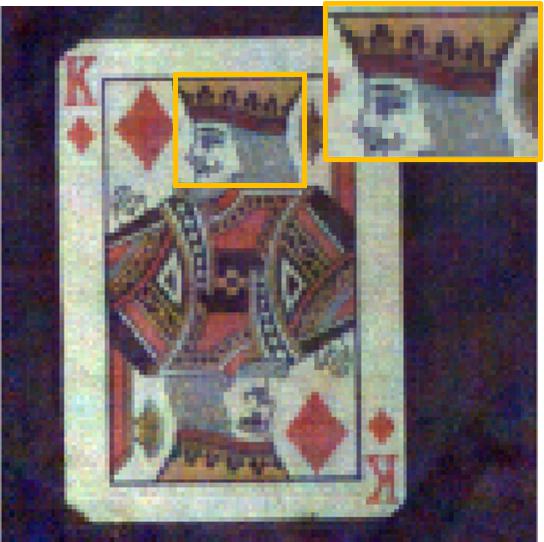} \\
			\includegraphics[width=\figwidth,keepaspectratio]{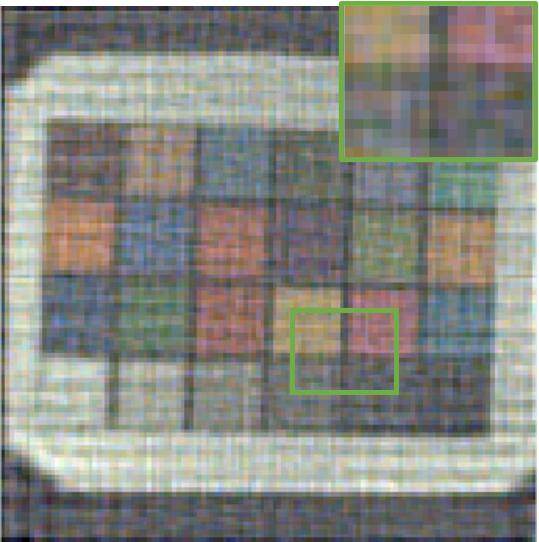} &
			\includegraphics[width=\figwidth,keepaspectratio]{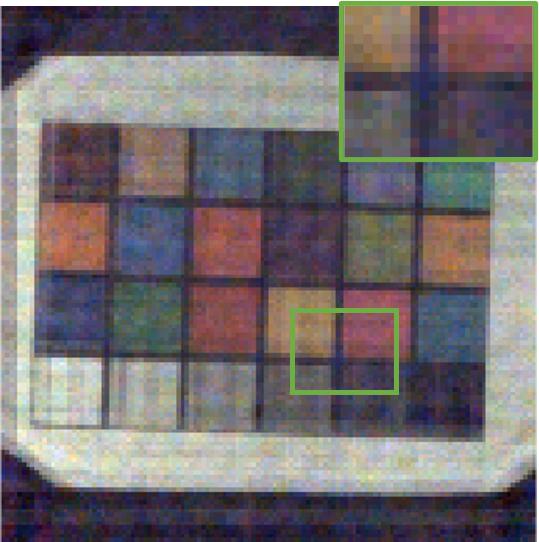} &
			\includegraphics[width=\figwidth,keepaspectratio]{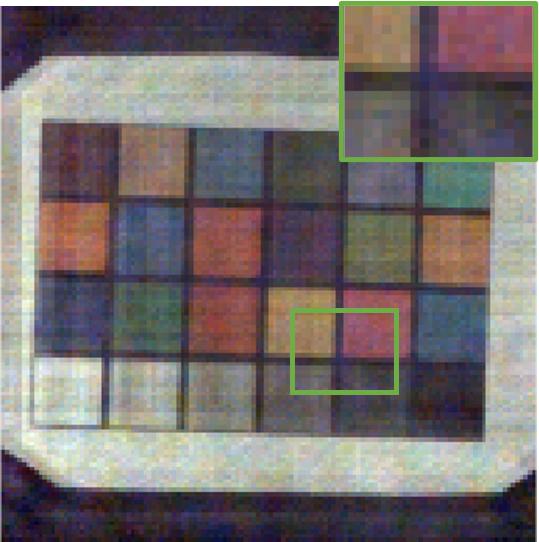} &
			\includegraphics[width=\figwidth,keepaspectratio]{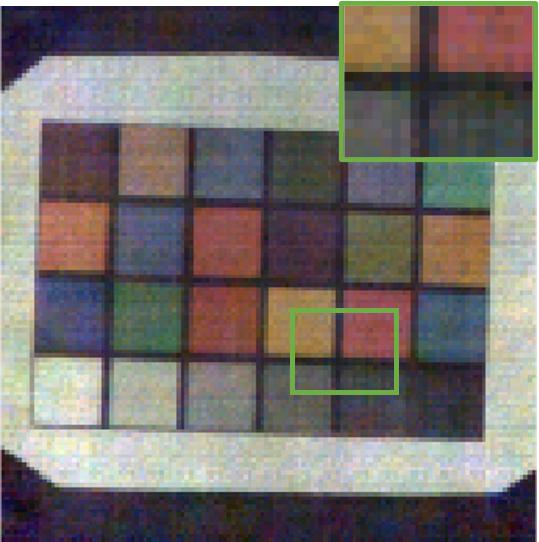} \\
			\includegraphics[width=\figwidth,keepaspectratio]{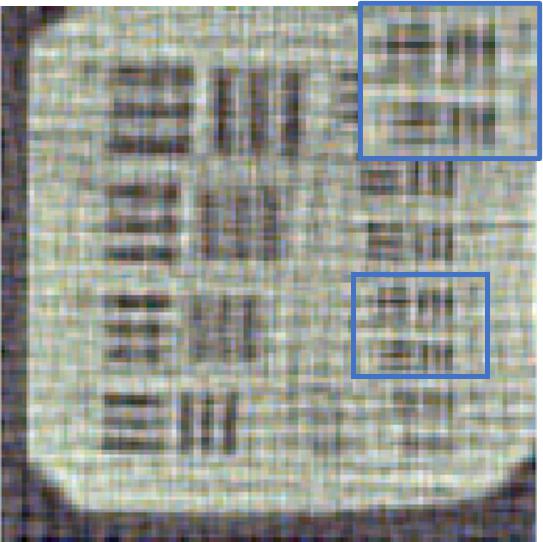} &
			\includegraphics[width=\figwidth,keepaspectratio]{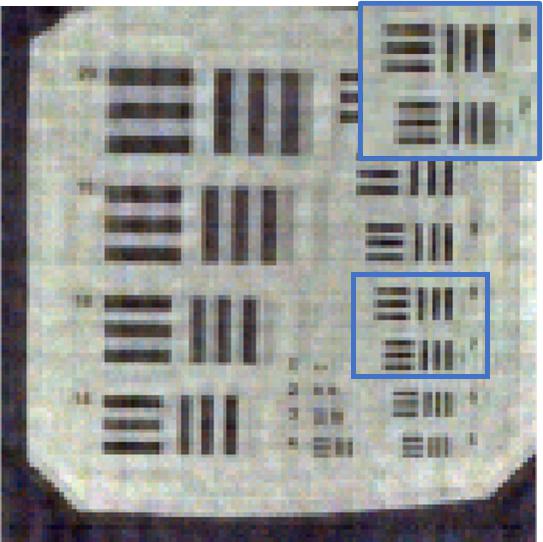} &
			\includegraphics[width=\figwidth,keepaspectratio]{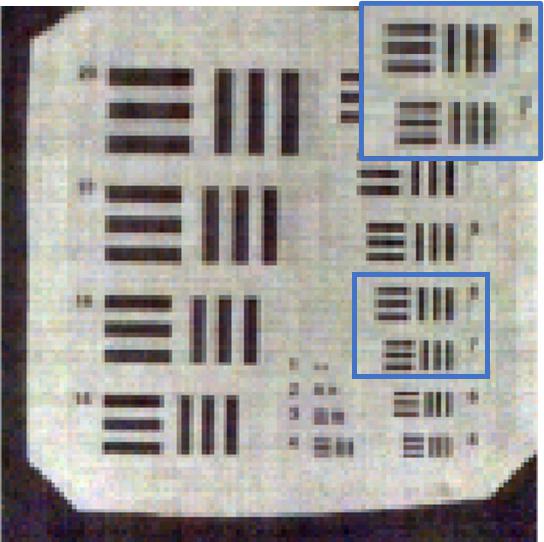} &
			\includegraphics[width=\figwidth,keepaspectratio]{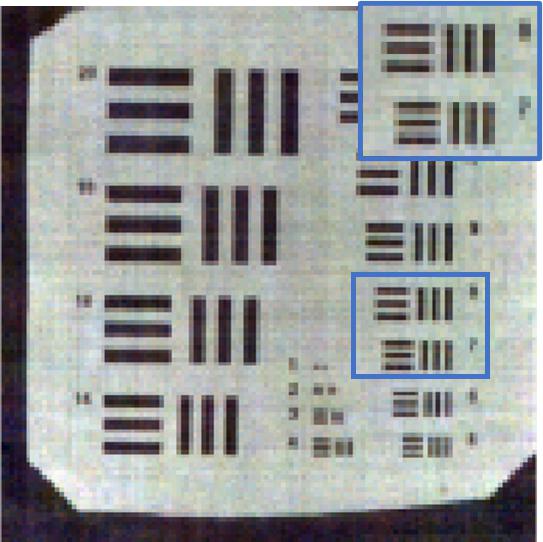} 
			\\
			\includegraphics[width=\figwidth,keepaspectratio]{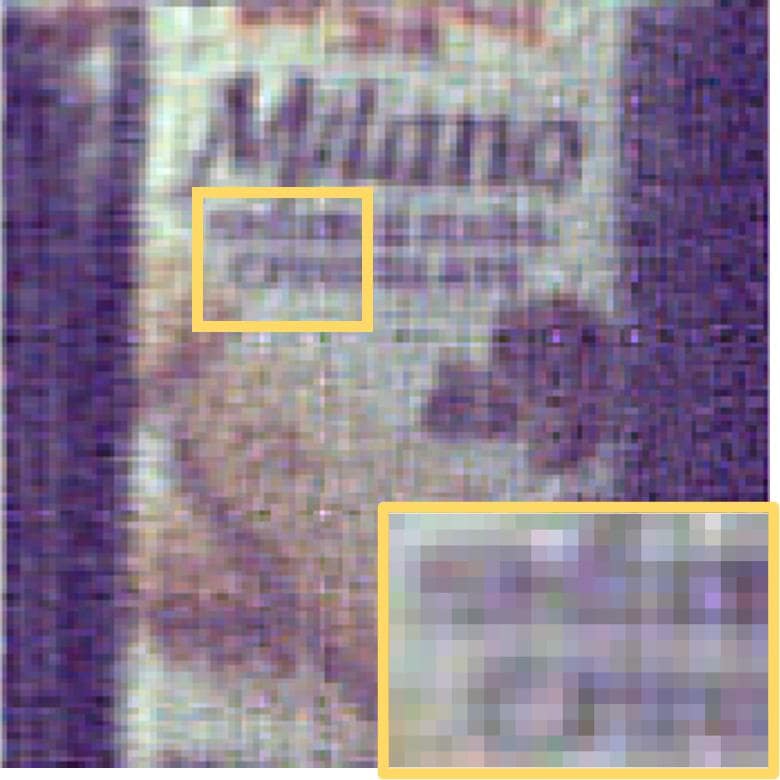} &
			\includegraphics[width=\figwidth,keepaspectratio]{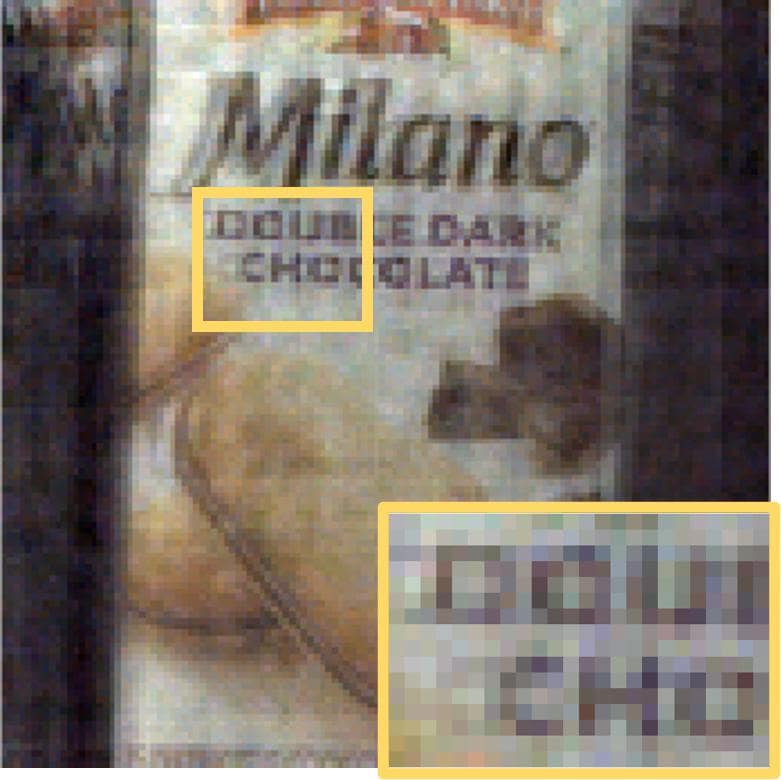} &
			\includegraphics[width=\figwidth,keepaspectratio]{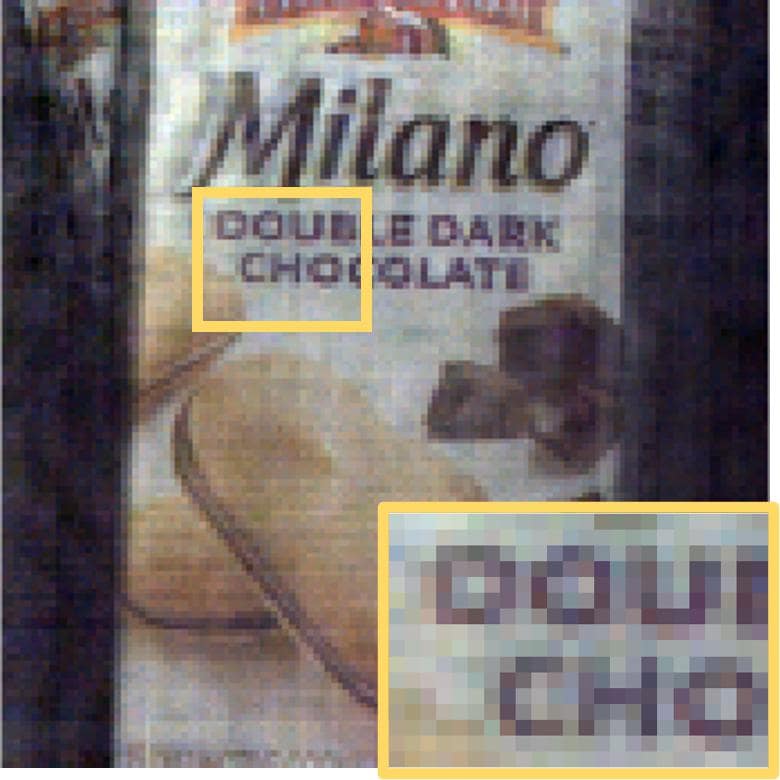} &
			\includegraphics[width=\figwidth,keepaspectratio]{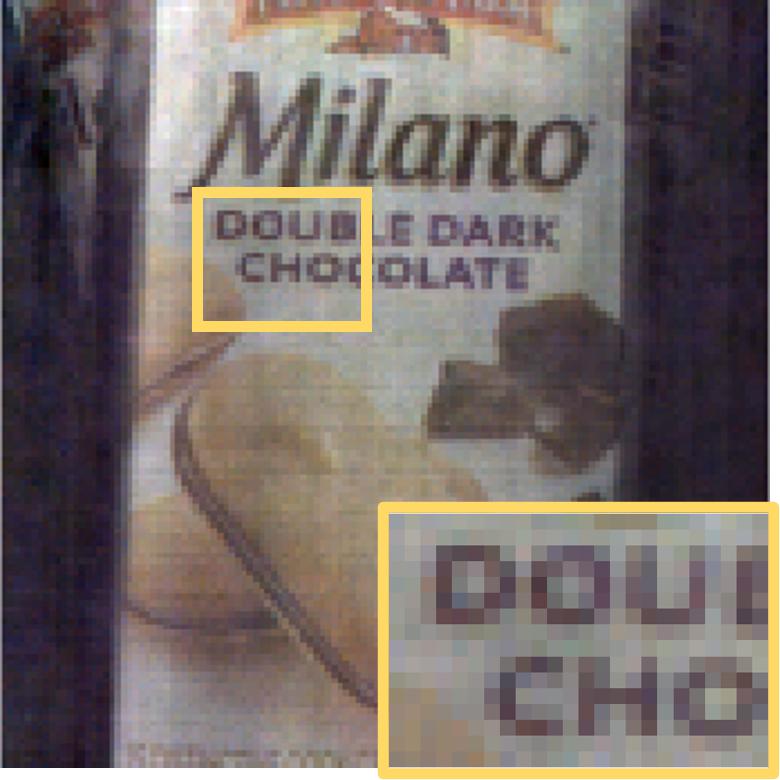}\\
			
			\includegraphics[width=\figwidth,keepaspectratio]{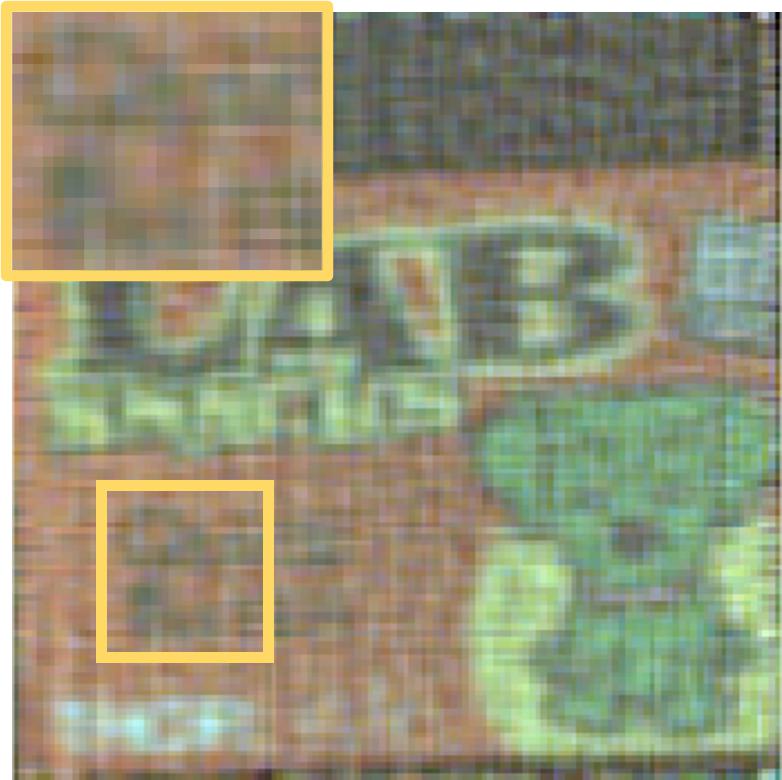} &
			\includegraphics[width=\figwidth,keepaspectratio]{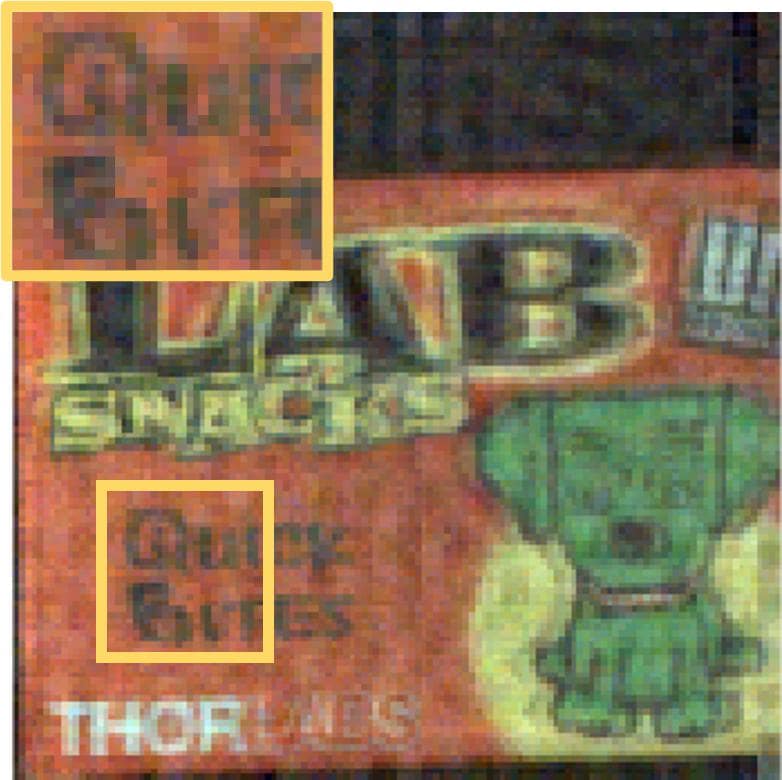} &
			\includegraphics[width=\figwidth,keepaspectratio]{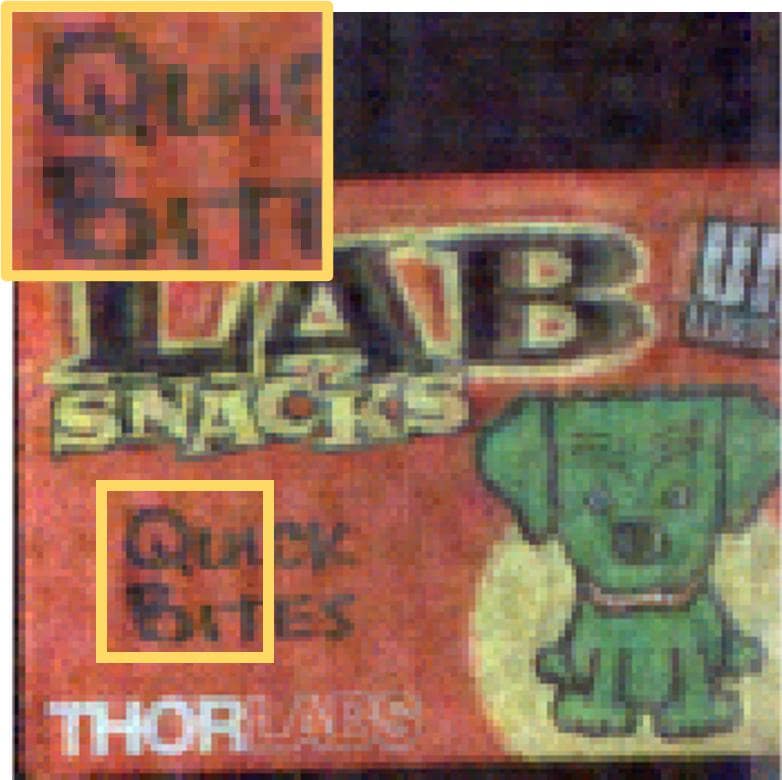} &
			\includegraphics[width=\figwidth,keepaspectratio]{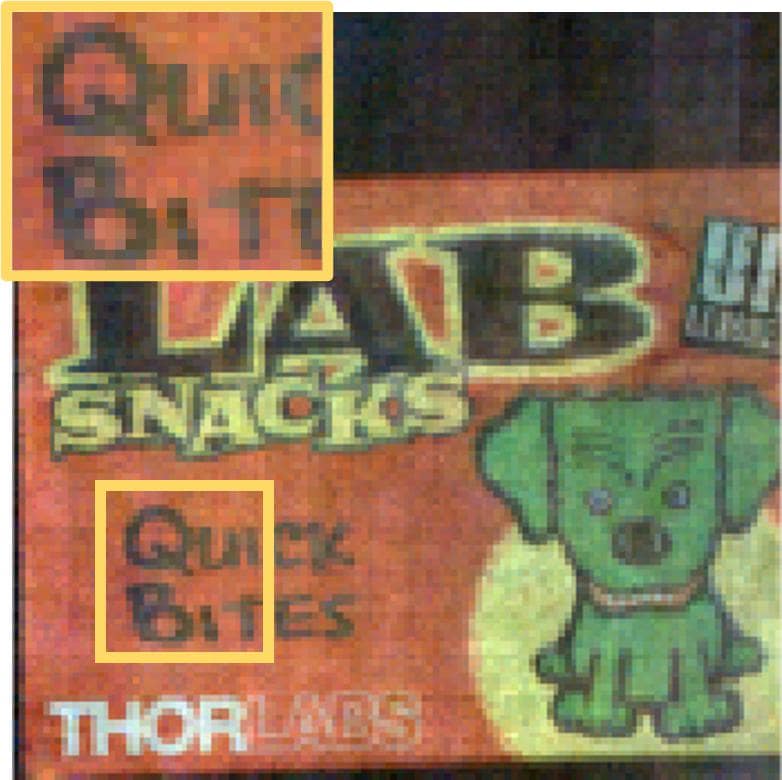}\\
			
		\end{tabular}
		\caption{Experimental results on five test scenes with different numbers of shifting dots patterns.
			The reconstruction results using multiple coded illumination patterns outperform the standard lensless camera with uniform illumination. The quality of reconstruction gradually increases as the number of illumination patterns increases. }
		\label{fig:experiment_results_pattern_nums}
	\end{figure}

	\renewcommand{\figwidth}{0.24\linewidth}
	\begin{figure}[t]
		\setlength\tabcolsep{1pt}
		\centering
		\footnotesize
		\begin{tabular}{cccc}
			Uniform &
			64 Random &
			64 Shifting dots &
			64 Hadamard 
			\\
			\includegraphics[width=\figwidth,keepaspectratio]{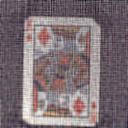} &
			\includegraphics[width=\figwidth,keepaspectratio]{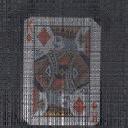} &
			\includegraphics[width=\figwidth,keepaspectratio]{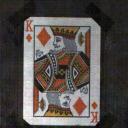} &
			\includegraphics[width=\figwidth,keepaspectratio]{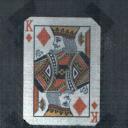} \\
			\includegraphics[width=\figwidth,keepaspectratio]{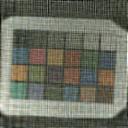} &
			\includegraphics[width=\figwidth,keepaspectratio]{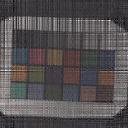} &
			\includegraphics[width=\figwidth,keepaspectratio]{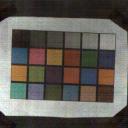} &
			\includegraphics[width=\figwidth,keepaspectratio]{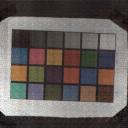} 
			\\
			\includegraphics[width=\figwidth,keepaspectratio]{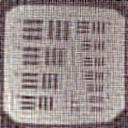} &
			\includegraphics[width=\figwidth,keepaspectratio]{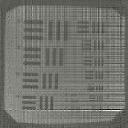} &
			\includegraphics[width=\figwidth,keepaspectratio]{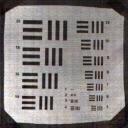} &
			\includegraphics[width=\figwidth,keepaspectratio]{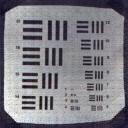} \\
			\includegraphics[width=\figwidth,keepaspectratio]{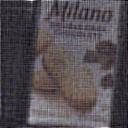} &
			\includegraphics[width=\figwidth,keepaspectratio]{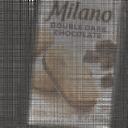} &
			\includegraphics[width=\figwidth,keepaspectratio]{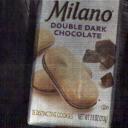} &
			\includegraphics[width=\figwidth,keepaspectratio]{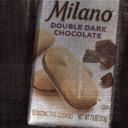}\\		
			\includegraphics[width=\figwidth,keepaspectratio]{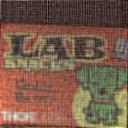} &
			\includegraphics[width=\figwidth,keepaspectratio]{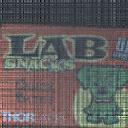} &
			\includegraphics[width=\figwidth,keepaspectratio]{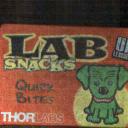} &
			\includegraphics[width=\figwidth,keepaspectratio]{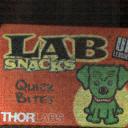}\\
			
		\end{tabular}
		\caption{Experimental results on five test scenes with different types of illumination patterns. The orthogonal patterns (shifting dots, hadamard) outperform random patterns and uniform illumination.
		}
		\label{fig:experiment_results_pattern_types}
	\end{figure} 
	
	\subsection{Effect of Illumination Patterns}
	We present experimental results of our method on five different scenes in Fig.~\ref{fig:experiment_results_pattern_nums} and Fig.~\ref{fig:experiment_results_pattern_types}. The first three rows are printed card, color board, and a resolution target on sheets of paper. Since these scenes are planar, they fit the 2D scene assumption of our model perfectly. The last two rows are real objects with depth variations and may cause depth mismatch in the reconstruction.
	
	From the results in Fig.~\ref{fig:experiment_results_pattern_nums}, we observe that a single uniform illumination provides poor reconstruction. 
	As we increase the number of illumination patterns, the resolution of the reconstruction images improves significantly. 
	For example, in the third row, the horizontal and vertical features that cannot be resolved with the uniform illumination are easily resolved with the 49 patterns. Also, we observe in the fourth row that, the letters on the cookie bag that are completely unrecognizable with uniform illumination become very clear with 49 illumination patterns. The last two rows also suggest that our method is robust to small variations in depth. 
	From the results in Fig.~\ref{fig:experiment_results_pattern_types}, the repeated orthogonal patterns(shifting dots and Hadamard) outperform other patterns, as expected from the singular values in  Fig.~\ref{fig:singular_values}.
	
	Note that the neighboring pixels have a similar response on the sensor in the lensless imaging system; therefore, neighboring pixels are harder to resolve compared to two pixels that are far from each other. The shifting dots pattern is a dot array where the illuminated pixels are maximally separated; therefore, two neighboring pixels in the scene are not illuminated at the same time. 
	Also, as we increase the number of shifting dots patterns, the distance between adjacent illuminated pixels also increases, and that provides better reconstruction. Hadamard patterns provide similar separation because of their orthogonal structure.

\begin{figure}
		\centering
		\begin{subfigure}[t]{0.4\linewidth} 
			% \centering
			\includegraphics[width=1\linewidth]{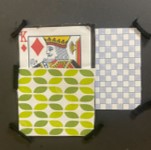}
			\caption{Test scene}
		\end{subfigure}
		\begin{subfigure}[t]{0.4\linewidth} 
			% \centering
			\includegraphics[width=1\linewidth]{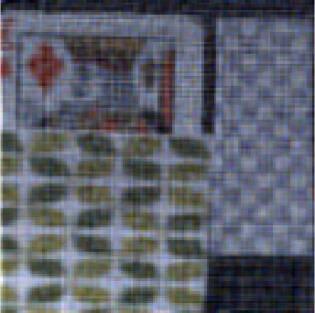}
			\caption{1 Uniform}
		\end{subfigure}
		
		\begin{subfigure}[t]{0.4\linewidth} 
			% \centering
			\includegraphics[width=1\linewidth]{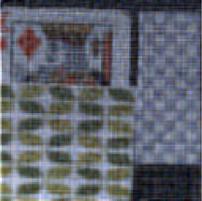}
			\caption{49 Uniform}
		\end{subfigure}
		\begin{subfigure}[t]{0.4\linewidth}
			% \centering
			\includegraphics[width=1\linewidth,keepaspectratio]{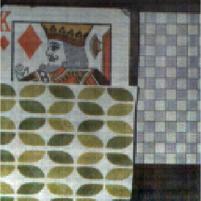}
			\caption{49 Shifting dots}
		\end{subfigure}
		\caption{Sample results for imaging performance with one uniform, 49 uniform, and 49 shifting dots illumination patterns. Capturing 49 shots requires the same data acquisition time, but the results for 49 shifting dots patterns are significantly better than 49 uniform patterns.}
		\label{fig:experiment_results_equal_exposure}
	\end{figure}

	Note that if we increase the number of measurements with coded illumination, the system conditioning and the quality of reconstructed images improve. On the other hand, if we increase the number of measurements for the uniform illumination pattern, the system conditioning and the quality of reconstructed images remains unchanged. To show this effect, we provide experimental results comparing the quality of reconstruction with multiple measurements captured under uniform and coded illumination. We present the results using 49 uniform and 49 shifting dots illumination patterns in Fig.~\ref{fig:experiment_results_equal_exposure}. Note that the 49 uniform and 49 shifting dots illuminations use the same amount of exposure/capture time. We observe that shifting dots illumination provides significantly better reconstruction compared to uniform illumination. 
	We observe that capturing 49 shots with uniform illumination reduces noise slightly compared to one uniform illumination, but 49 shifting dots provide significantly superior results. Note that multiple shots with uniform illumination can provide slightly better SNR because noise variance reduces due to averaging. The improvement with shifting dots patterns is the result of the better conditioning of the system that multiple uniform illumination shots cannot provide.  
	
	In summary, the ill-conditioned system matrices with uniform illumination pattern cause various artifacts. Capturing measurements from coded illumination improves the conditioning of the overall system and the resulting reconstructed images contain much fewer artifacts and noise. More illumination patterns demand more acquisition time, which enforces a trade-off between the quality of reconstruction and data acquisition time.

	\subsection{Effective Resolution and MTF}
	We use the resolution target image in Fig.~\ref{fig:target_line_plots} to analyze the spatial resolution of our method empirically. The distance between two printed white stripes in group 20 (upper-left) is 7.5mm (0.13 lp/mm) and the distance between two white stripes in group 5 is 1.9mm (0.52 lp/mm). 
	The resolution of our imaging system is directly determined by the sampling interval of the illumination patterns. We place the target scene 40cm away from the camera and projector. The resolution of the MP-CL1 projector is $1280\times720$ and the overlap between the camera FOV and the maximum illuminating area at 40cm throw distance is $22\times13$cm. The projector pixel pitch is 0.17mm, which determines the achievable resolution of our method with the MP-CL1 projector. In our experiments, the width of every reconstructed pixel is 0.93mm, and the angular sampling of our system is 0.27$^{\circ}$. We can select smaller sampling intervals pitch by dividing the illuminating area into more pixels, but the minimal size cannot be smaller than projector pixel.

	We present the modulation transfer function (MTF) of the reconstruction from different numbers of illumination patterns in Fig.~\ref{fig:target_line_plots}. MTF measures how well we can discern the intensity of bright and dark pixels in line pairs under different conditions \cite{williams1989MTF}. To plot the MTF, we manually select every group of horizontal and vertical line pairs after subtracting a fixed DC background, then average every group along the columns and rows, respectively, and finally compute the contrast percentage as  $\frac{I_\text{max}-I_\text{min}}{I_\text{max}+I_\text{min}}\times 100$. We observe from the MTF plots that the contrast ratio for all the resolution groups improves as we increase the number of illumination patterns. 
	We also plot the intensity values of the horizontal line stripes from multiple groups in  Fig.~\ref{fig:target_line_plots}, where the peaks and valleys correspond to the white and black stripes of the horizontal line pairs. We observe that the line pairs of group 12 can be distinguished clearly while the line pairs of group 3 cannot be distinguished. 

	\renewcommand{\figwidth}{0.3\linewidth}
	\begin{figure}[t]
		\centering
		\centering
		\begin{subfigure}[t]{\figwidth}
			\includegraphics[width=1\linewidth]{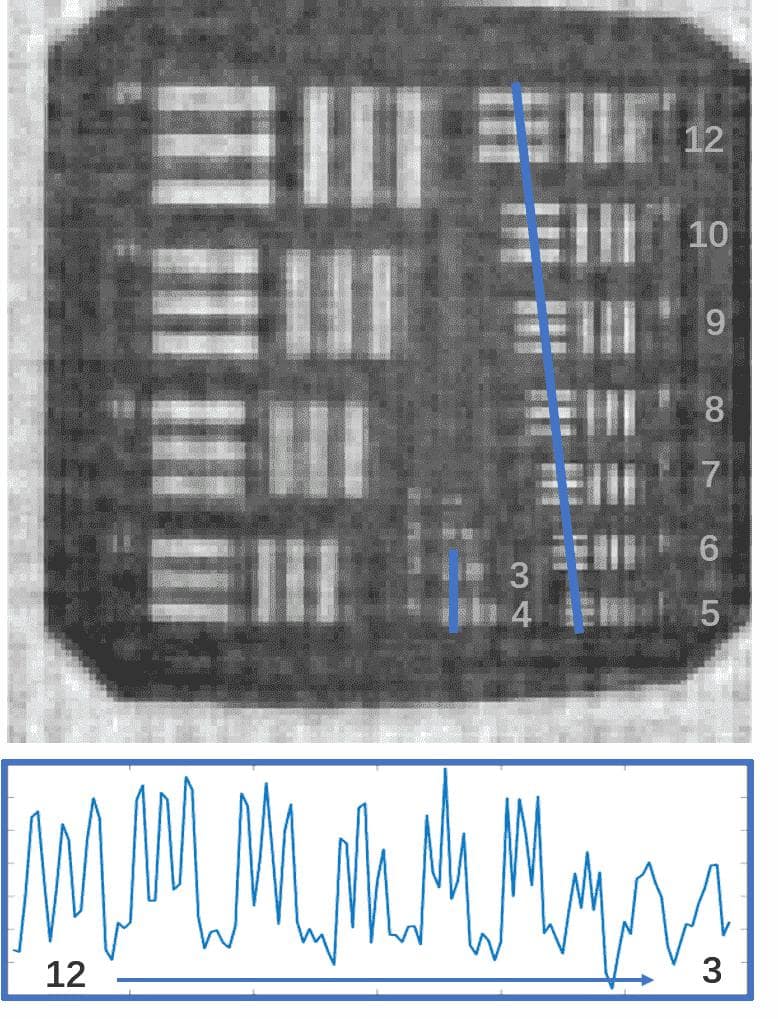}
			\caption{ 9 patterns }
		\end{subfigure}
		~
		\begin{subfigure}[t]{\figwidth}
			\includegraphics[width=1\linewidth]{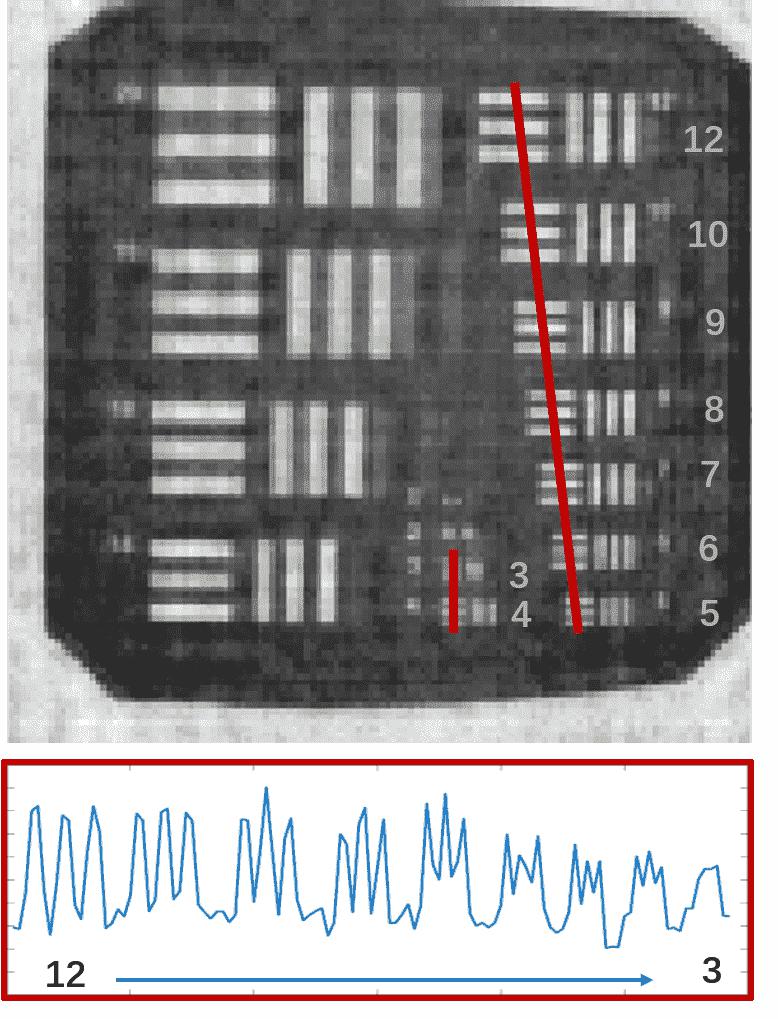}
			\caption{ 16 patterns }
		\end{subfigure}
		~
		\begin{subfigure}[t]{\figwidth}
			\includegraphics[width=1\linewidth]{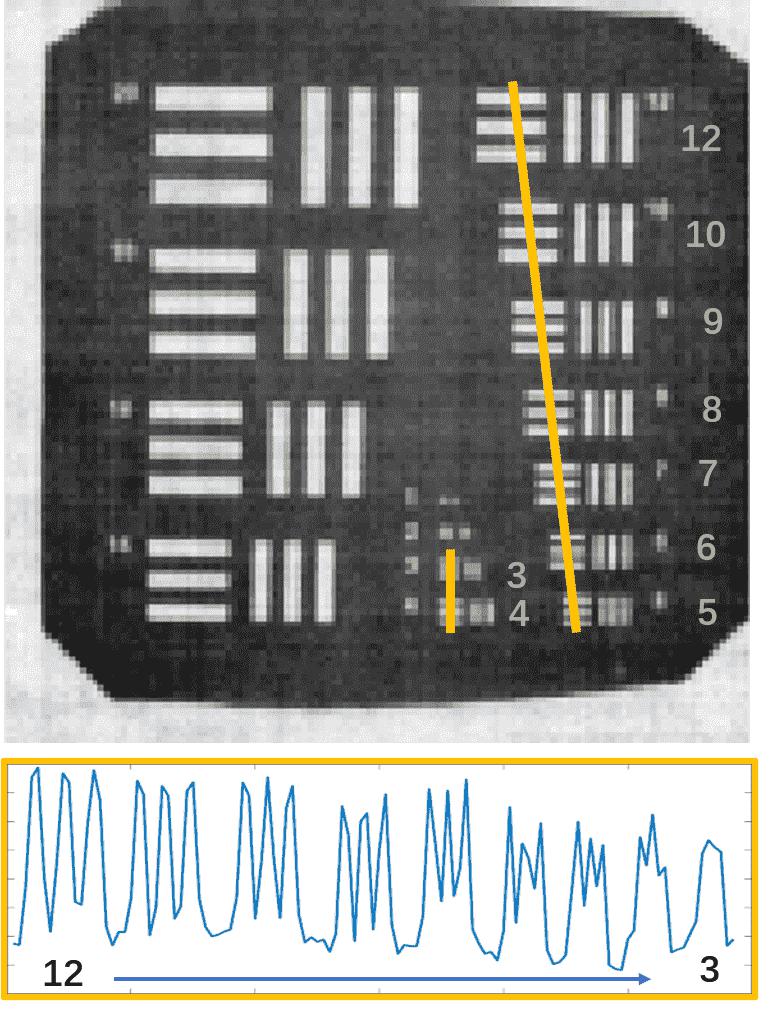}
			\caption{49 patterns }
		\end{subfigure} \\
		\begin{subfigure}[t]{1\linewidth}
			\centering
			\includegraphics[width=0.85\linewidth,keepaspectratio]{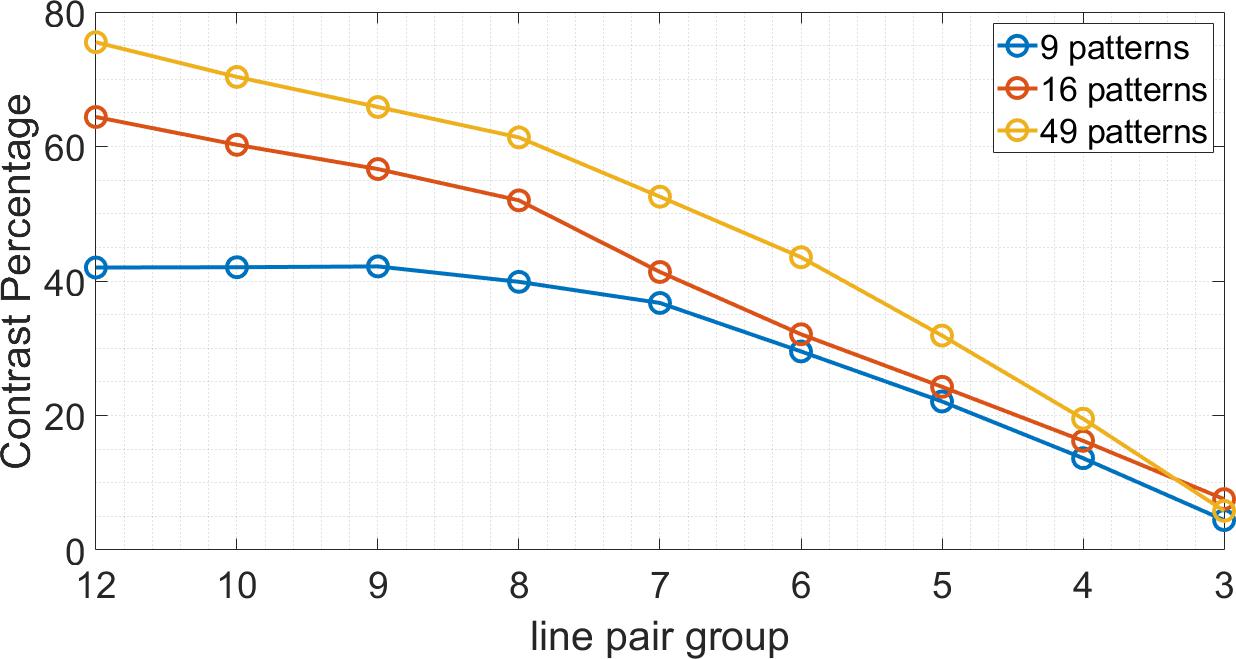}
			\caption{Modulation transfer function plot. The vertical axis shows the contrast ratio in percentage.  
			}
		\end{subfigure}
		\caption{Resolution analysis of coded illumination. Top images in (a,b,c) show resolution target reconstructed using 9, 16, and 49 shifting dots patterns. Bottom plots in (a,b,c) show the intensity of a line from group 12 to group 3. The MTF (modulation transfer function) plot for different numbers of shifting dots illumination patterns is shown in (d). To compute MTF, we manually select each group of horizontal and vertical line pairs after subtracting a fixed DC background, then average every group along the columns and rows, respectively, and finally compute the contrast ratio. 
		}
		\label{fig:target_line_plots}
	\end{figure}

	\renewcommand{\figwidth}{0.23\linewidth}
	\begin{figure}[t]
		\setlength\tabcolsep{1pt}
		\centering
		\footnotesize
		\begin{tabular}{cc|cc}
			$128\times128$ &
			$64\times64$ &
			$128\times128$ &
			$64\times64$
			\\
			\rotatebox{90}{\parbox{1.8cm}{\centering uniform}}
			\includegraphics[width=\figwidth,keepaspectratio]{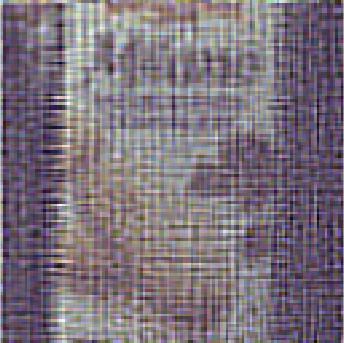} &
			\includegraphics[width=\figwidth,keepaspectratio]{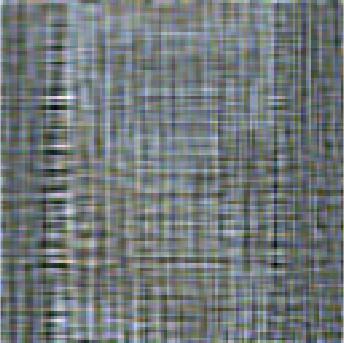} &
			\includegraphics[width=\figwidth,keepaspectratio]{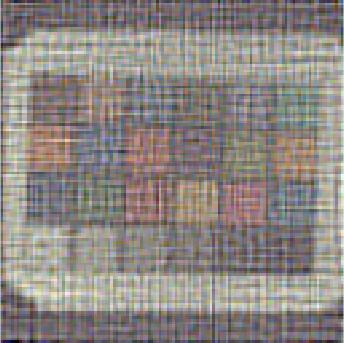} &
			\includegraphics[width=\figwidth,keepaspectratio]{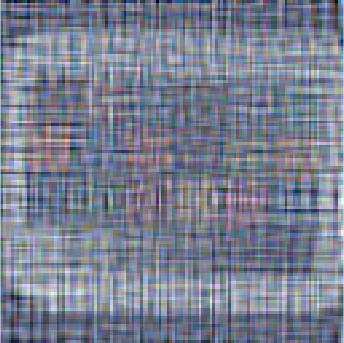} 
			\\
			\rotatebox{90}{\parbox{1.8cm}{\centering 16 shift}} \includegraphics[width=\figwidth,keepaspectratio]{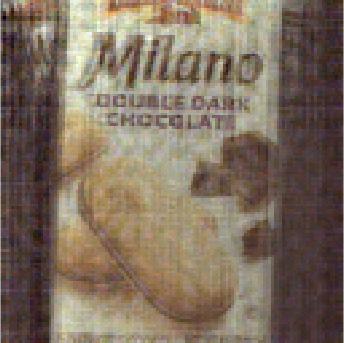} &
			\includegraphics[width=\figwidth,keepaspectratio]{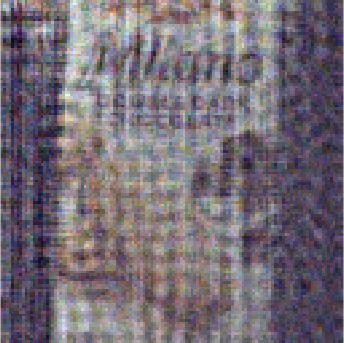} & \includegraphics[width=\figwidth,keepaspectratio]{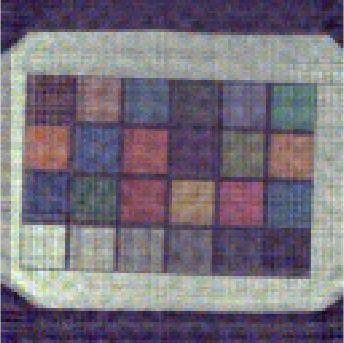} &
			\includegraphics[width=\figwidth,keepaspectratio]{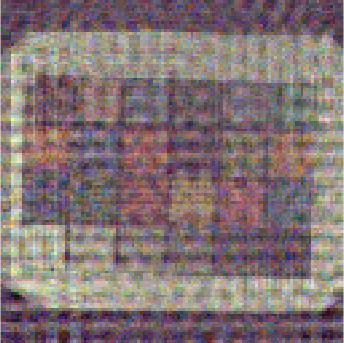} 
			\\
			\rotatebox{90}{\parbox{1.8cm}{\centering 25 shift}} 
			\includegraphics[width=\figwidth,keepaspectratio]{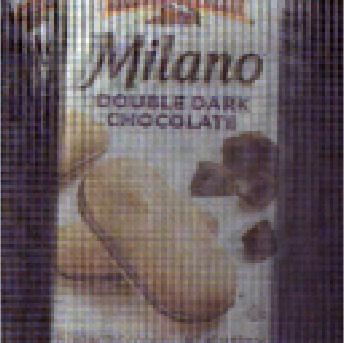} &
			\includegraphics[width=\figwidth,keepaspectratio]{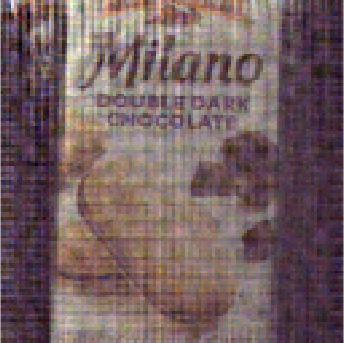} & \includegraphics[width=\figwidth,keepaspectratio]{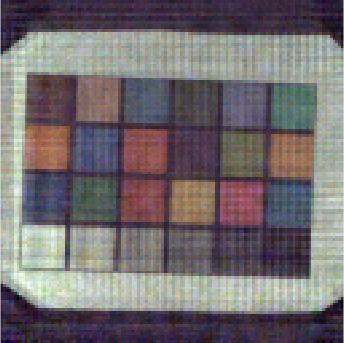} &
			\includegraphics[width=\figwidth,keepaspectratio]{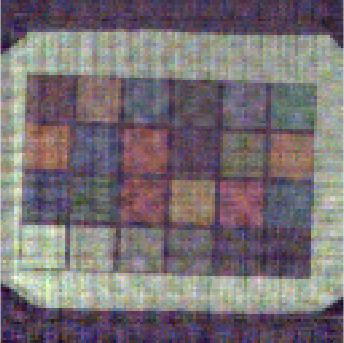} 
			\\ 
			\rotatebox{90}{\parbox{1.8cm}{\centering 49 shift}} \includegraphics[width=\figwidth,keepaspectratio]{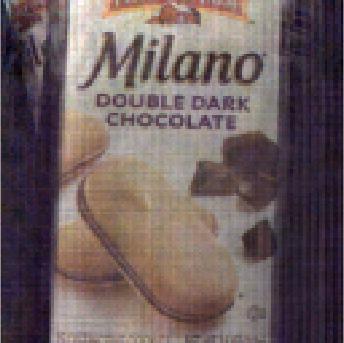} &
			\includegraphics[width=\figwidth,keepaspectratio]{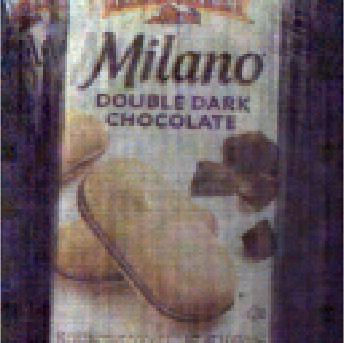} & \includegraphics[width=\figwidth,keepaspectratio]{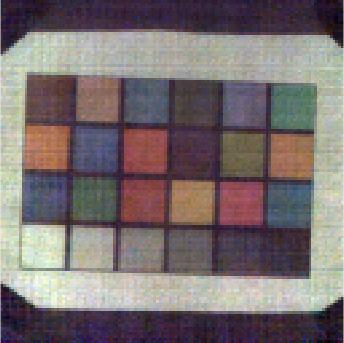} &
			\includegraphics[width=\figwidth,keepaspectratio]{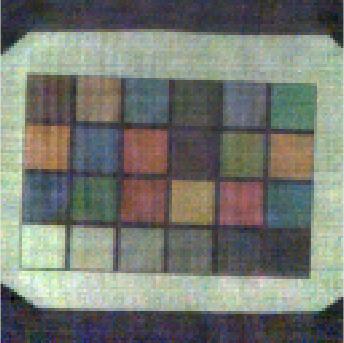} 
		\end{tabular}
		\caption{Experimental results for reconstructing the $128\times 128$ images from $64\times64$ and $128\times128$ measurements. Single (uniform) illumination-based method fails to recover images as the number of measured pixels reduce. Our method with 49 shifting dots patterns recovers near-perfect reconstruction at different levels of binning (compression) factors. }
		\label{fig:experiment_results_sensor_binning}
	\end{figure} 
	
	\subsection{Compressive Sensor Measurements}
	One potential application of coded illumination with lensless imaging is to expand the space of sensor measurements without increasing the number of sensor pixels. This can be especially useful when sensors capture fewer measurements than the number of pixels we seek to recover. 
	Compressive sensing often deals with such scenarios where the number of sensor pixels is smaller than the number of unknown scene pixels \cite{candes2006compressive,donoho2006compressive}. The system becomes under-determined and the reconstruction quality becomes worse. In lensless imaging systems, the number of sensor measurements is often larger than the number of reconstruction pixels, but that comes at the expense of low resolution of reconstructed images or larger and more costly sensors.

	To validate the robustness of our method in the case of an under-determined system, we performed an experiment by binning the sensor measurements at increasing factors. The results are presented in  Fig.~\ref{fig:experiment_results_sensor_binning}. In our experiments, we capture $512\times512$ measurements and bin them to $256\times 256$, $128\times 128$, or $64\times64$ pixels by averaging the neighboring pixels in post-processing. The reconstruction image has $128\times128$ pixels, which implies the $64\times64$ sensor measurements yield an under-determined system with one illumination. 
	We present the result of two scenes with different binning factors. We observe that as we bin a larger number of pixels, the system with a single uniform illumination becomes ill-conditioned and the quality of reconstruction degrades. On the other hand, the system with 49 shifting dots patterns provides stable reconstruction even when the measurements are binned to $64\times64$ pixels. 

	\subsection{Deep Network-based Denoising vs Reconstruction}
	Deep learning-based methods have been widely used for image recovery and enhancement tasks \cite{khan2020flatnet,kristina2019learning ,chang2019DeepOptics3D}. For instance, UNet \cite{olaf2015unet} is used for image denoising and removing artifacts from reconstructed images \cite{zheng2021simple,khan2020flatnet}. Deep learning is also used as the priors to improve the reconstruction results \cite{bora2017generator}, especially when the number of measurements is small. UNet-based networks have been used to recover photorealistic images from lensless measurements in \cite{khan2020flatnet}. In our experiments, we observed that deep learning-based methods that learn to recover images directly from sensor measurements perform very well on simulated data but provide catastrophic results on real data. 
	In contrast, the simple least-squares method we discussed in \eqref{eq:closed_form} provides stable results in all cases because coded illumination provides a well-conditioned system. We present simulation and experimental results below.

	We performed some experiments to compare the performance of four methods with coded and uniform illumination: least squares (LS) in \eqref{eq:closed_form}, LS with a trained UNet that is used as a refinement network, LS with pretrained refinement network in FlatNet \cite{khan2020flatnet}, and end-to-end trained FlatNet. 
	We present results for simulated sensor measurements in Fig.~\ref{fig:simu_flatnet_unet} and on real  data captured using our prototype in Fig.~\ref{fig:exp_flatnet_unet}.

	The description of the four methods is as follows. We used 1000 natural RGB images from \cite{chaladze2017dataset} to train the deep networks. 
	\begin{itemize}[leftmargin=4mm,topsep=0pt,itemsep=0ex,partopsep=1ex,parsep=1pt]
		\item \textbf{LS.} Images are reconstructed using the least-square (LS) method with $\ell_2$-regularization \eqref{eq:closed_form}. 
		\item \textbf{LS + trained UNet.} We trained a UNet using 1000 training images that learns to map the LS reconstruction into a refined image. 
		\item \textbf{Trained FlatNet.} FlatNet~\cite{khan2020flatnet} contains two steps: inversion step and refinement step that are trained jointly in an end-to-end manner. The inversion step maps the sensor measurements into a coarse image estimate, while the refinement step maps the coarse estimate to a cleaner image. 
		We train both steps of the FlatNet in an end-to-end manner with simulated measurements for uniform and coded illuminations. We created the synthetic data by simulating the noisy lensless measurements for training images using the calibrated transfer matrices $\Phi_L, \Phi_R$ described in Sec.\ref{sec:exp_setup}. Note that we need to train a separate network for different types/number of illumination patterns.   
		\item \textbf{LS + pretrained FlatNet.}  The refinement network in FlatNet has same architecture as our trained UNet. For the sake of comparison, we also used the pretrained refinement network from FlatNet to refine the LS solution. 
	\end{itemize}

	Figure~\ref{fig:simu_flatnet_unet} shows images reconstructed from simulated measurements of selected test images. 
	We observe that LS, LS+UNet, and trained FlatNet  provide good reconstruction for simulated data using uniform illumination. In particular, images from trained FlatNet have best visual appearance simulations Fig.~\ref{fig:simu_flatnet_unet}(d). For coded illumination, the quality of reconstruction improves for all the methods. The pretrained refinement network from FlatNet provides overly smoothed images and distorts spatial details. 
	
	Figure~\ref{fig:exp_flatnet_unet} shows images reconstructed from measurements captured by real prototype. 
	Trained FlatNet  fails to recover images from real data captured with uniform illumination in Fig.~\ref{fig:exp_flatnet_unet}. In contrast, LS and LS+UNet provide a low-resolution reconstruction with uniform illumination. 
	For coded illumination, trained FlatNet reconstructions have high quality that is similar to the results provided by LS and LS + trained UNet. 
	Figure~\ref{fig:exp_flatnet_unet} further shows that UNet can partially reduce the artifacts and noise in the images reconstructed from the uniform illumination, but fails to improve the spatial resolution. In contrast, coded illuminations significantly improves the resolution and quality of the reconstructed images.
	Additional results can be found in the supplementary material.

	Our main takeaway from these experiments is that  deep networks cannot always recover missing details if the system is ill conditioned. In particular, deep network-based refinement/denoising can improve the appearance of images but cannot recover missing details. End-to-end trained networks, which recover images directly from multiplexed sensor measurements, perform well when train and test settings are identical, but provide catastrophic results in case of any mismatch. Coded illuminations improve the conditioning of the system, which benefits all the recovery methods. 
	
	\renewcommand{\figwidth}{0.24\linewidth}
	\begin{figure}[t]
		\setlength\tabcolsep{1pt}
		\centering
		\footnotesize
		\begin{tabular}{cccc}
			(a) &
			(b) &
			(c) &
			(d)
			\\
			\rotatebox{90}{\parbox{2.1cm}{\centering uniform}}
			\includegraphics[width=\figwidth,keepaspectratio]{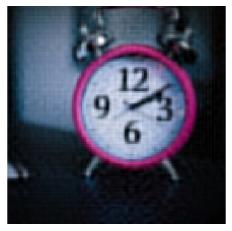} &
			\includegraphics[width=\figwidth,keepaspectratio]{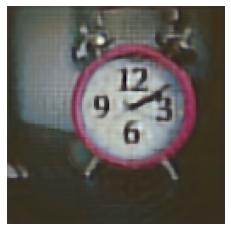} &
			\includegraphics[width=\figwidth,keepaspectratio]{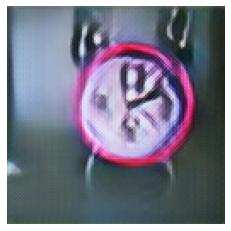} &
			\includegraphics[width=\figwidth,keepaspectratio]{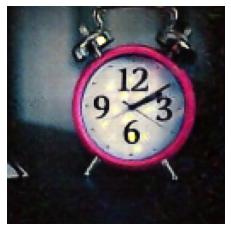} 
			\\
			SSIM: 0.803 &
			0.501 &
			0.781 &
			0.903 
			\\
			\rotatebox{90}{\parbox{2.1cm}{\centering 49 shifting dots}}
			\includegraphics[width=\figwidth,keepaspectratio]{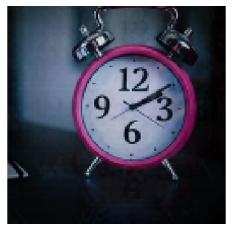} &
			\includegraphics[width=\figwidth,keepaspectratio]{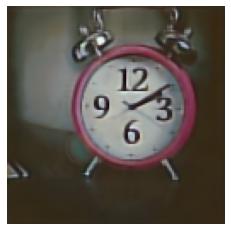} &
			\includegraphics[width=\figwidth,keepaspectratio]{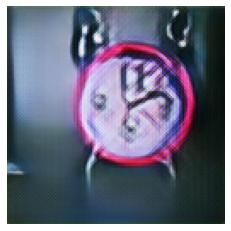} &
			\includegraphics[width=\figwidth,keepaspectratio]{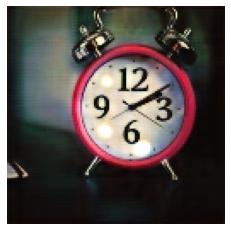} 
			\\
			SSIM: 0.936 &
			0.562 &
			0.831 &
			0.904
			\\ 
			\rotatebox{90}{\parbox{2.3cm}{\centering uniform}}
			\includegraphics[width=\figwidth,keepaspectratio]{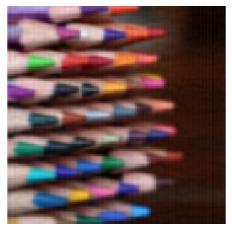} &
			\includegraphics[width=\figwidth,keepaspectratio]{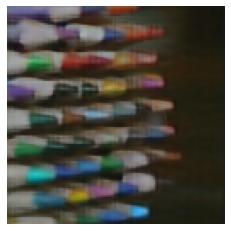}&
			\includegraphics[width=\figwidth,keepaspectratio]{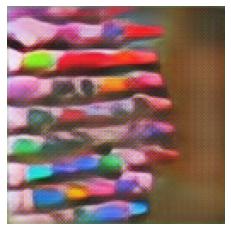} &
			\includegraphics[width=\figwidth,keepaspectratio]{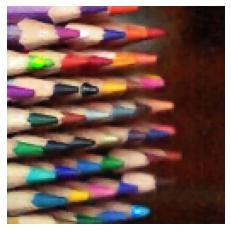} 
			\\
			SSIM:0.942 &
			0.851 &
			0.474 &
			0.950 
			\\
			\rotatebox{90}{\parbox{2.3cm}{\centering 49 Shifting dots}}
			\includegraphics[width=\figwidth,keepaspectratio]{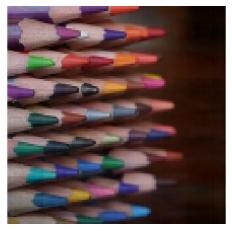} &
			\includegraphics[width=\figwidth,keepaspectratio]{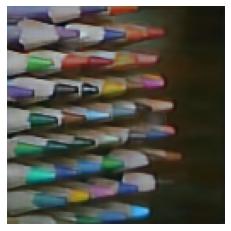}&
			\includegraphics[width=\figwidth,keepaspectratio]{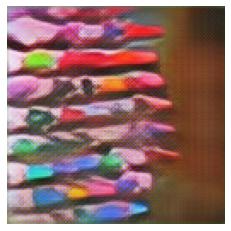} &
			\includegraphics[width=\figwidth,keepaspectratio]{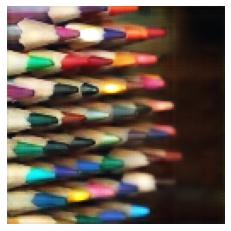} 
			\\
			SSIM:0.966 &
			0.898 &
			0.469 &
			0.936  
		\end{tabular}
		\caption{Reconstruction results for simulated measurements with 49 uniform and shifting dots illumination patterns. 
			Images in four columns show (a) LS solution, (b) LS solution with trained UNet refinement, (c) LS solution with pretrained FlatNet refinement, and (d) trained FlatNet that reconstructs image directly from measurements. For each image, we show the SSIM value underneath. 
		}
		\label{fig:simu_flatnet_unet}
	\end{figure}

	\renewcommand{\figwidth}{0.24\linewidth}
	\begin{figure}[h]
		\setlength\tabcolsep{1pt}
		\centering
		\footnotesize
		\begin{tabular}{cccc}
			(a) &
			(b) &
			(c) &
			(d)
			\\
			\rotatebox{90}{\parbox{2.1cm}{\centering uniform}}
			\includegraphics[width=\figwidth,keepaspectratio]{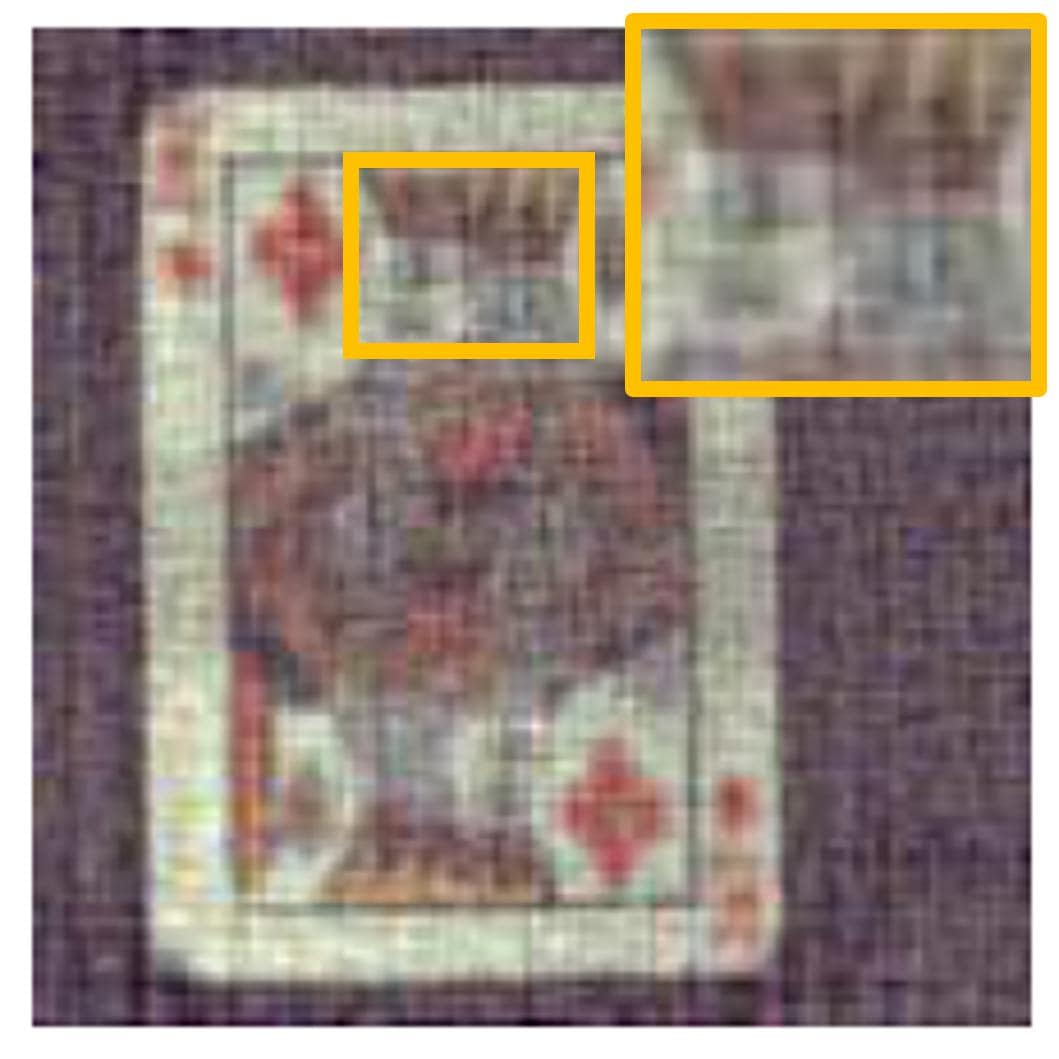} &
			\includegraphics[width=\figwidth,keepaspectratio]{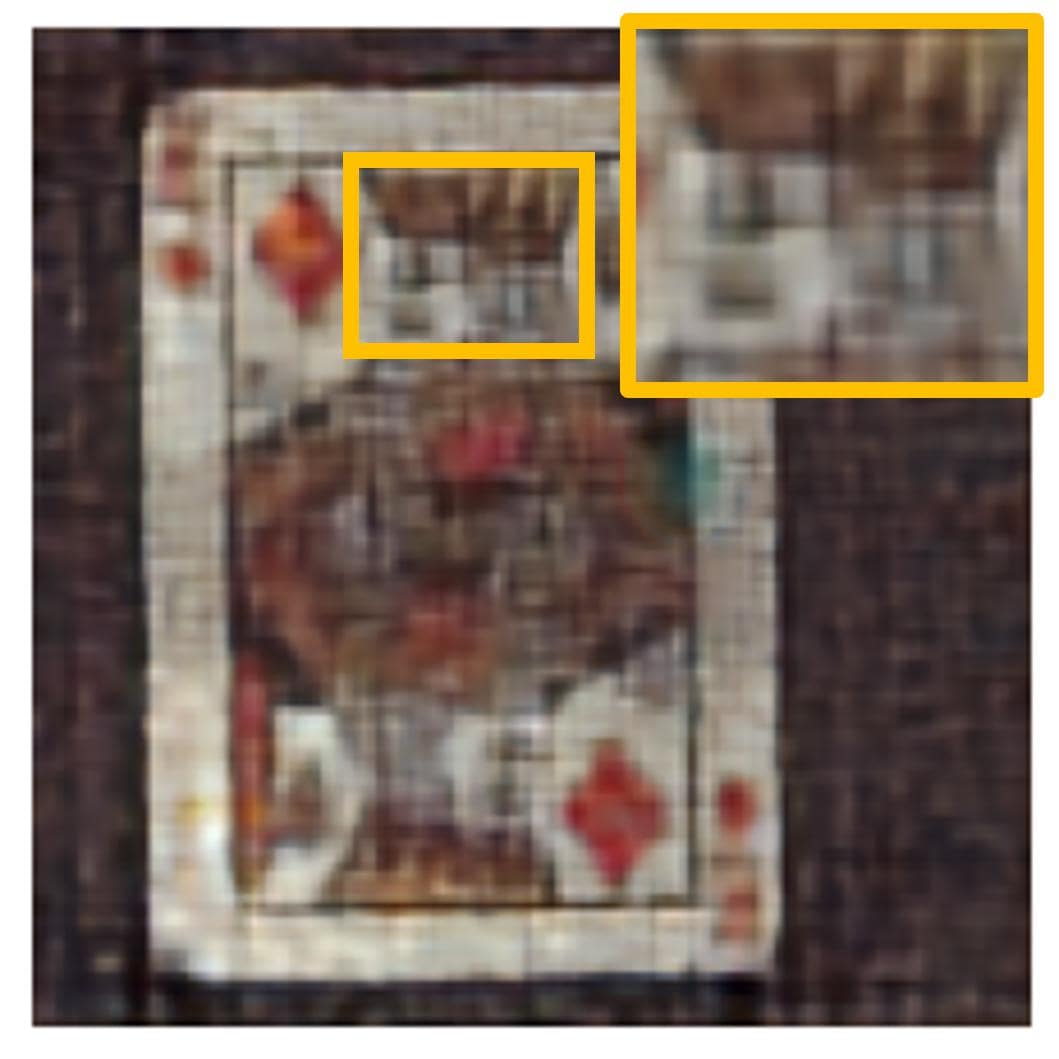} &
			\includegraphics[width=\figwidth,keepaspectratio]{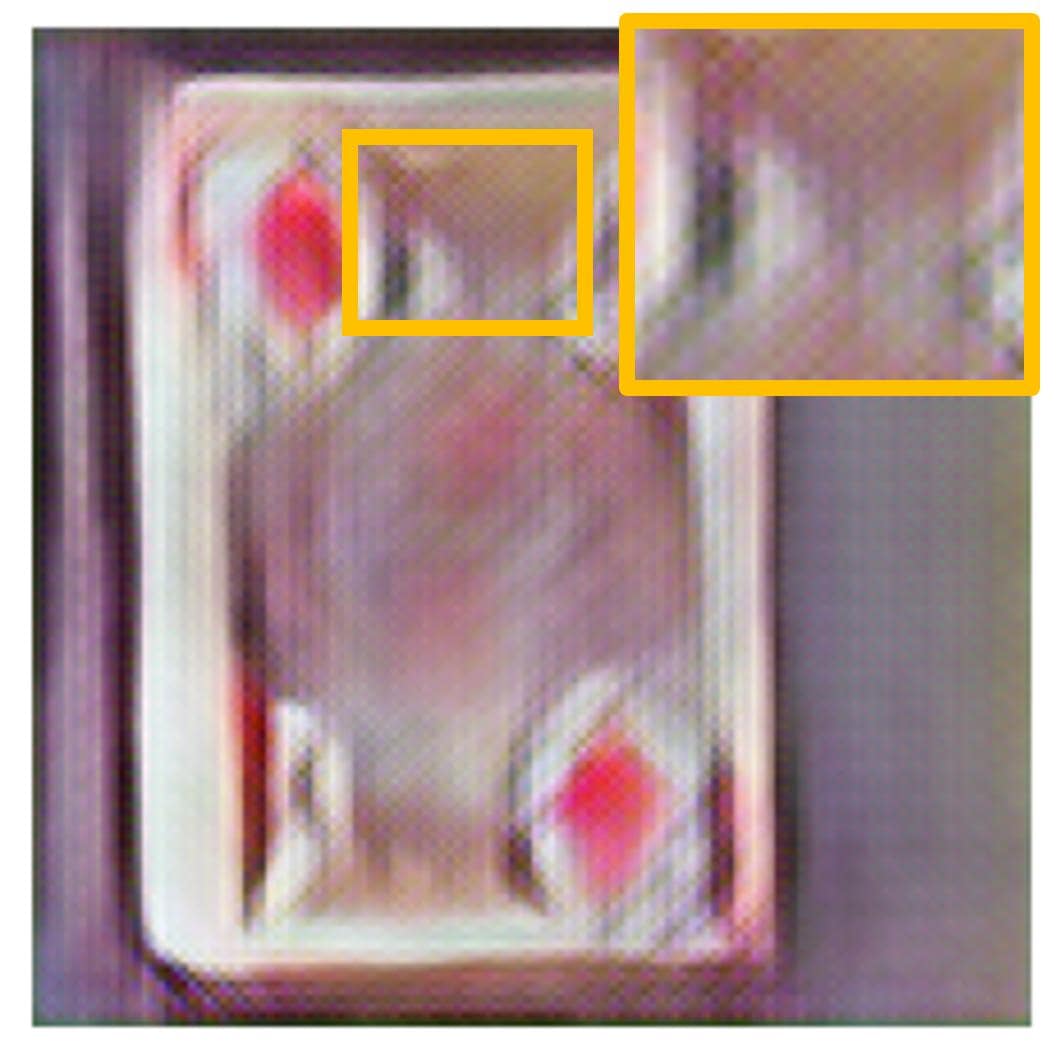} &
			\includegraphics[width=\figwidth,keepaspectratio]{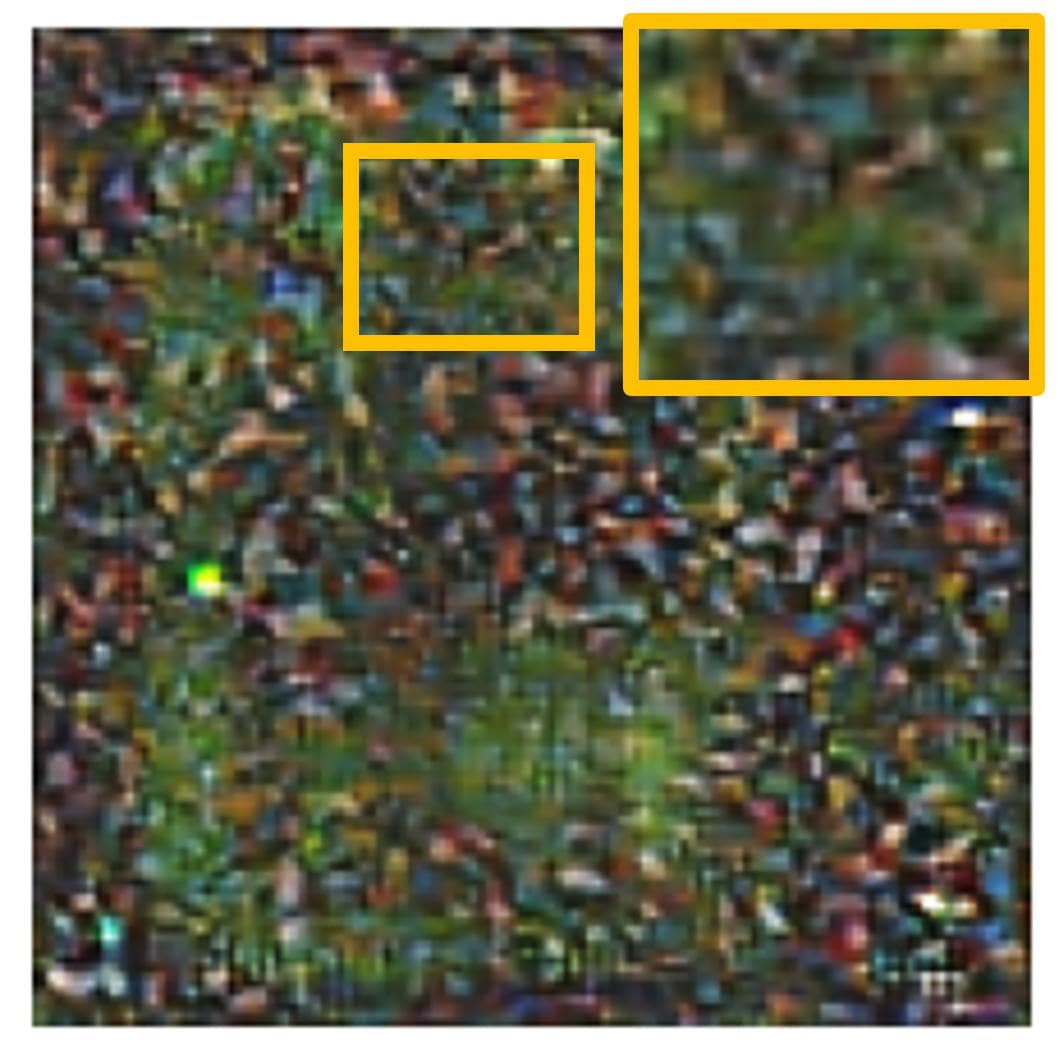} 
			\\
			\rotatebox{90}{\parbox{2.1cm}{\centering 49 shifting dots}}
			\includegraphics[width=\figwidth,keepaspectratio]{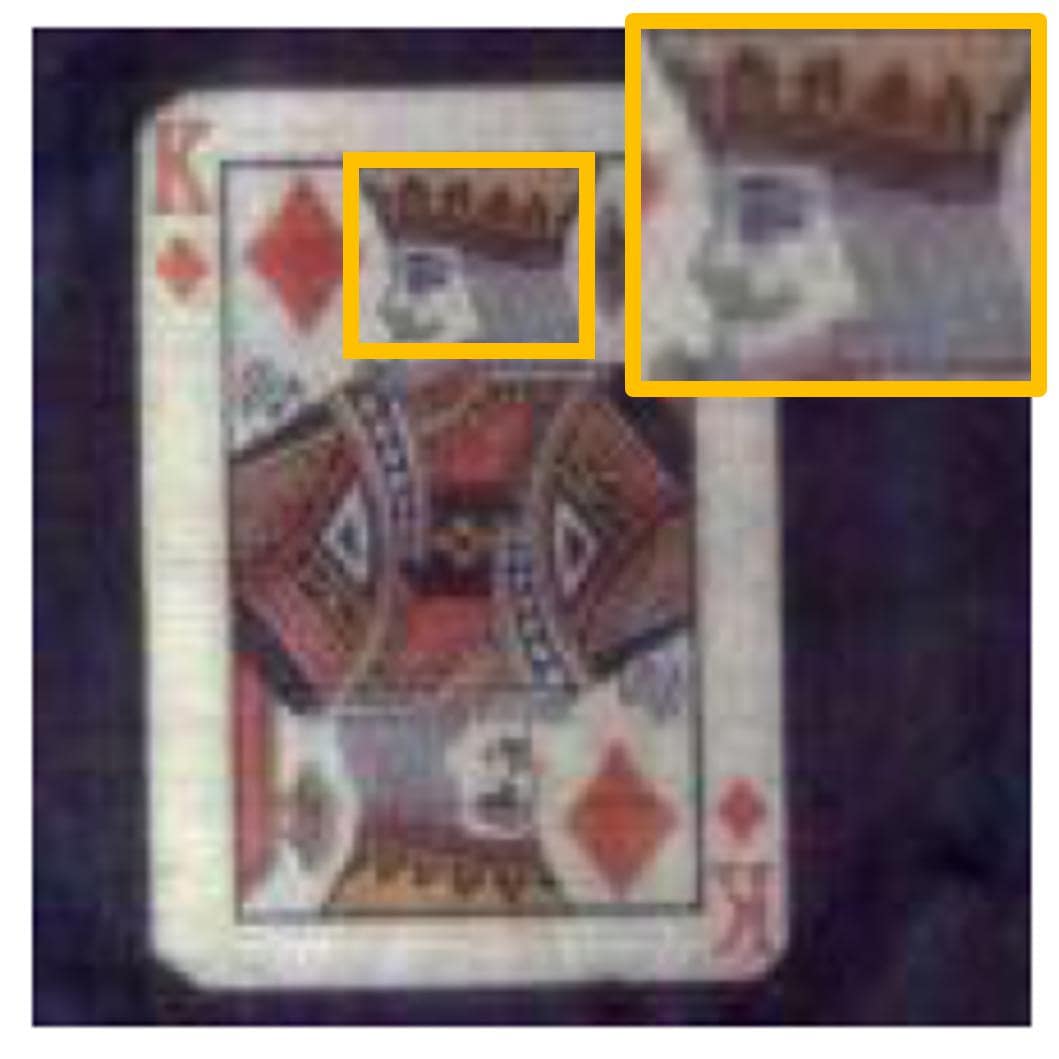} &
			\includegraphics[width=\figwidth,keepaspectratio]{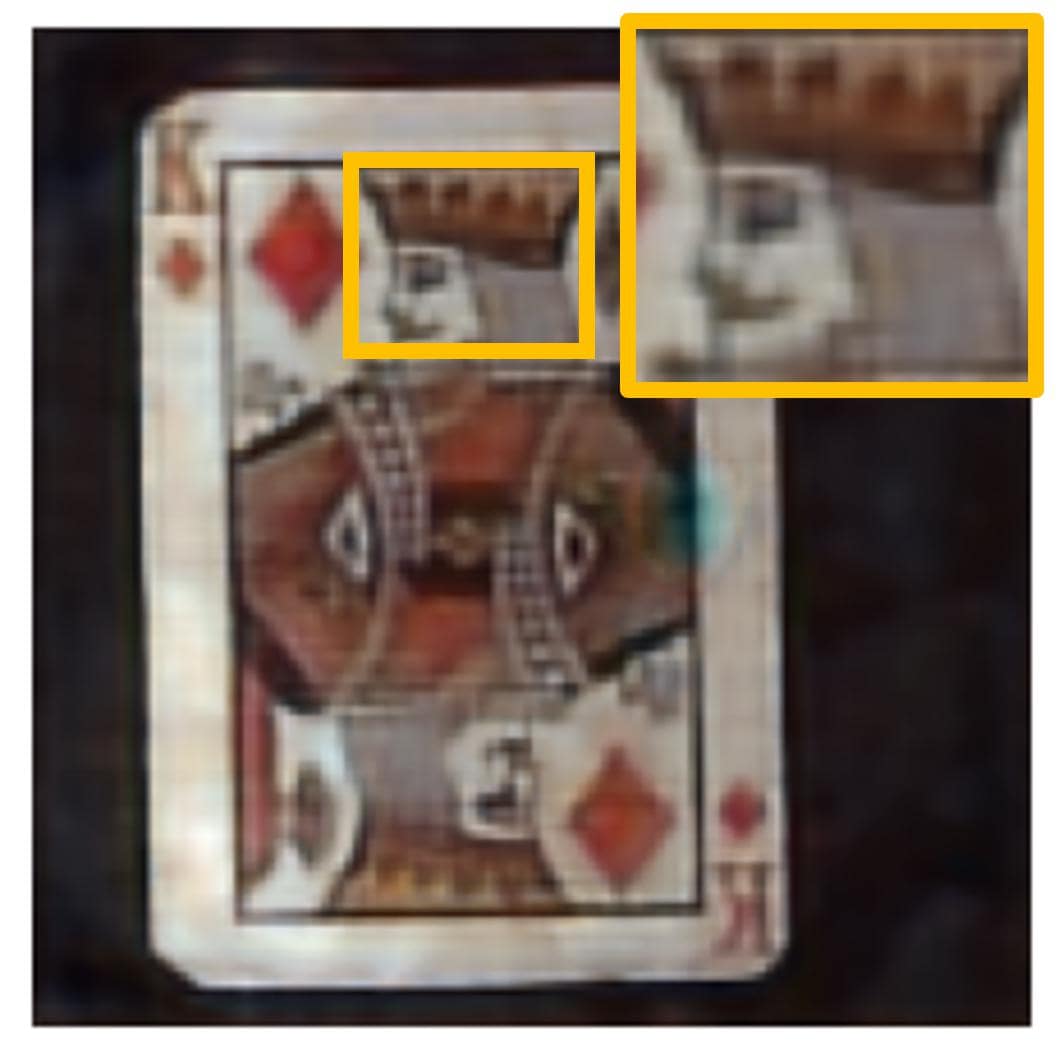} &
			\includegraphics[width=\figwidth,keepaspectratio]{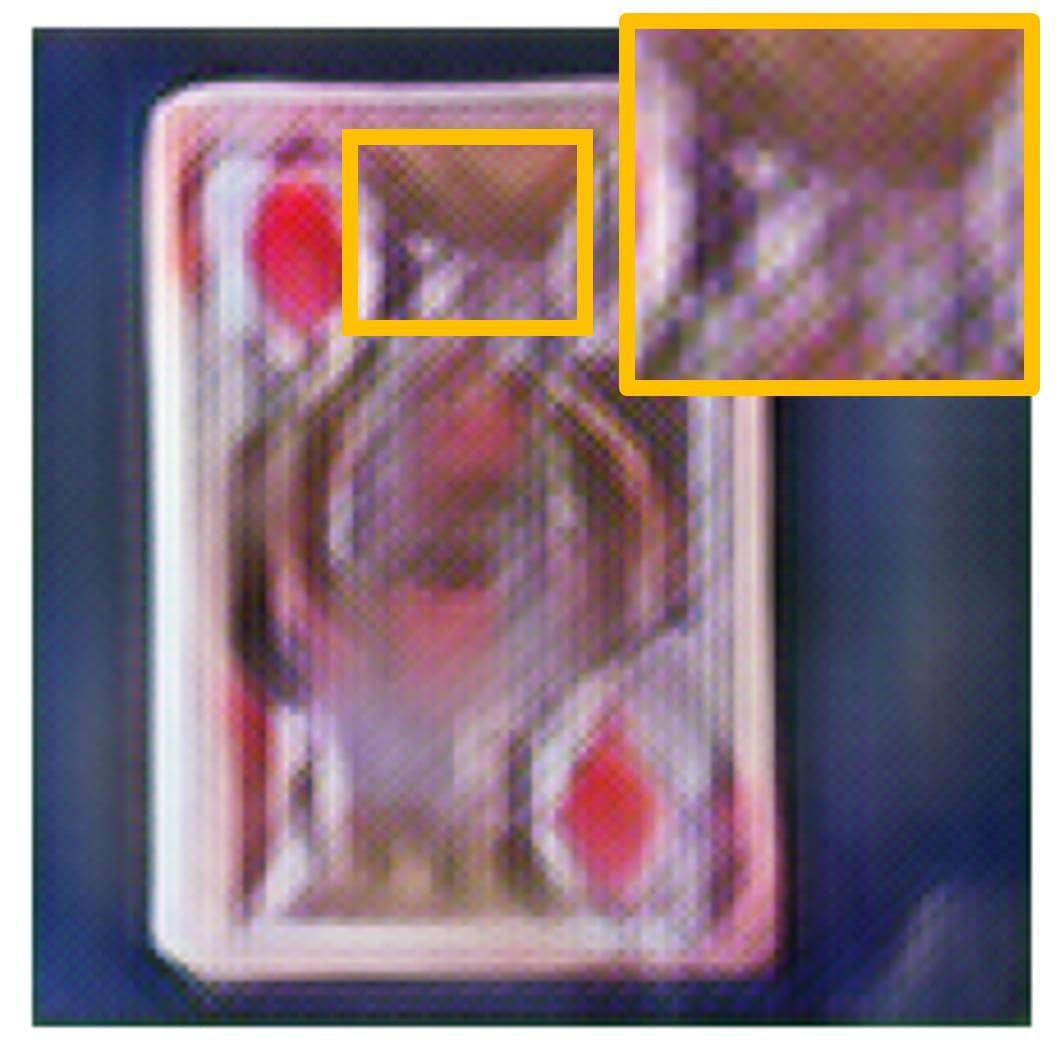} &
			\includegraphics[width=\figwidth,keepaspectratio]{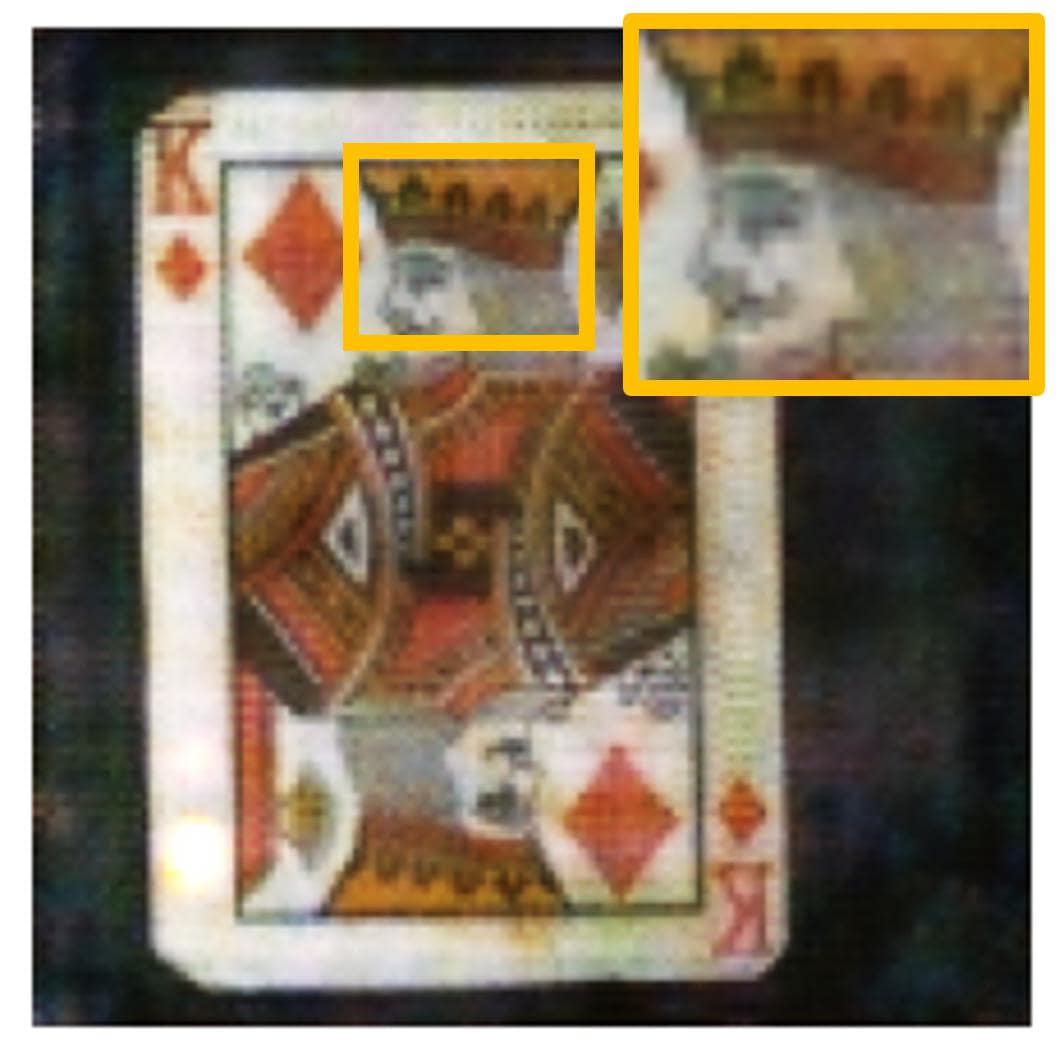} 
			\\ 
			\rotatebox{90}{\parbox{2.1cm}{\centering uniform}}
			\includegraphics[width=\figwidth,keepaspectratio]{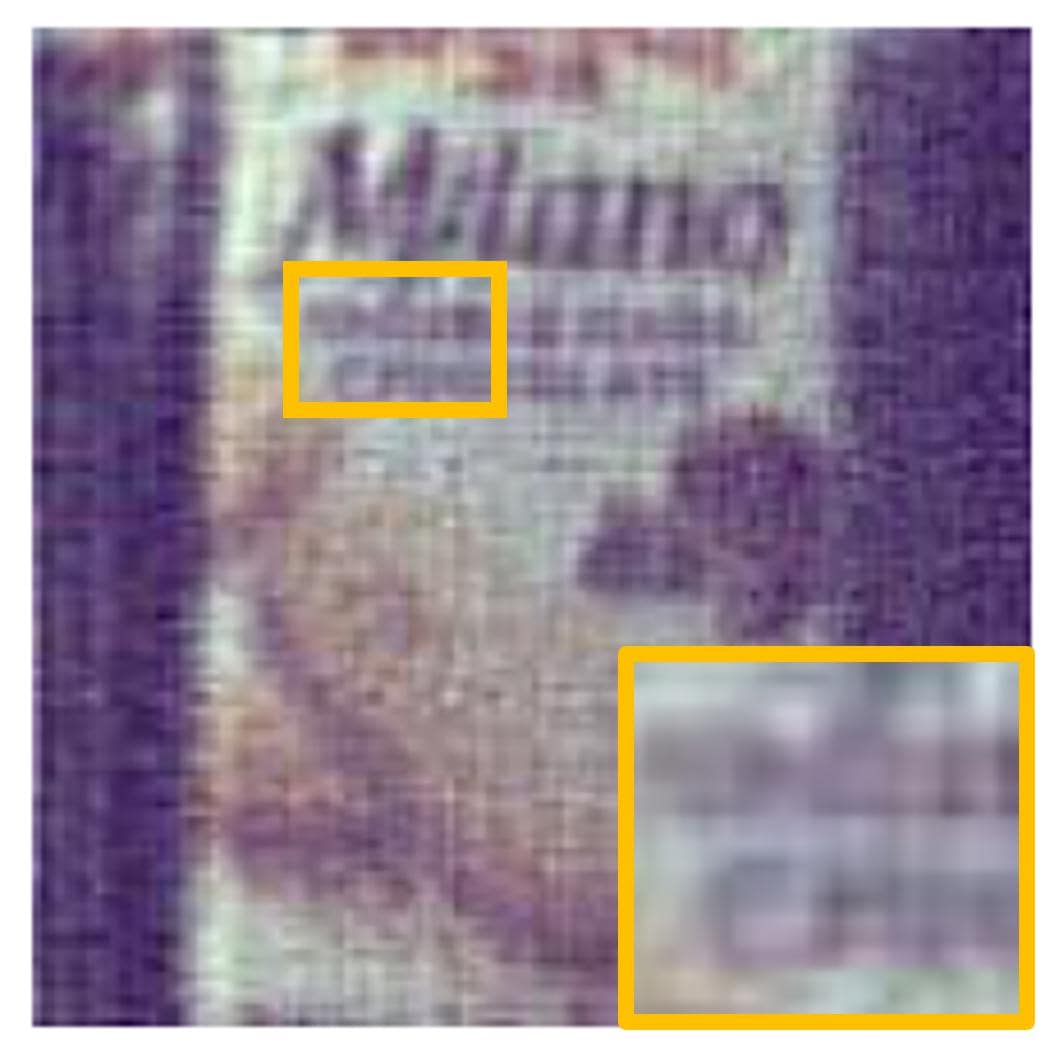} &
			\includegraphics[width=\figwidth,keepaspectratio]{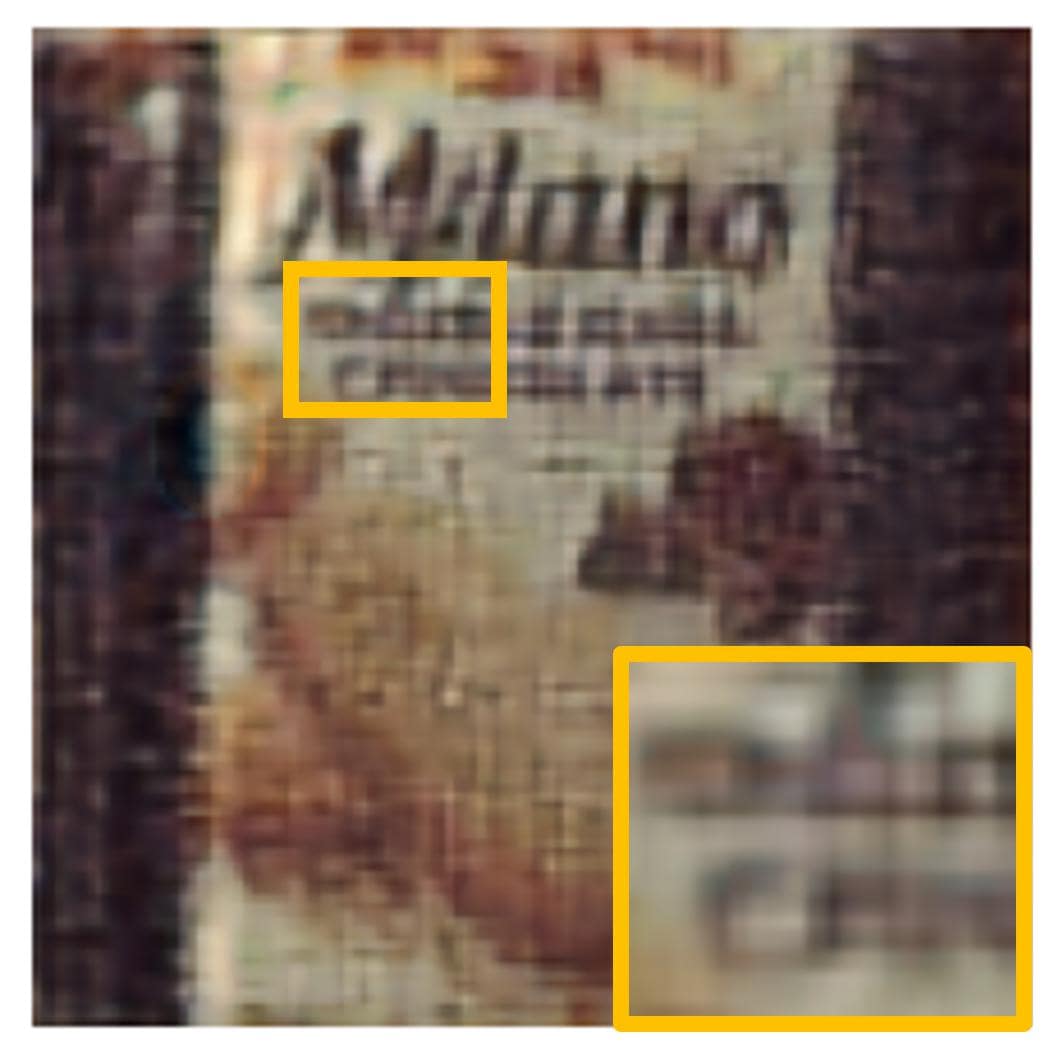} &
			\includegraphics[width=\figwidth,keepaspectratio]{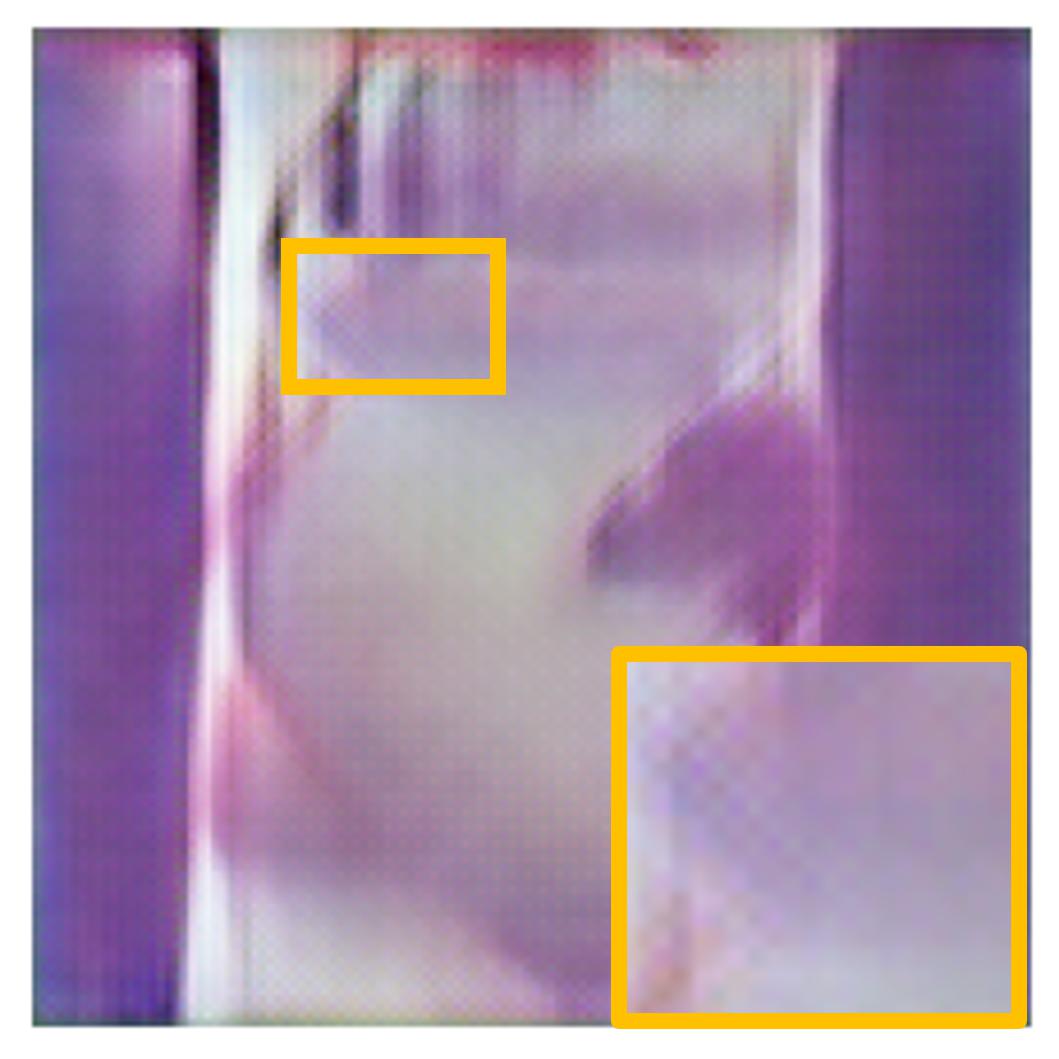} &
			\includegraphics[width=\figwidth,keepaspectratio]{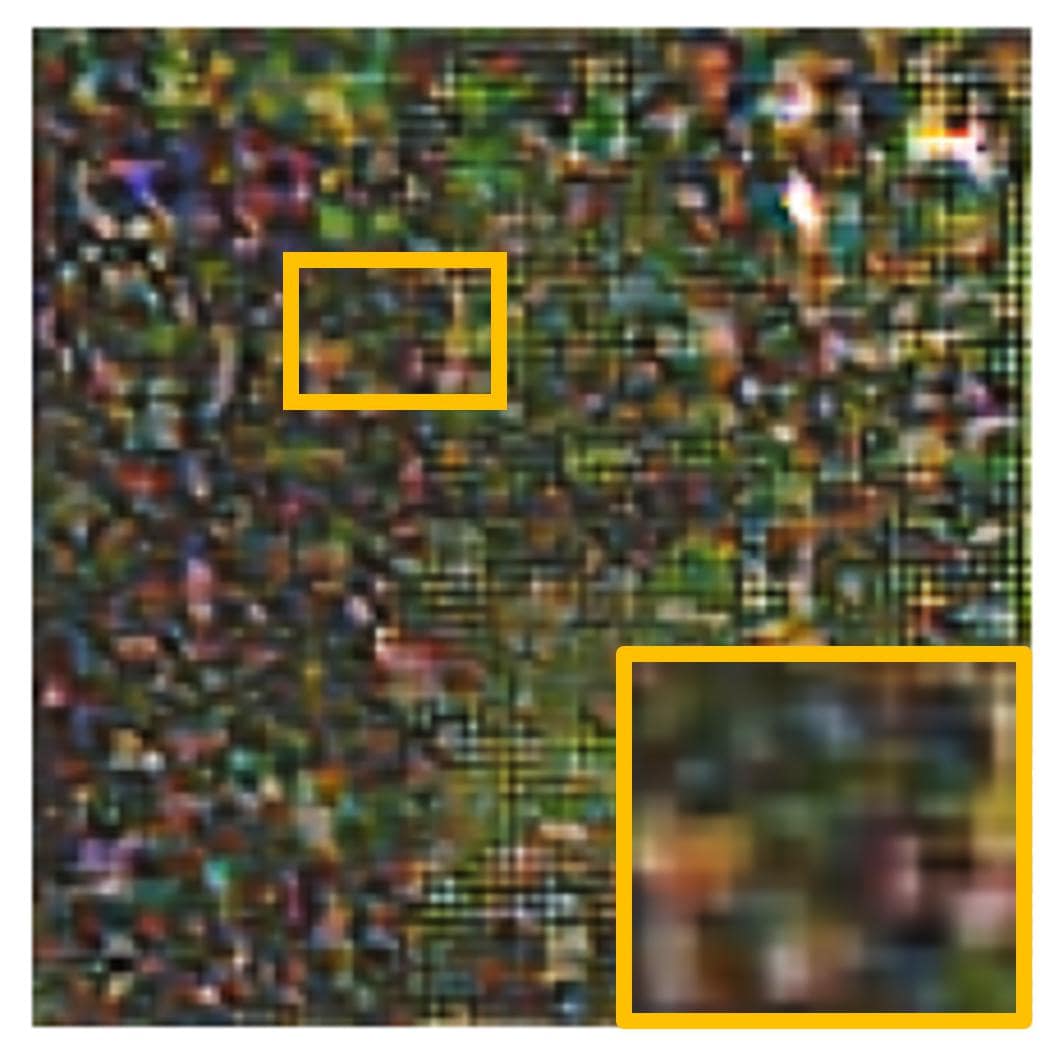} 
			\\
			\rotatebox{90}{\parbox{2.1cm}{\centering 49 shifting dots}}
			\includegraphics[width=\figwidth,keepaspectratio]{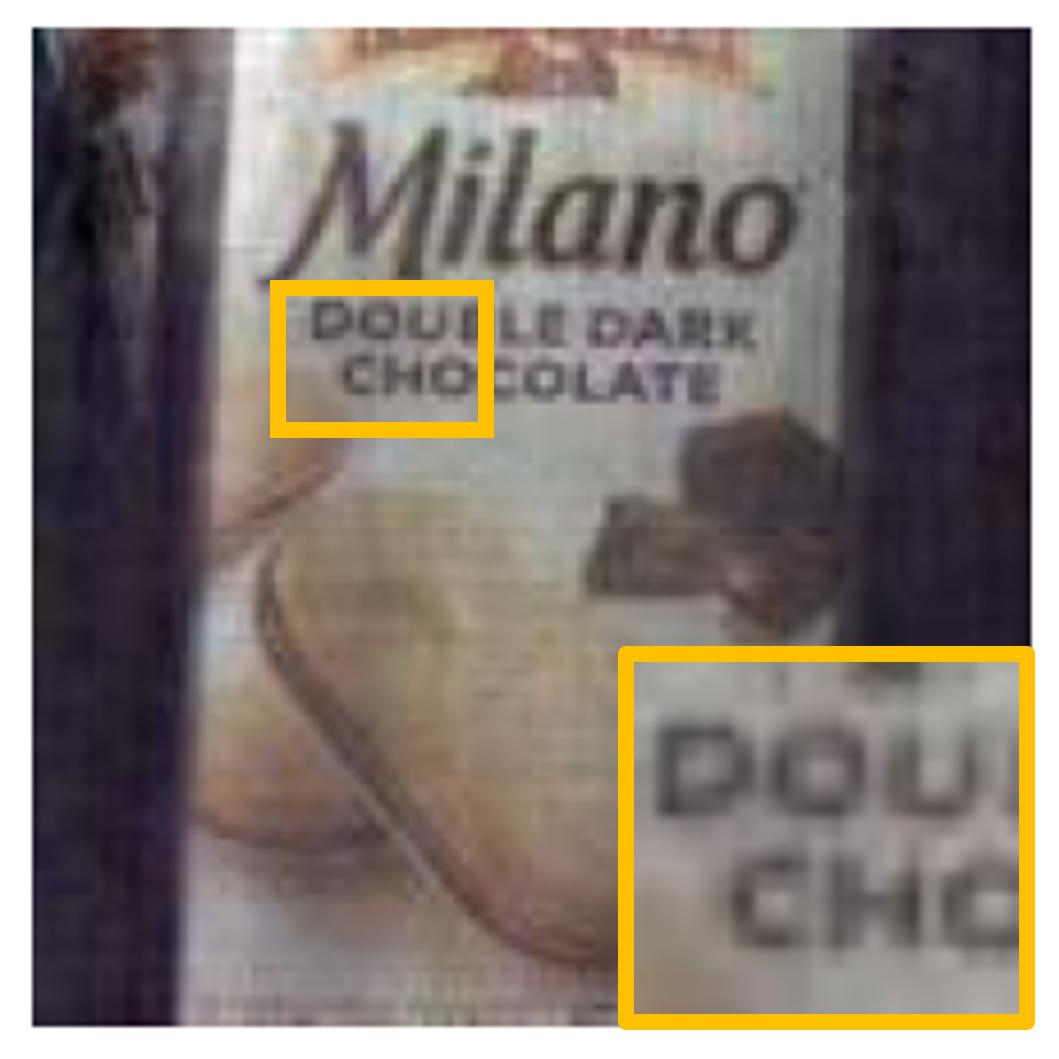} &
			\includegraphics[width=\figwidth,keepaspectratio]{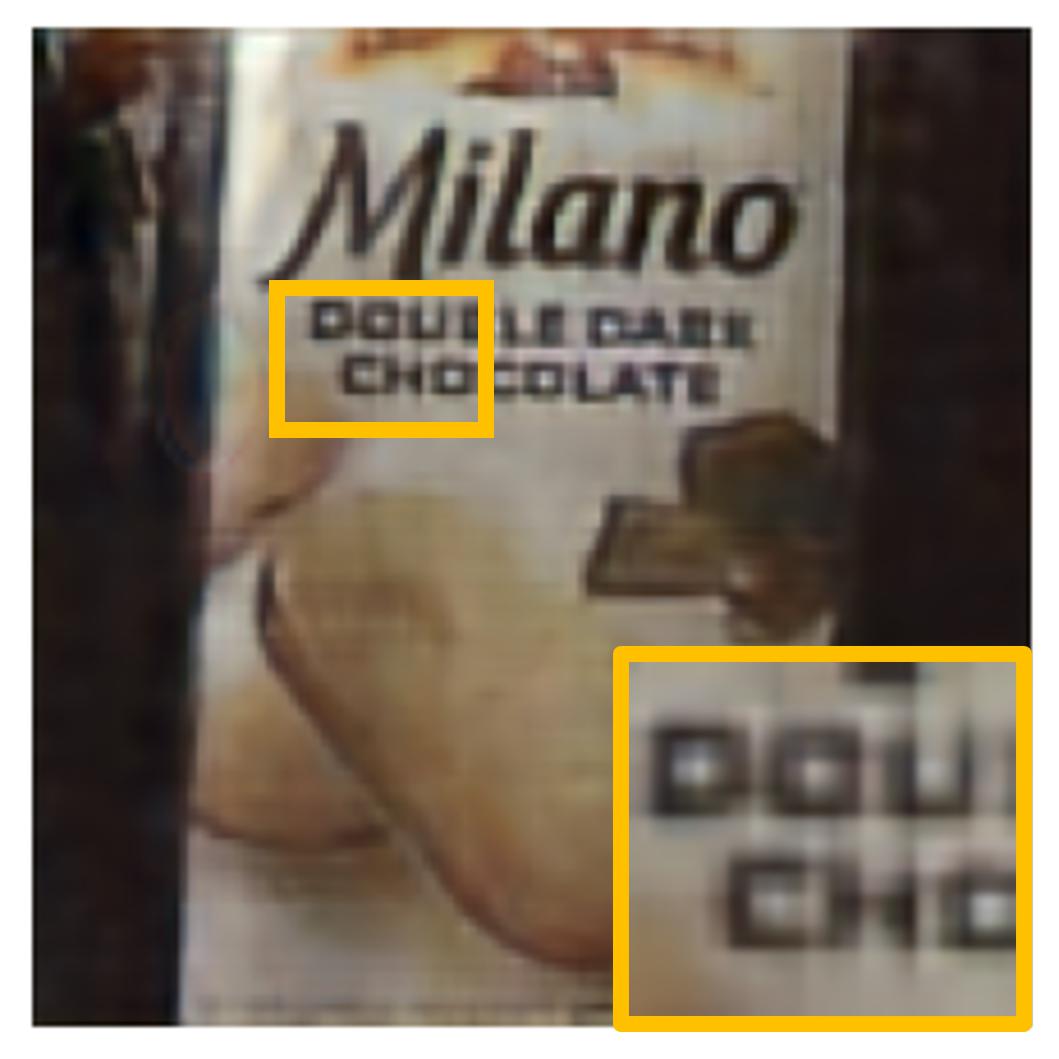} &
			\includegraphics[width=\figwidth,keepaspectratio]{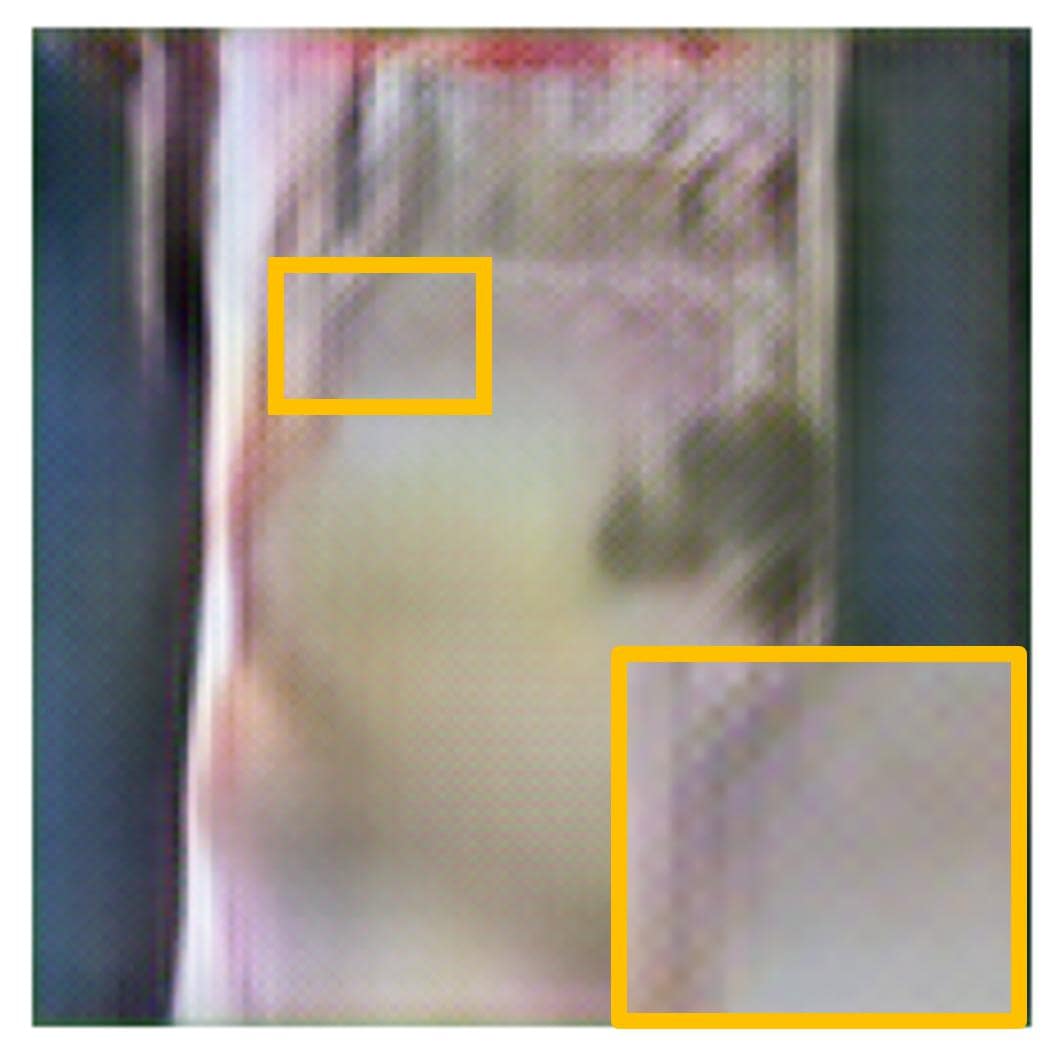} &
			\includegraphics[width=\figwidth,keepaspectratio]{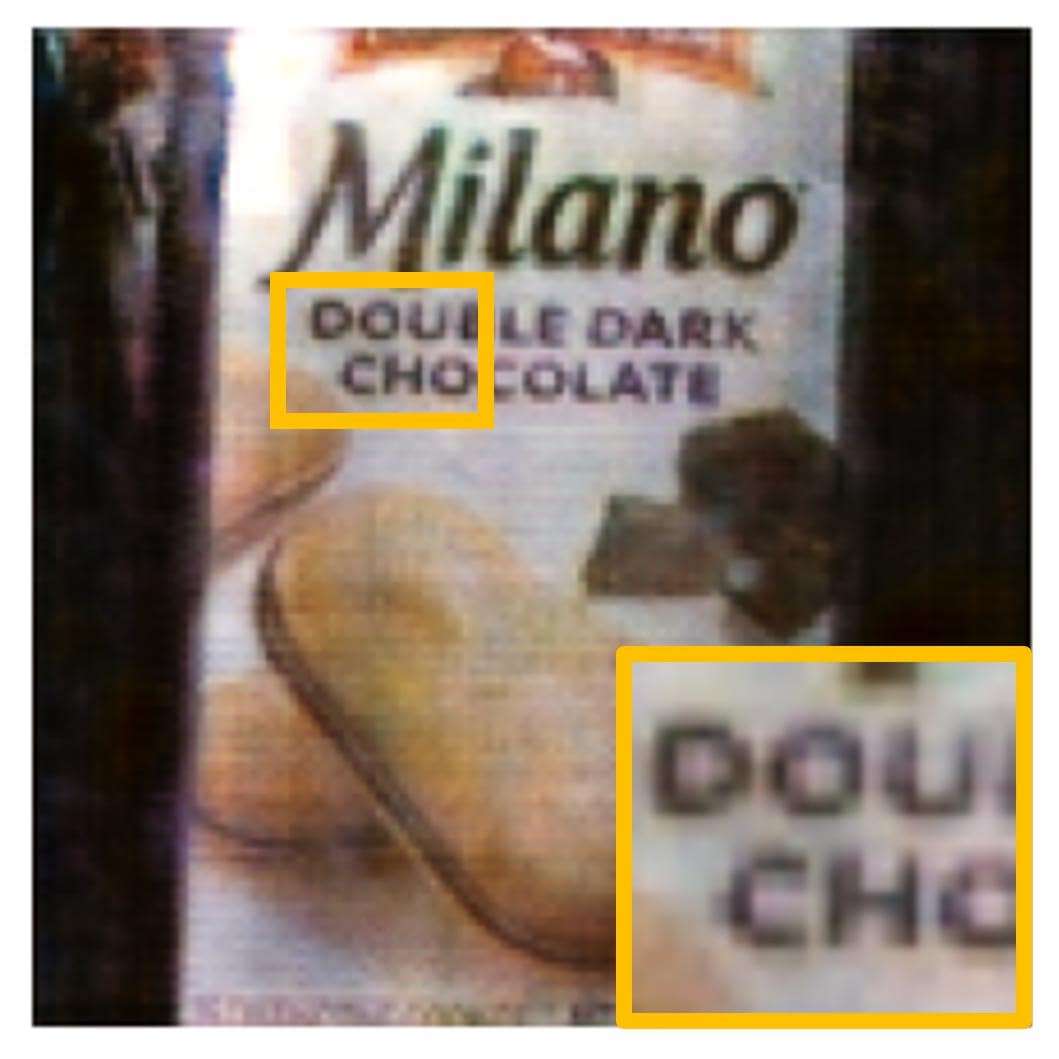} 
			\\ 
		\end{tabular}
		\caption{Reconstruction results for real-hardware measurements with 49 uniform and shifting dots illumination patterns. Images in four columns show (a) LS solution, (b) LS solution with trained UNet refinement, (c) LS solution with pretrained FlatNet refinement, and (d) trained FlatNet that  reconstructs image directly from measurements. 
		}
		\label{fig:exp_flatnet_unet}
	\end{figure} 
	
	%% Discussion 
	\section{Discussion}
	We propose a framework for combining coded illumination with lensless imaging. We present extensive simulation and real experiment results to demonstrate that we can get significantly improved reconstruction with multiple coded illumination patterns compared to original uniform illumination. 
	
	\subsubsection*{Advantages}
	\begin{itemize}[leftmargin=4mm,topsep=0pt,itemsep=0ex,partopsep=1ex,parsep=1pt]
		\item Qualitative experiment results show that our method has robust performance even if the original system is severely ill-conditioned (small sensor-mask distance and small number of measurements). 
		\item In space-limited applications such as under-the-display sensing, where the sensor-to-mask distance has to be small, our proposed method can offer significantly better reconstruction compared to uniform illumination. 
		\item Our proposed reconstruction algorithm is efficient in both space and time. For shifting dots (and any other orthogonal) pattern over $k\times k$ blocks, the system can be transformed into $k^2$ independent small problems that can be solved in parallel for faster and better reconstruction. 
		\item Shifting dots patterns have a simple structure but provide the best results in our experiments. 
		Shifting dots patterns can be easier to implement because all the patterns are simply the shifted copies of the same base pattern. 
		\item Programmable orthogonal patterns would provide similar gains as shifting dots with potentially better performance in strong ambient light (at the expense of complex hardware). 
	\end{itemize} 
	
	\subsubsection*{Limitations} 
	\begin{itemize}[leftmargin=4mm,topsep=0pt,itemsep=0ex,partopsep=1ex,parsep=1pt]
		\item The resolution of reconstructed images are limited by the resolution of illumination patterns. 
		\item Coded illumination hardware can bring additional cost and complexity to the system design. 
		\item Our current system cannot capture dynamic scenes because our model requires multiple measurements of every scene under different illumination patterns. 
		\item The shifting dots patterns are simple and easy to implement, but they reduce the light throughput. 
		\item The camera and projector are co-located in our setup; therefore coded illumination does not provide a direct advantage toward depth estimation. Reconstruction of 3D and dynamic scenes with this framework is possible and requires additional work \cite{zheng2022coded3d}. 
		\item Similar to other active illumination systems, the performance of our method will suffer under strong ambient lights (see supplementary).
		\item Our imaging model does not account for non-Lambertian objects, and any such object can cause artifacts in the reconstruction (see supplementary). 
		
	\end{itemize}

	\subsection{Error Analysis in Multi-shot System}\label{sec:noise}
	In this section, we discuss the effects of noise and system conditioning on the reconstruction error. 
	In our coded illumination method (especially, with shifting dots), each frame of the multi-shot system can have low SNR  compared to single-shot system, given the same capture time.  While the SNR of each frame is worse, the multi-shot systems offer much better conditioning, which in turn provides better reconstruction. 
	
	Let us represent the measurements captured by a reference lensless camera under uniform illumination in the following vectorized form: $y = Ax+e$, where ${A}$ denotes the system matrix, ${x}$ denotes the unknown image, and ${e}$ denotes the measurement noise. 
	We denote the measurement signal to noise ratio of this reference system as $\text{SNR}_{ref}$. Note that under the assumption that photon count can be approximated as $\mathcal{N}(\lambda t, \lambda t)$, $\text{SNR}_{ref} \propto \sqrt{\lambda t}$, where $\lambda$ represents rate of photon arrival and $t$ represents exposure time). 
	
	Suppose $A$ is full rank but ill-conditioned with singular values $s_1,\ldots, s_N$, and the least-squares solution is $\hat x$. If we capture $T$ measurements with uniform illumination (or increase exposure time of a single measurement by a factor of $T$), then the measurement SNR will increase by a factor of $\sqrt{T}$ but the singular values remain unchanged. A simple derivation (see details in \cite{schechner2007multiplexlighting}) can show that mean signal to mean reconstruction error ratio (SER) can be written as 
	\begin{equation}
	\text{SER}_{unif} = \frac{\mathbb{E}\|x\|_2}{\mathbb{E}\|x-\hat x\|_2} =  \frac{\sqrt{T}\, \text{SNR}_{ref}}{\sqrt{\text{Tr}({A}^\top {A})^{-1}}} =  \frac{\sqrt{T}\, \text{SNR}_{ref}}{\rho_{ref}}, 
	\end{equation}
	where $\rho_{ref} = \text{Tr}({A}^\top {A})^{-1}= \sqrt{\sum_{i=1}^N 1/s_i^2}$. 
	The reconstruction error is dominated by the small singular values as they will increase the denominator. 
	
	On the other hand, using coded illuminations will provide us a modified system matrix $\tilde A$ with singular values $\tilde s_1,\ldots, \tilde s_N$ that decay at a slower rate (as shown in Fig. \ref{fig:singular_values}). The SNR of coded-illumination measurements can be different from uniform illumination; for example, if we turn on every $k$th pixel in the illumination pattern, the measurement SNR will reduce by a factor of $\sqrt{k}$. In our equivalent design with shifting dots, we would capture $T = k$ measurements with shifting dots. %
	Therefore, the mean signal to mean reconstruction error ratio (SER) with $T$ shifting dots illumination patterns can be written as 
	\begin{equation}
	\text{SER}_{dots} = \frac{\mathbb{E}\|x\|_2}{\mathbb{E}\|x-\hat x\|_2} = \frac{\text{SNR}_{ref}}{\rho_{dots}},
	\end{equation}
	$\rho_{dots} = \text{Tr}(\tilde{A}^\top \tilde{A})^{-1}= \sqrt{\sum_{i=1}^N 1/\tilde s_i^2}$. 
	Note that if we use $T$ Hadamard patterns instead of shifting dots, then the SNR per measurement would not reduce, and the error bound above will not have the  $\sqrt{T}$ factor in the denominator. 
	
	In summary, suppose we use same total exposure time (or number of equal-exposure measurements) for uniform and shifting dots patterns, then 
	\begin{equation}\label{eq:SER_inequality}
	\frac{\text{SER}_{dots}}{\text{SER}_{unif}} = \frac{1}{T} \frac{\rho_{ref}}{\rho_{dots}}.
	\end{equation}
	In practice, the coded illumination-based system offers significantly better conditioning compared to reference system; that is, $\rho_{ref} \gg \rho_{dots}$. 
	
	We also show experimental results for  multi-shot coded illumination system and uniform pattern system with the same amount of exposure time in Fig.~\ref{fig:experiment_results_equal_exposure}(c,d). We can observe that the 49 shifting dots outperform the reconstruction using uniform illumination.

	\subsection{Coded Illumination with Nonseparable Systems}

	All the lensless cameras can be modeled as a linear system; therefore, the broader finding that adding coded illumination improves the conditioning of the system is valid across all the lensless designs. Our particular choice of separable masks and illumination patterns is indeed specific to FlatCam-inspired designs, where we gain  computational advantages as the solution can be written in a separable closed form.

	In principle, we can add coded illumination to convolutional models such as DiffuserCam as $\mathbf{Y}_i=\varphi*(\mathbf{P}_i\odot\mathbf{X})$, where $\varphi$ denotes PSF for the mask, $\mathbf{X}$ denotes the unknown image, $\mathbf{P}_i$ denotes $i$th illumination pattern, and $\mathbf{Y}_i$ denotes the corresponding sensor measurements. We do not have a closed-form solution for such an imaging model; therefore, reconstruction involves using iterative solver. 
	
\begin{figure}[t]
		\setlength\tabcolsep{1pt}
		\renewcommand{\arraystretch}{1} % Default value: 1
		\footnotesize
		\centering
		\begin{subfigure}[b]{0.9\linewidth}
			% \centering
			\begin{tabular}{cccc}
				uniform &
				49 shifting dots &
				uniform &
				49 shifting dots
				\\ \includegraphics[width=0.24\linewidth,keepaspectratio]{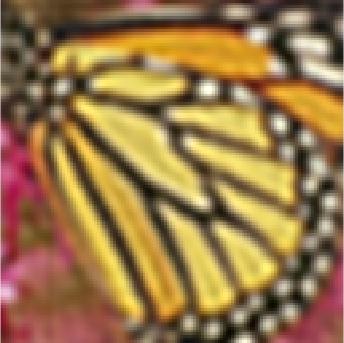} &
				\includegraphics[width=0.24\linewidth,keepaspectratio]{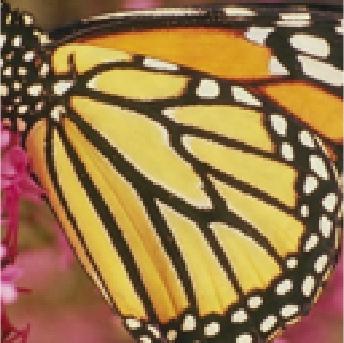}  &
				\includegraphics[width=0.24\linewidth,keepaspectratio]{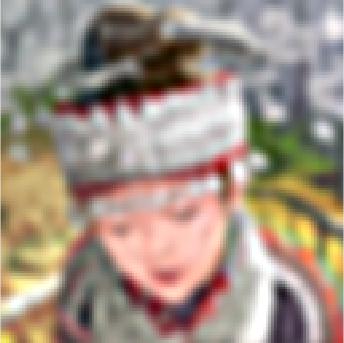} &
				\includegraphics[width=0.24\linewidth,keepaspectratio]{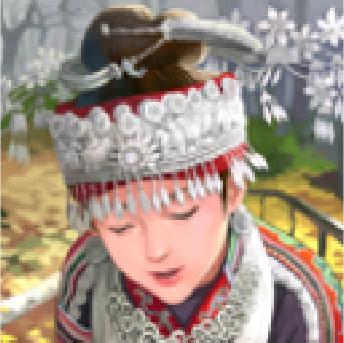} 
				% 		&
				\\
				PSNR: 18.96dB &
				38.46dB &
				21.71dB &
				42.53dB
				
				\\
				% 		\rotatebox{90}{\parbox{1.5cm}{\centering  DiffuserCam }}
				\includegraphics[width=0.24\linewidth,keepaspectratio]{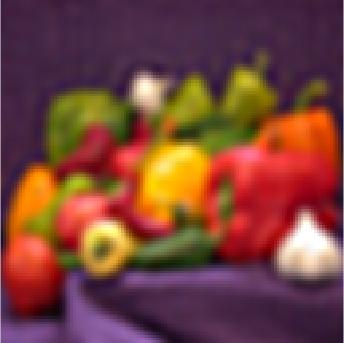} &
				\includegraphics[width=0.24\linewidth,keepaspectratio]{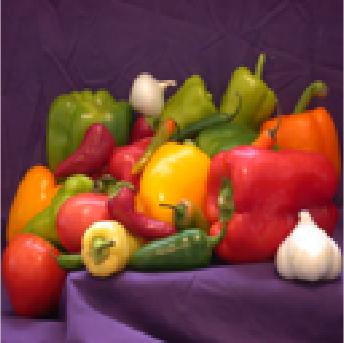}  &
				\includegraphics[width=0.24\linewidth,keepaspectratio]{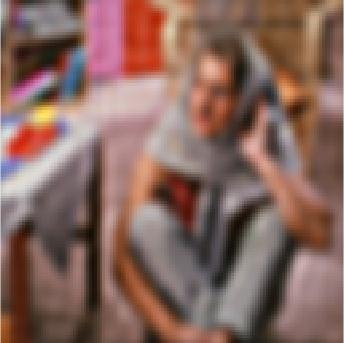} &
				\includegraphics[width=0.24\linewidth,keepaspectratio]{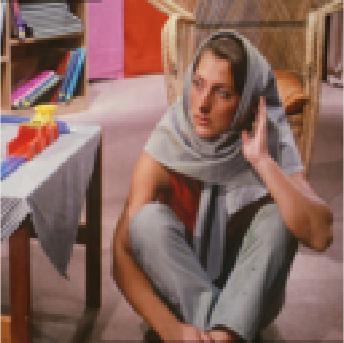} 
				% 		&
				\\
				PSNR: 19.23dB &
				44.51dB &
				24.56dB &
				44.52dB
			\end{tabular}
			\caption{Example reconstructed images using DiffuserCam model \cite{antipa2018diffusercam}. }
		\end{subfigure}
		~
		\begin{subfigure}[b]{0.49\linewidth}
			% \centering
			\includegraphics[width=1\linewidth,keepaspectratio]{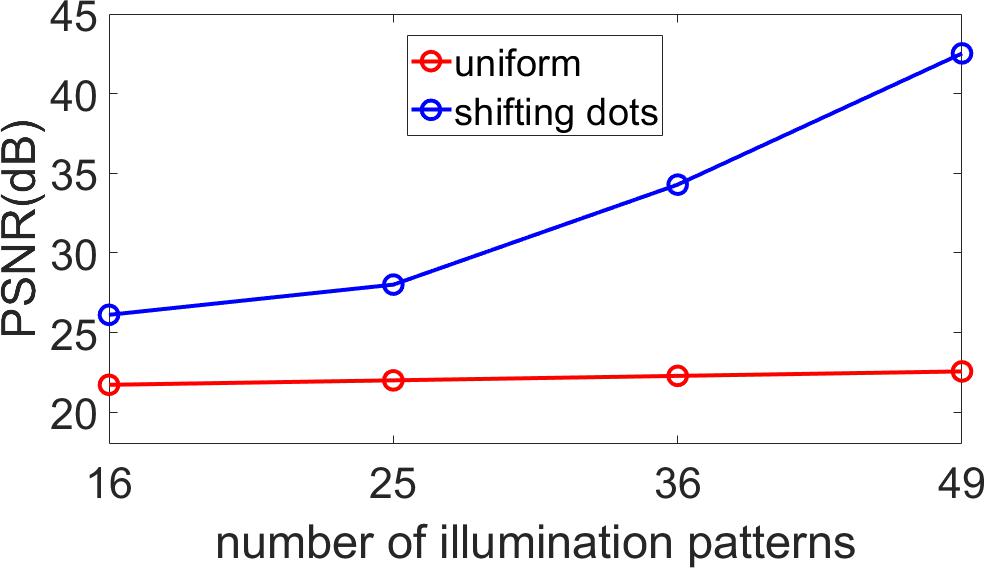}
			\caption{Average PSNR values. }
		\end{subfigure}
		\begin{subfigure}[b]{0.49\linewidth}
			% \centering
			\includegraphics[width=1\linewidth,keepaspectratio]{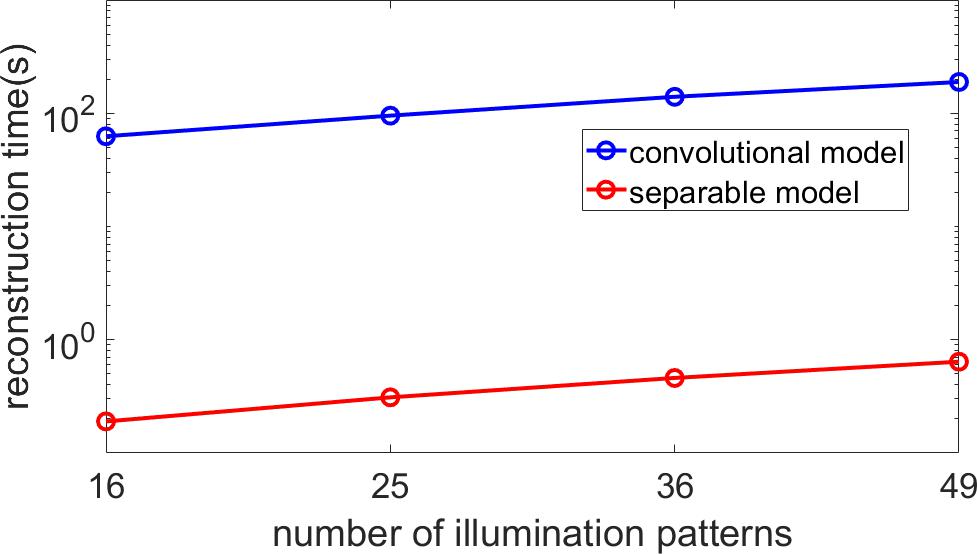} % xlabel 
			\caption{Average reconstruction time. }
			\label{fig:conv_recon_time}
		\end{subfigure}
		\caption{Simulation results with uniform illumination and 49 shifting dots coded illumination patterns using the convolution model in DiffuserCam~\cite{antipa2018diffusercam}. (a) Sample reconstructed images in (a) and average reconstruction PSNR vs number of patterns in (b) show that coded illumination patterns improve the imaging performance of DiffuserCam. (c) Average reconstruction time for the convolutional model is an order of magnitude larger than the separable model that has a simple closed-form solution. 
		}
		\label{fig:simu_diffusercam}
	\end{figure}

	To validate our argument, we also provide simulation results in Fig.~\ref{fig:simu_diffusercam} combining our coded illumination system with the convolution model proposed in DiffuserCam\cite{antipa2018diffusercam}. The PSFs used in the simulation are also from the DiffuserCam data. 
	We observe in Fig.~\ref{fig:simu_diffusercam} that the coded illumination pattern also improves the condition of the DiffuserCam.
	
	Combining a convolution model with coded illumination has some additional practical challenges that do not arise in the separable model we presented earlier in the paper. Two main challenges we encounter are (1) alignment of spatial grids and (2) reconstruction complexity. 
	The alignment process is necessary to avoid any model mismatch. The sampling grid in the convolutional model cannot be arbitrary, as it is deteremined by the PSF and sensor pixel pitch. The alignment requires additional (tedious) calibration steps to estimate relative sizes/displacements of illumination pixels and scene pixels. We leave the calibration and hardware implementation tasks for future work. In contrast, in our paper, we used a separable model that can be calibrated using the illumination pattern itself (as long as illumination grid yields a separable response on the sensor). Coded illumination-based convolutional models do not have a simple closed-form solution because the combined operator cannot be diagonalized with Fourier transform. We can implement the forward and adjoint operators using Fourier transform, but the overall reconstruction procedure is much slower than the separable model we used in our paper, as shown in Fig.~\ref{fig:conv_recon_time}.

	% \noindent \textbf{Acknowledgments.} The authors acknowledge Rongjia Zhang's help with initial simulations. This work is supported in part by NSF Awards CCF-2046293, CMMI-2133084 and AFOSR Award FA9550-21-1-0330.
	
	\noindent \textbf{Supplementary material.} Additional simulation and experimental results are in the supplementary document. The codes are available at \url{https://github.com/CSIPlab/codedcam}.

	\bibliographystyle{IEEEtran}
	\bibliography{TCI_bib}
	
%	\begin{IEEEbiography}[{\includegraphics[width=1in,height=1.25in,clip,keepaspectratio]{figures/yucheng.jpg}}]{Yucheng Zheng} (Student Member, IEEE)~ received the B.Sc. degree in electrical engineering from the Nanjing University of Aeronautics and Astronautics, Nanjing, China in 2017. He is currently working toward the Ph.D. degree at the University of California, Riverside, CA, USA. His current research interests include computational imaging, computer vision and signal processing.
%	\end{IEEEbiography}
%	\vfill
%	\begin{IEEEbiography}[{\includegraphics[width=1in,height=1.25in,clip,keepaspectratio]{figures/salman.jpg}}]{M. Salman Asif} (Senior Member, IEEE)~received his B.Sc. degree from the University of Engineering and Technology, Lahore, Pakistan, and his M.S and Ph.D. degrees from the Georgia Institute of Technology, Atlanta, Georgia, USA. He is currently an Associate Professor at the University of California Riverside, USA. Prior to that he worked as a Postdoctoral Researcher at Rice University and a Senior Research Engineer at Samsung Research America, Dallas. He has received NSF CAREER Award, Google Faculty Award, Hershel M. Rich Outstanding Invention Award, and UC Regents Faculty Fellowship  and Development Awards. His research interests include computational imaging, signal/image processing, computer vision, and machine learning. 
%	\end{IEEEbiography}

\clearpage

\onecolumn

\noindent \begin{center} {\large  {Coded Illumination for Improved Lensless Imaging\\ Supplementary Material}} \end{center}

\begin{appendices}

In this document we provide the following additional simulation and experimental results: 
\begin{itemize}
	\item Simulations to evaluate effects of different number and types of illumination patterns,  sensor-to-mask distances, and numbers of sensor pixels. 
	\item Experiments to evaluate performance with  different number and types of illumination patterns, with different number of sensor pixels, under ambient lights, for non-lambertian scenes, and with TV-$\ell_1$ regularization.
	\item Simulations and experiments to compare performance of deep networks for refinement and reconstruction using uniform and coded illumination. 
	
\end{itemize}

\section{Simulation Results}
We follow the same simulation setup as described in the main text. To validate the performance of the proposed algorithm, we simulated a lensless imaging system where a separable coded-mask is placed on top of an image sensor, the sensor-mask distance is 2mm. We simulate a $128\times128$ planar scene that is 40cm away from the sensor, and the width of scene is 12cm. We use a separable MLS mask, where the size of each feature is 60$\mu$m. The length of each sensor pixel in the simulation is 11.2$\mu$m and the total number of captured sensor pixels in one frame is $512\times512$. We use these settings as default and change one parameter at a time while performing different simulation tests.
We first test the performance of our system for different type and number of illumination patterns. 
The conditioning of the imaging system is affected by the sensor-mask distance and the number of sensor pixels. When the sensor-mask distance is small, the PSFs of neighbouring pixels are similar to each other and the system is likely to be ill-conditioned. When the number of sensor pixels is small, we may have fewer measurements than the unknown, which results in an under-determined system. We show the performance of our method with different sensor-mask distances and different numbers of sensor pixels. We use the shifting-dots illumination patterns for these experiments.

\subsection{Effect of Illumination Patterns}

\begin{figure*}[t]
	\centering
	\includegraphics[width=\linewidth]{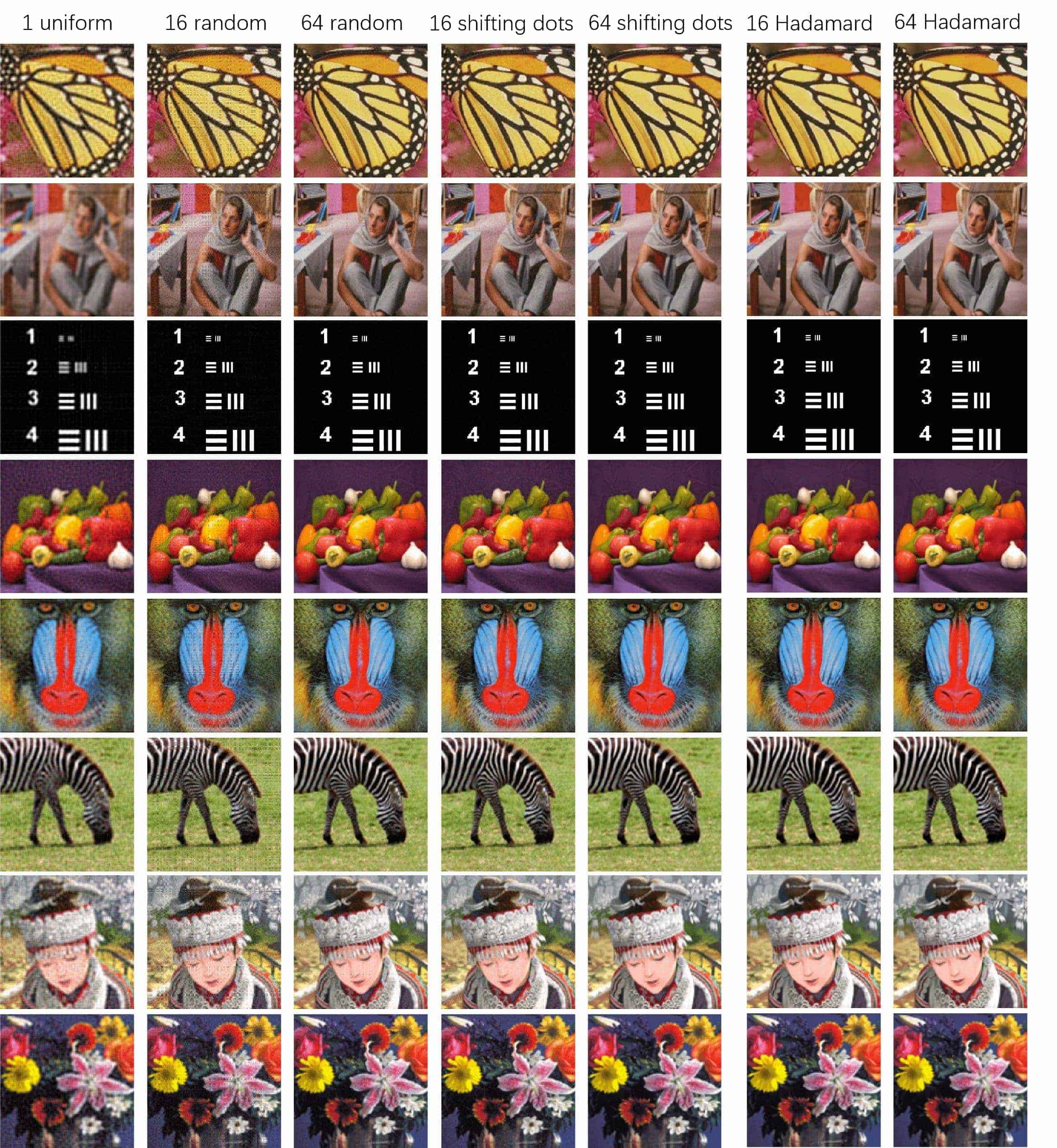}
	\caption{The simulation results for different \textbf{types} and \textbf{numbers} of illumination patterns, including uniforn pattern, 16 and 64 random, shifting dots, and Hadamard patterns. The quality of reconstruction improves as the number of illumination patterns increases. The shifting dots patterns and Hadamard patterns outperform random and uniform patterns. 
	}
	\label{fig:results_illumPattern}
\end{figure*}

We show the simulation results using different types and number of patterns in Fig.~\ref{fig:results_illumPattern}. 
We show the results reconstructed from measurements using uniforn pattern, 16 and 64 random, shifting dots, and Hadamard illumination patterns respectively. We observe that more illumination patterns significantly improve the reconstruction quality with respect to noise attenuation and image resolution. We also observe that the shifting dots results outperform the random illumination results. We also observe that the Hadamard patterns have the best performance. The shifting dots and Hadamard patterns have the same singular values response, but the Hadamard patterns have a larger light throughput. 
Intuitively, the neighbouring pixels tend to contribute very similar sensor measurements so it is hard to distinguish neighbouring pixels using a single uniform illumination. By applying shifting dots patterns to data acquisition, we separate the pixels that are captured in the same frame.   

\clearpage
\subsection{Effect of Sensor-to-Mask Distance}
\begin{figure*}[ht]
	\centering
	\includegraphics[width=\linewidth]{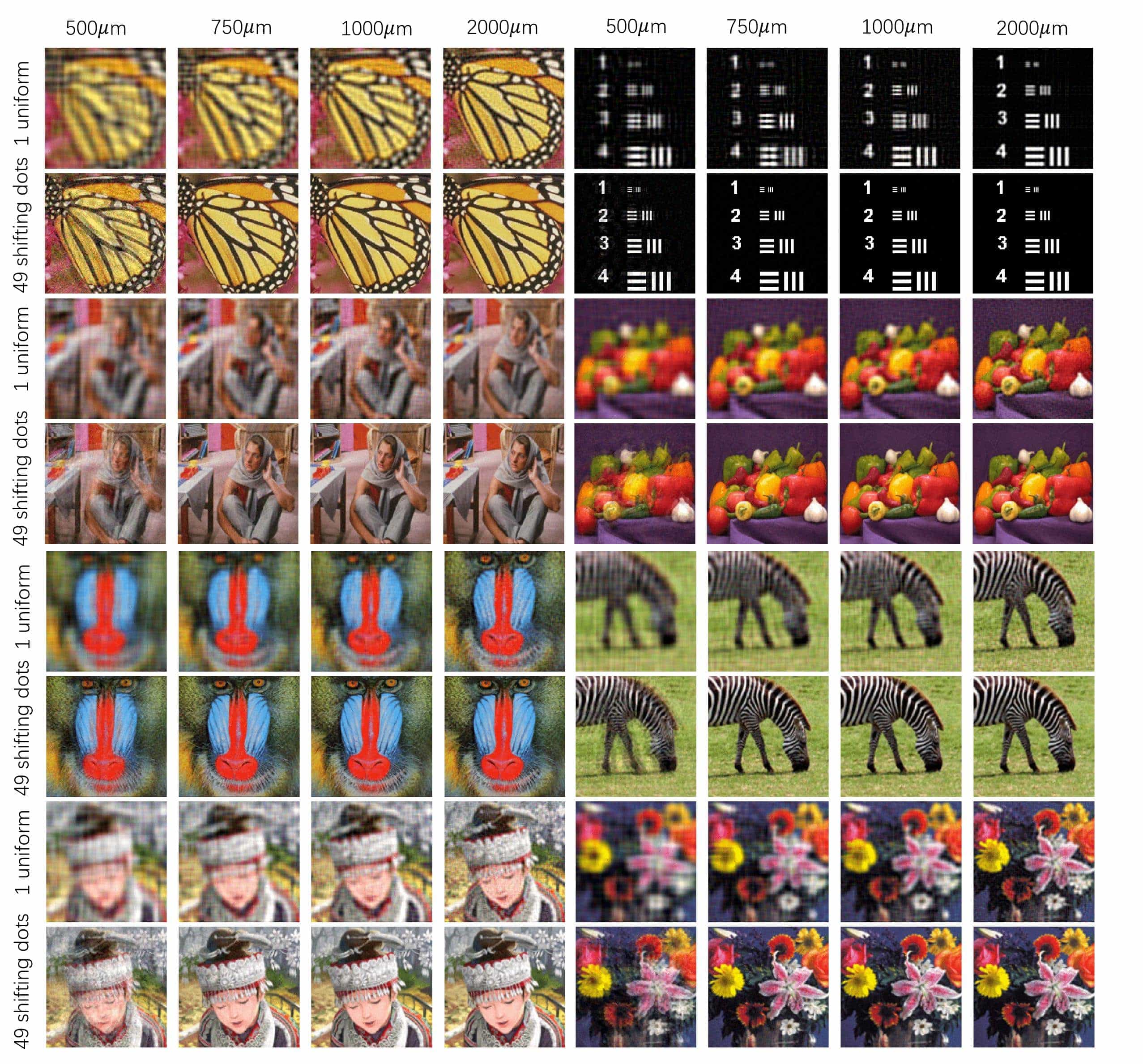}
	\caption{The simulation results for \textbf{sensor-to-mask distances} at 500$\mu$m, 750$\mu$, 1000$\mu$m, and 2000$\mu$m. The reconstruction quality improves as the sensor-to-mask distance increases. With 49 shifting dots patterns, the reconstruction quality is robust even if the sensor-to-mask distance is small. }
	\label{fig:results_sensor2mask}
\end{figure*}

The sensor mask distance of lensless imaging system greatly influences the conditioning of the system. We present simulation results using different sensor-mask distances.
We keep the sensor size fixed at $512\times512$, while the sensor-to-mask distance increases from $500\mu$m to $750\mu$m, $1000\mu$m and $2000\mu$m. The examples of reconstruction for different test images are shown in Fig. ~\ref{fig:results_sensor2mask}. We observe in the image array that as the sensor to mask distance increases, the reconstruction quality improves. Compared to the original results using single uniform illumination pattern (first row), using multiple coded illumination patterns (second row) offers us a significant improvement.

\clearpage
\subsection{Effect of Number of Sensor Pixels}
\begin{figure*}[ht]
	\setlength\tabcolsep{1pt}
	\centering
	\includegraphics[width=\linewidth]{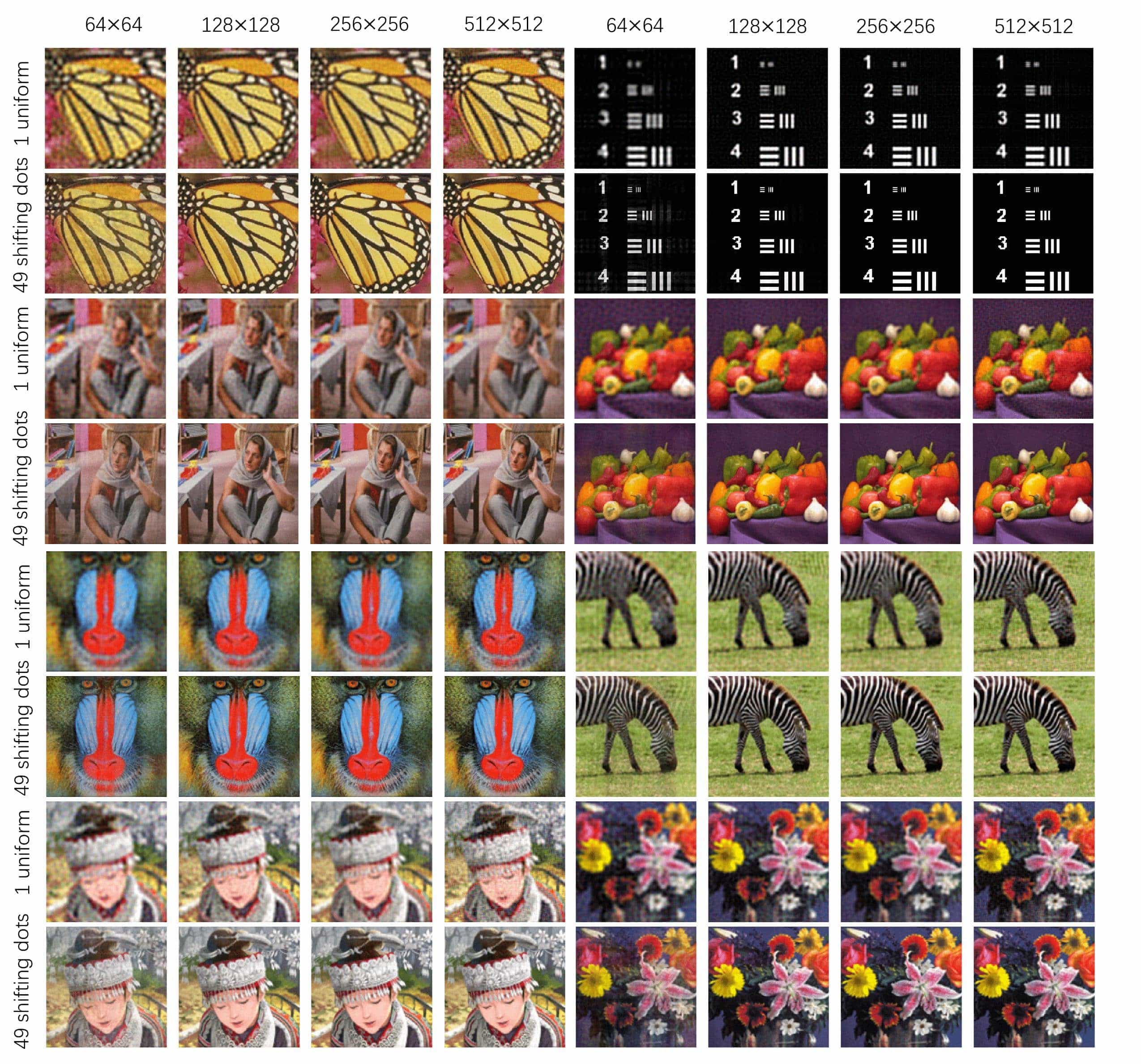}
	\caption{Simulation results of testing scenes using different \textbf{numbers of sensor pixels}.  Each frame of the simulated sensor measurements have $512\times 512$ entries, which are binned into $256\times 256, 128\times 128$, and $64\times 64$ pixels. The resolution of the scene is $128\times128$. }
	\label{fig:results_numsensorpix}
\end{figure*}

The number of measurements also influences the conditioning of system. In particular, when measurements are fewer than the unknowns, the system become under-determined. We present simulation results with $64\times64$, $128\times128$, $256\times256$, and $512\times 512$ sensor pixels while keeping the sensor-to-mask distance 2000$\mu$m. The resolution of target image is $128\times128$, which means we have both under-determined case and over-determined cases in our test.
We show example reconstruction in Fig. ~\ref{fig:results_numsensorpix}. We observe that, under the uniform illumination, the image quality improves as we increase the number of sensor pixels. Using multiple illumination patterns provides significantly better quality in all cases.

\clearpage
\section{Experimental Results}
We present some additional results from the data captured with our prototype, as described in the main text. 

\subsection{Additional Experiment Results}
\begin{figure*}[t]
	\centering
	\includegraphics[width=1\linewidth]{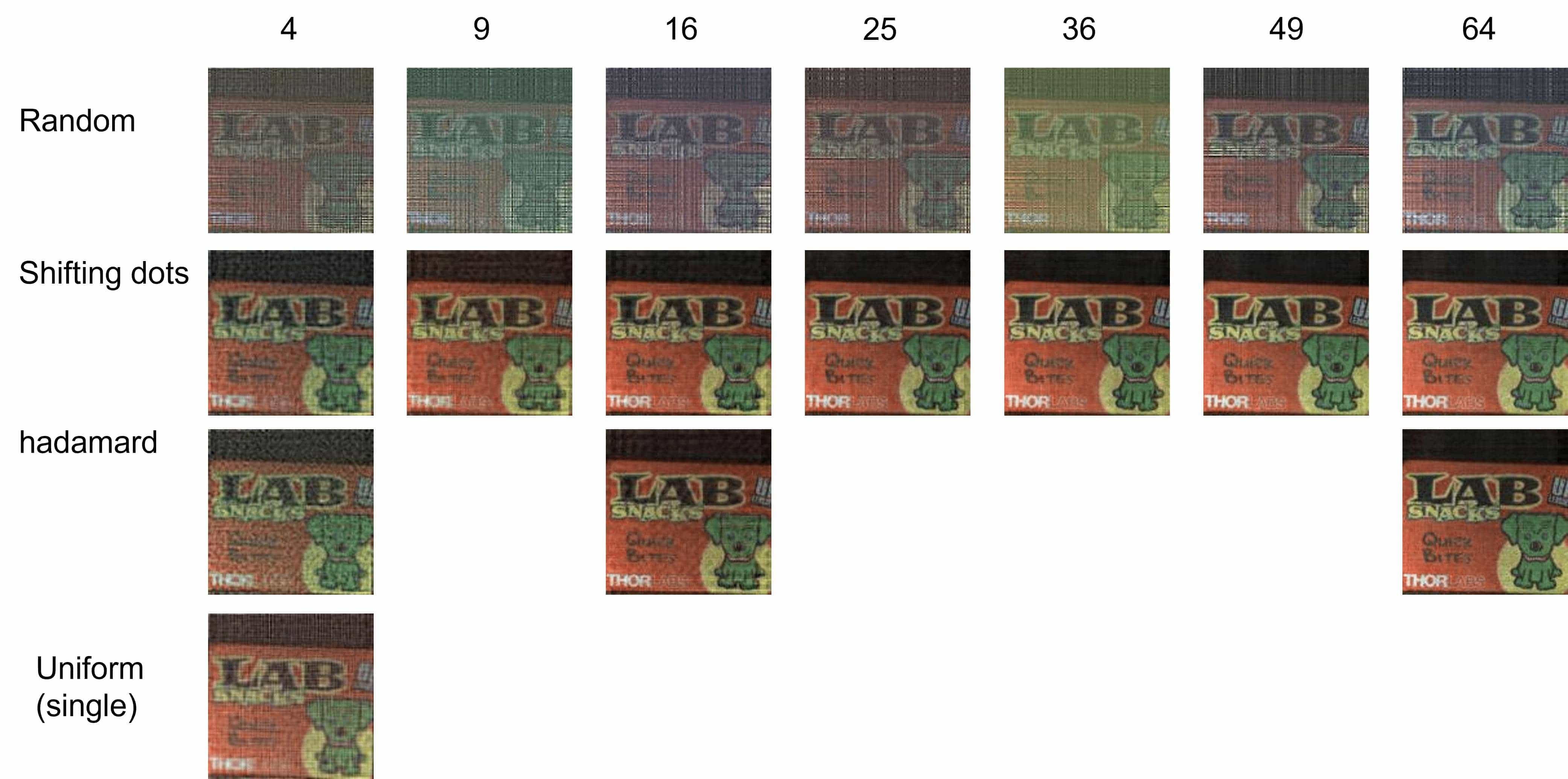}
	\includegraphics[width=1\linewidth]{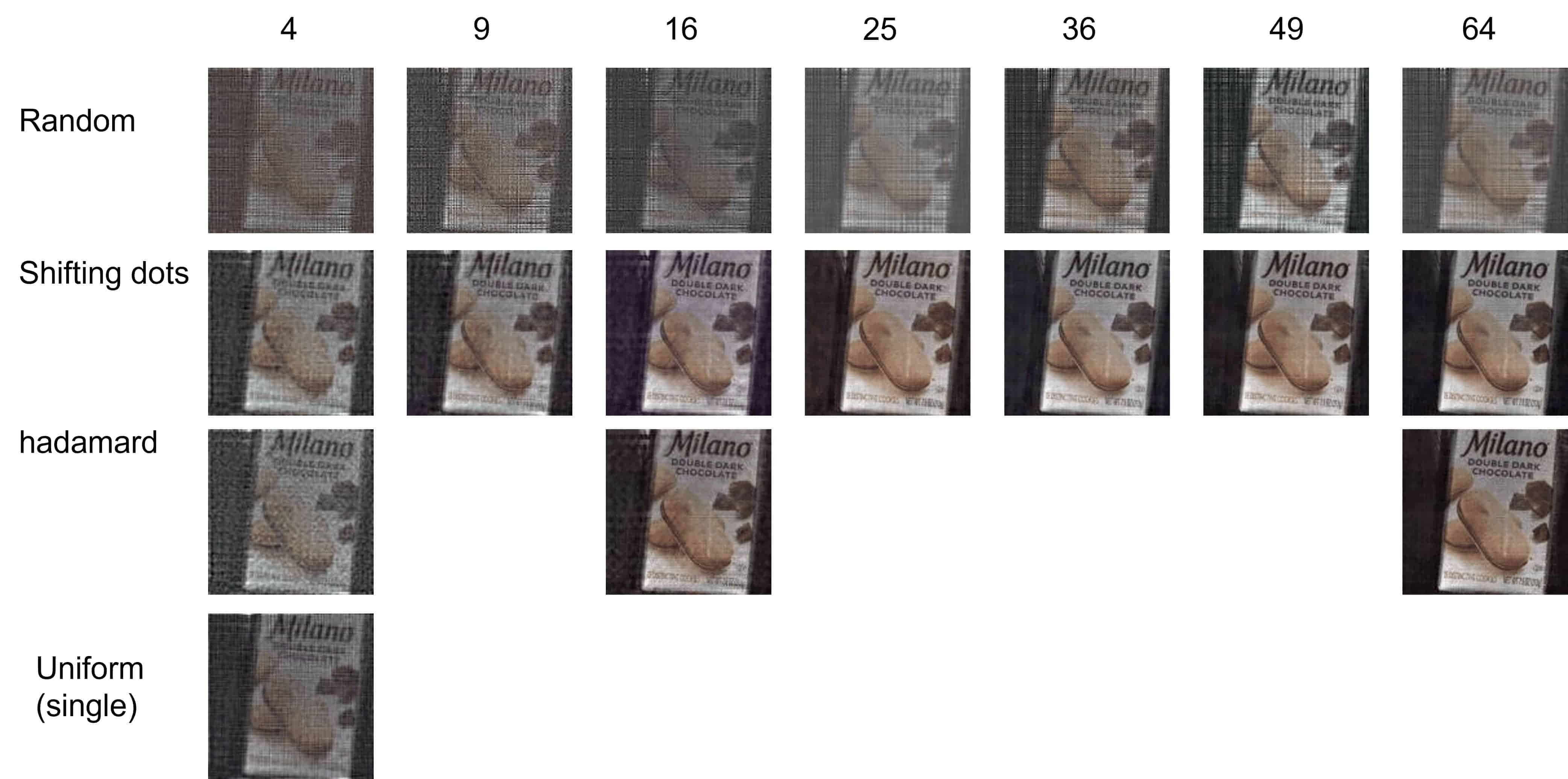}
	\caption{Experimental results of real scenes with different numbers and types of illumination patterns. The test scenes are real objects with some depth variation. Our method with coded illumination is robust with depth variation. }
	\label{fig:experiment_results_additionalpatterns2}
\end{figure*}

We present hardware results on five test scenes in Fig.~\ref{fig:experiment_results_additionalpatterns1} and Fig.~\ref{fig:experiment_results_additionalpatterns2}, including planar scene "K card", "color chart", "resolution chart", which are printed on pieces of paper and real object scenes "Thorlabs box", "cookie bag", which have some depth variation. We test uniform illumination, different numbers of random and orthogonal patterns (shifting dots and hadamard).

\begin{figure*}[h]
	\centering
	\includegraphics[width=1\linewidth]{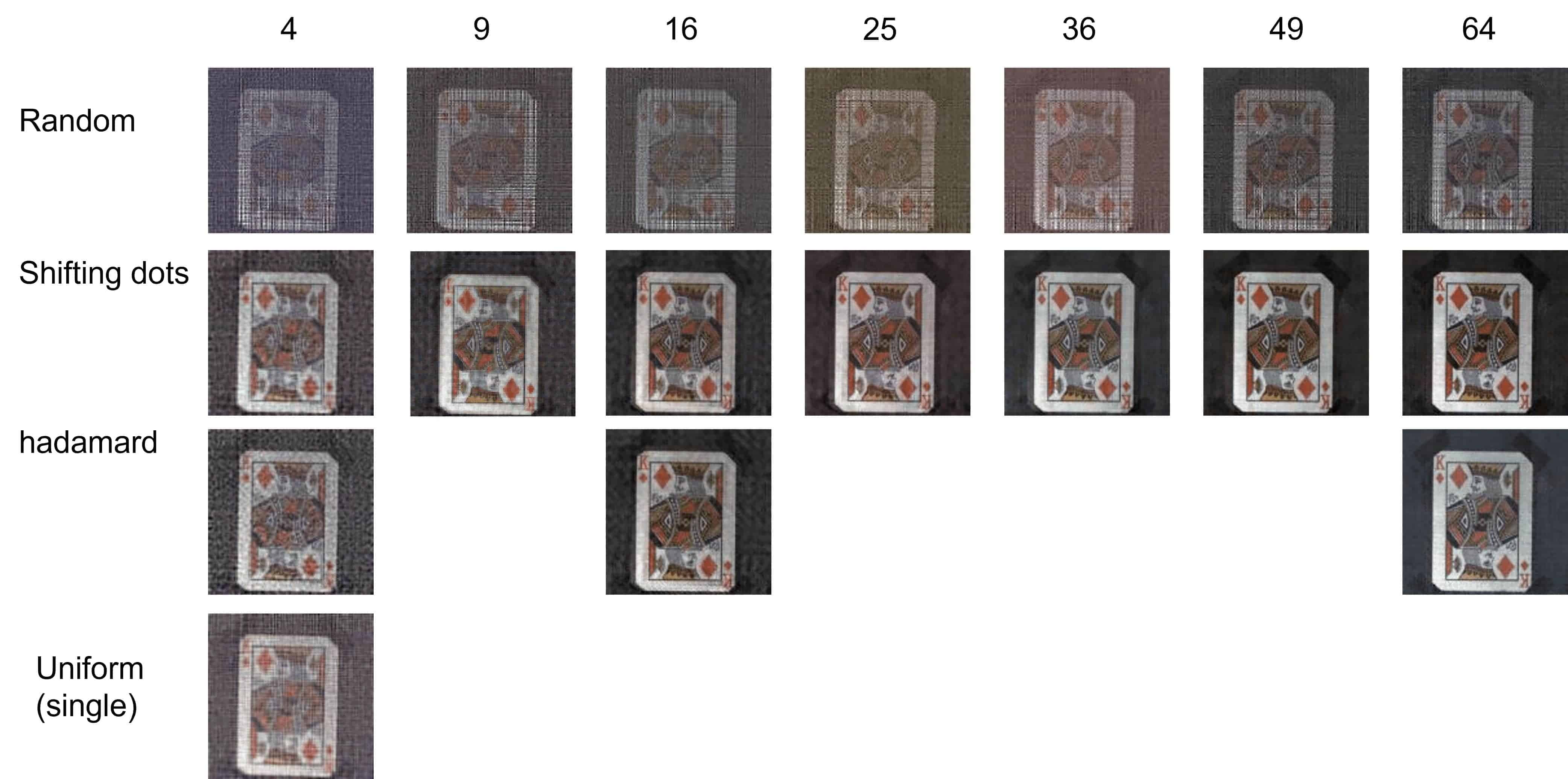}
	\includegraphics[width=1\linewidth]{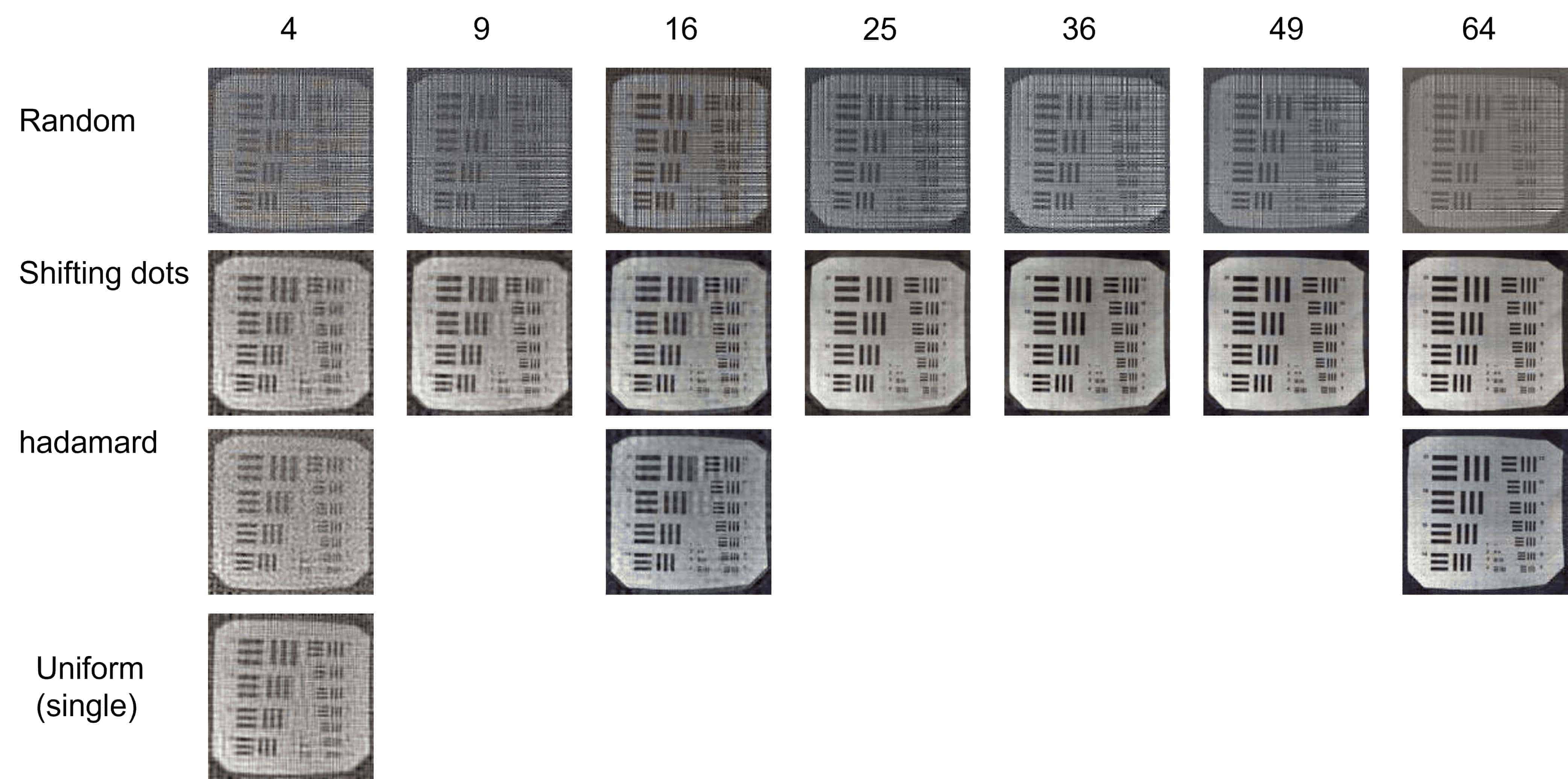}
	\caption{Experimental results of real scenes with different numbers and types of illumination patterns. The test scenes are planar scene printed on pieces of paper. Orthogonal patterns outperform all the other patterns. }
	\label{fig:experiment_results_additionalpatterns1}
\end{figure*}

\begin{figure*}[h]
	\centering 
	\includegraphics[width=1\linewidth]{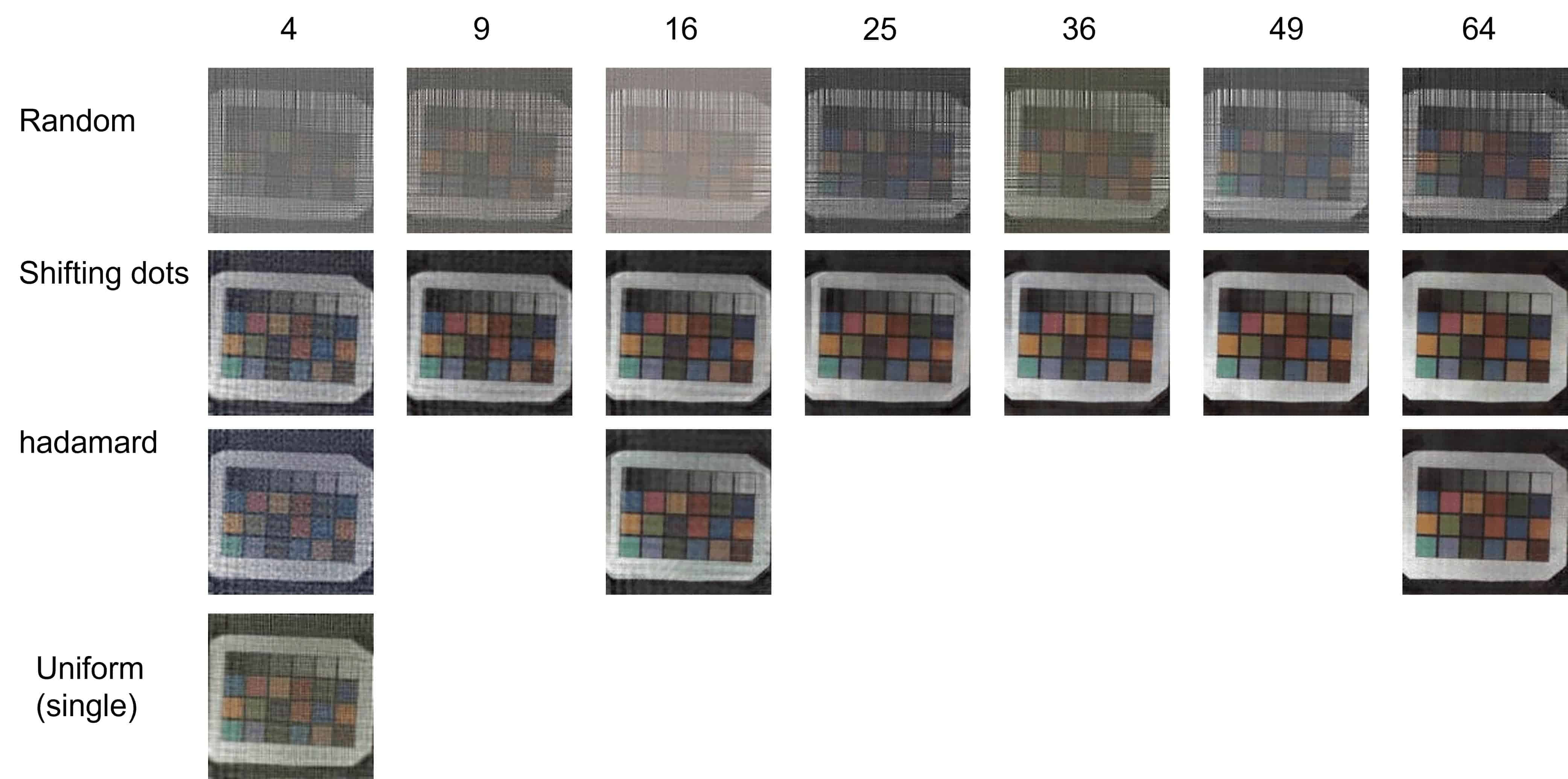}
	\caption{Experimental results of real scenes with different numbers and types of illumination patterns. The test scenes are planar scene printed on pieces of paper. Orthogonal patterns outperform all the other patterns. }
	\label{fig:experiment_results_additionalpatterns1}
\end{figure*}
\clearpage
\subsection{Compressive Sensor Measurements}
In lensless imaging, the number of sensor measurements often surpasses the number of reconstruction pixels for stable recovery. This may result in low resolution of lensless imaging or more costly sensor. The application of coded illumination can expand the number of measurements without increasing the number of sensor pixels. We present additional experimental results for different binning factor in Fig.~\ref{fig:experiment_results_sensor_binning}. In these experiments, we capture $512\times512$ measurements and bin them to $64\times64, 128\times128$ and $256\times256$ by averaging the neighbouring pixels in post-processing, the reconstruction image has $128\times128$ pixels, the $128\times128$ and $64\times64$ columns are under-determined. We show the result of three additional scenes under different binning, in which we observe that the results of using 49 shifting dots illumination patterns remain recognizable and stable even if the measurements are binned to $64\times64$ sensors, while the results corresponding to the conventional (single) uniform illumination degrades quickly as the binning factor increases. 

\begin{figure*}[htb]
	\setlength\tabcolsep{1pt}
	\centering
	\includegraphics[width=.7\linewidth]{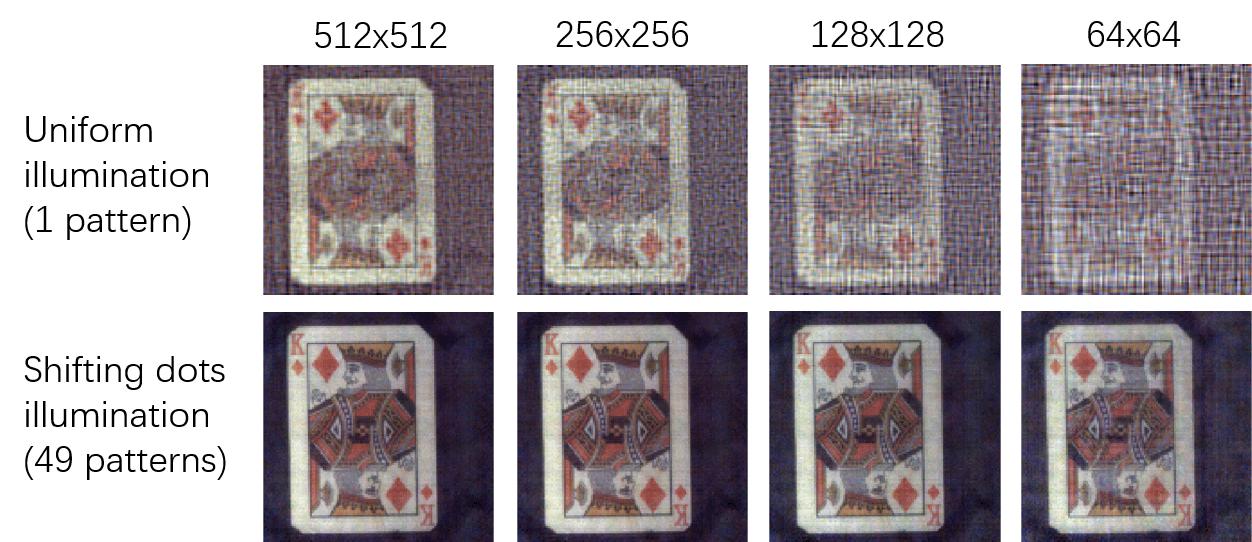}
	\includegraphics[width=.7\linewidth]{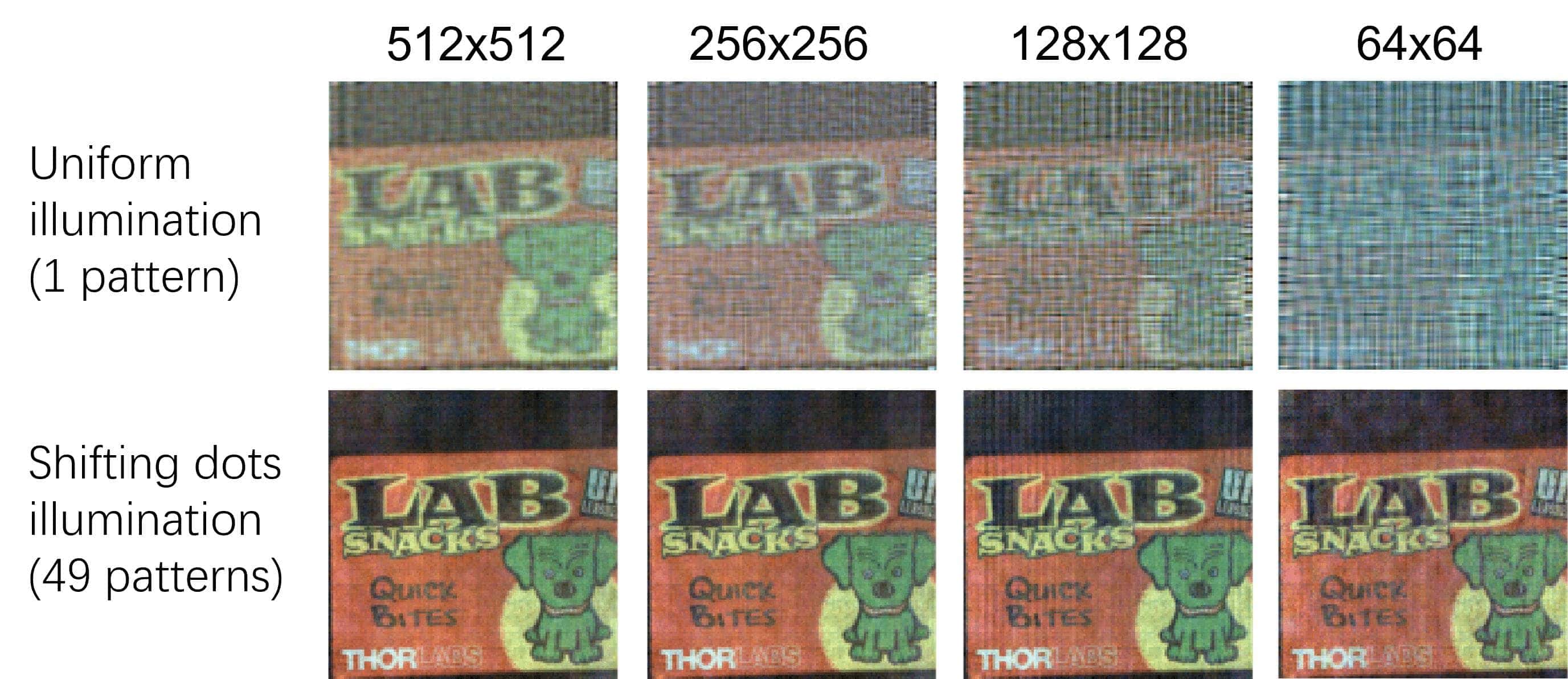} 
	\includegraphics[width=.7\linewidth]{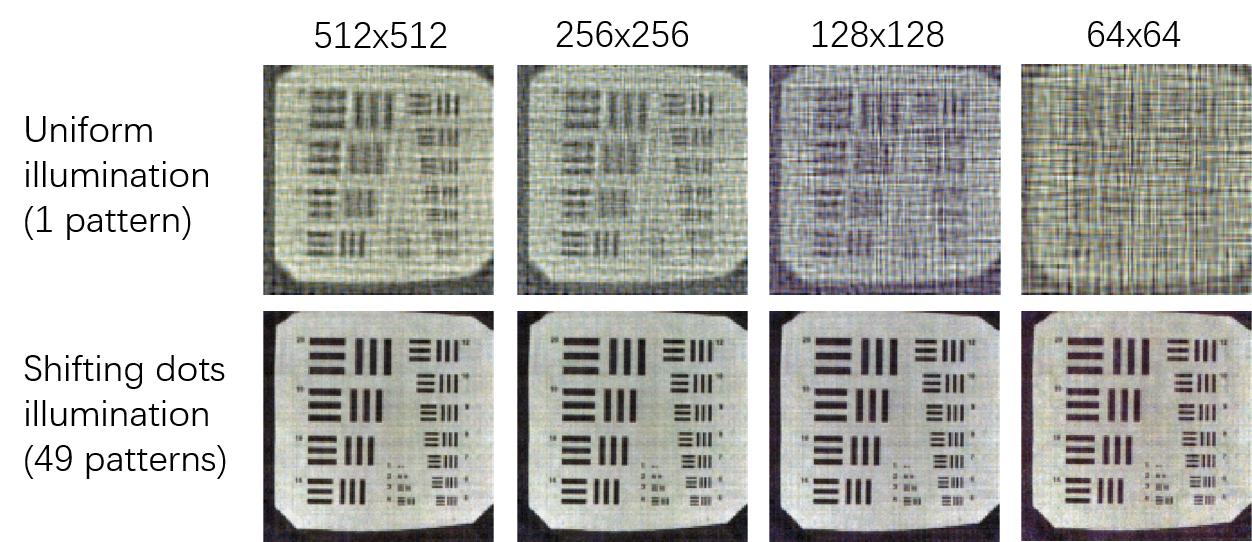} 
	
	\caption{Experimental results of real scenes with measurements at different binning. The original sensor provided $512\times 512$ measurements, which are binned into $256\times 256, 128\times 128$, and $64\times 64$ pixels. Since the shape of the reconstructed image is $128\times128$, the imaging system is under-determined when the sensor is binned to $64\times64$. Our method with 49 shifting dots patterns recovers near-perfect reconstruction at different levels of binning (compression) factors.  }
	\label{fig:experiment_results_sensor_binning}
\end{figure*}

\clearpage
\subsection{Performance with Ambient Lights}
We performed most of our experiments without ambient lights. Like any other active illumination system, strong ambient lights will affect the performance of our system. A common practice to cancel the ambient light effects is to subtract a reference image from the sensor measurements. Strong ambient light can result in a very small dynamic range of the lensless measurements after subtraction of the reference image. In the worst case, the ambient light can be so strong that measurements with different coded illumination are the same as measurements with one uniform illumination pattern.

\begin{figure}[h]
	\begin{subfigure}[t]{0.24\linewidth} 
		\centering
		\includegraphics[width=1\linewidth]{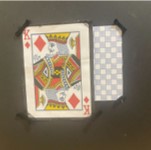}
		\caption{Test scene with bright light} 
	\end{subfigure}
	\begin{subfigure}[t]{0.24\linewidth} 
		\centering
		\includegraphics[width=1\linewidth]{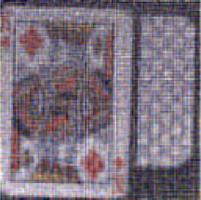}
		\caption{Uniform}
	\end{subfigure}
	\begin{subfigure}[t]{0.24\linewidth} 
		\centering
		\includegraphics[width=1\linewidth]{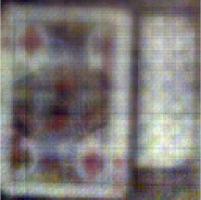}
		\caption{64 shifting dots}
	\end{subfigure}
	\begin{subfigure}[t]{0.24\linewidth}
		\includegraphics[width=1\linewidth]{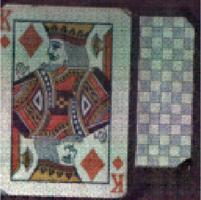}
		\caption{64 Hadamard patterns}
	\end{subfigure}
	\caption{Performance of different coded illumination patterns under strong ambient light. We show the dynamic range (DR) of measurements after reference image subtraction. (a) Test scene, (b) measurements captured with low ambient light (DR is from 0.11 to 0.26), (c) 64 shifting dots (DR is from 0 to 0.02), and (d) 64 Hadamard patterns ( DR is from 0.06 to 0.17). 
		The shifting dots patterns are dark compared to the strong ambient lights; therefore, its DR after reference image subtraction is small, which leads to poor reconstruction. In contrast, Hadamard patterns (that have exactly the same conditioning) provide much better results because they have a  larger DR after reference image subtraction. }
	\label{fig:experiment_results_ambient}
\end{figure}

We present one experimental result in Fig.~\ref{fig:experiment_results_ambient}, which we analyzed the effects of ambient lights on our system. 
We placed the scene in a strong ambient overhead light and captured measurements using one uniform, 64 shifting dots pattern, and 64 Hadamard pattern illuminations. 
The relative brightness of ambient light and projector affects the dynamic range of the measurements after reference image subtraction, which directly affects the reconstruction. 
Shifting dots pattern-based measurements have a small dynamic range (near the noise floor), which causes poor reconstruction. In contrast, Hadamard patterns that have the same conditioning for the underlying system as the shifting dots, offer much better dynamic range and the quality of reconstruction. {However, since Hadamard patterns contain both -1 and +1 values, it may take double capturing time compared to shifting dots. Therefore, shifting dots patterns are still preferred when the ambient lights are not strong. }

% reflective scenes
\clearpage
\subsection{Performance with non-Lambertian Objects}
Lensless imaging models (including this work) assume that all the objects in the scene have Lambertian surfaces. 
A strongly reflective surface in the scene would violate this assumption and the resulting model mismatch can cause severe artifacts in the reconstruction.
We present an experimental result to demonstrate this effect with different reflective objects in Fig~\ref{fig:experiment_results_reflective}. We created scenes with three reflective objects: gift card, metal blade, and CD. We observe that the results for gift card and metal blade are quite good while the results for a more reflective CD object has severe artifacts.

\renewcommand{\figwidth}{0.25\linewidth}
\begin{figure}[htb]
	\setlength\tabcolsep{1pt}
	\centering
	\footnotesize
	\begin{tabular}{ccccc}
		scene image &
		uniform &
		49 shifting dots &
		\\
		\includegraphics[width=\figwidth,keepaspectratio]{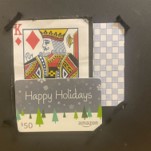} &
		\includegraphics[width=\figwidth,keepaspectratio]{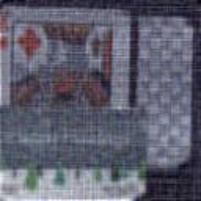} &
		\includegraphics[width=\figwidth,keepaspectratio]{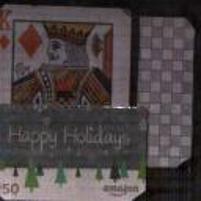}\\
		\includegraphics[width=\figwidth,keepaspectratio]{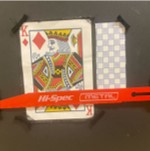} &
		\includegraphics[width=\figwidth,keepaspectratio]{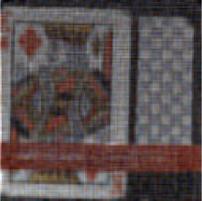} &
		\includegraphics[width=\figwidth,keepaspectratio]{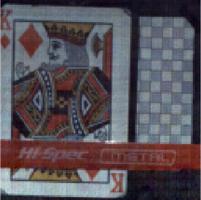}\\
		\includegraphics[width=\figwidth,keepaspectratio]{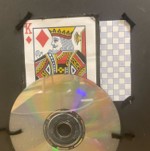} &	
		\includegraphics[width=\figwidth,keepaspectratio]{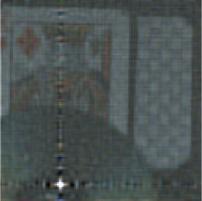} &
		\includegraphics[width=\figwidth,keepaspectratio]{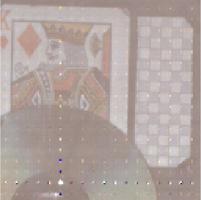}
	\end{tabular}
	\caption{Results of our method on three scenes with non-Lambertian surfaces: reflective gift card, metal blade, and CD. The CD is the most reflective of three. We observe good reconstruction for the gift card and the metal blade scene, whereas the reconstruction of the scene with CD contains severe artifacts.
	}
	\label{fig:experiment_results_reflective}
\end{figure}

\clearpage

\subsection{Effect of Regularization Functions}
We present additional results in Fig.~\ref{fig:experiment_results_TV} that compare  $\ell_2$-regularized least squares and and TV-$\ell_1$ regularization. TV-$\ell_1$ regularization 
is widely used for image recovery when the system is ill-conditioned 
As we discussed in the paper, the system conditioning improves significantly with coded illumination; therefore, we can get stable reconstruction using least squares method. We observe in Fig.~\ref{fig:experiment_results_TV}, the TV-$\ell_1$ provides slightly cleaner images, but they are not remarkably different from the $\ell_2$-regularized least squares results. 
Furthermore, the computational and storage advantages provided by the least squares formulation are not available with $\ell_1$ methods. 

\renewcommand{\figwidth}{0.24\linewidth}
\begin{figure*}[h]
	\setlength\tabcolsep{1pt}
	\centering
	\footnotesize
	\begin{tabular}{cccc}
		uniform &
		49 shifting dots &
		uniform &
		49 shifting dots 
		\\
		\rotatebox{90}{\parbox{1.5cm}{\centering  $\ell_2$ Regularized. }}
		\includegraphics[width=\figwidth,keepaspectratio]{figures/img01_pattern01.jpg} &
		\includegraphics[width=\figwidth,keepaspectratio]{figures/num49_img01_pattern03.jpg} &
		\includegraphics[width=\figwidth,keepaspectratio]{figures/img04_pattern01_num01.jpg} &
		\includegraphics[width=\figwidth,keepaspectratio]{figures/img04_pattern03_num49.jpg} \\
		\rotatebox{90}{\parbox{1.5cm}{\centering  TV-$\ell_1$ Regularized. }}
		\includegraphics[width=\figwidth,keepaspectratio]{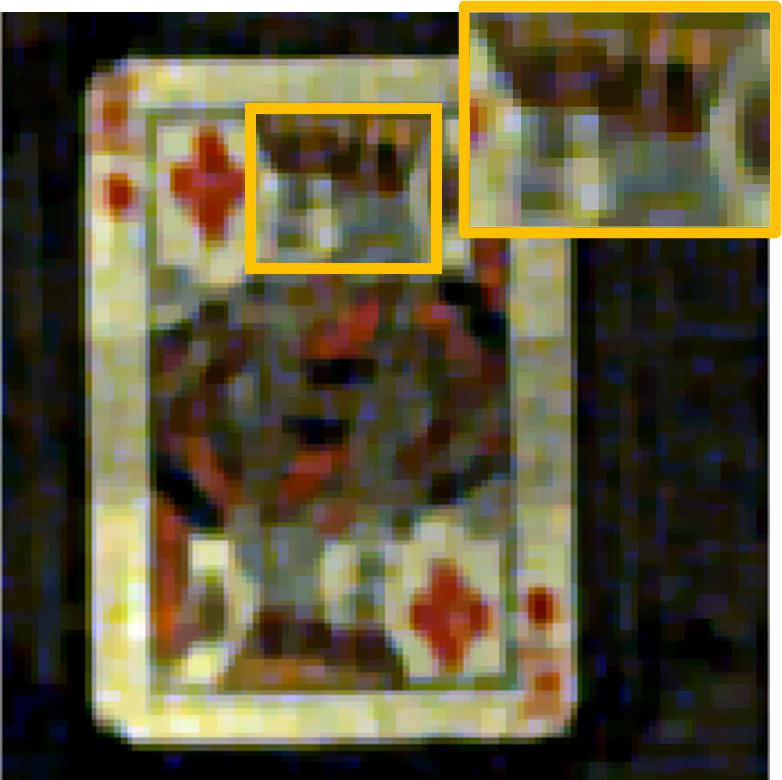} &
		\includegraphics[width=\figwidth,keepaspectratio]{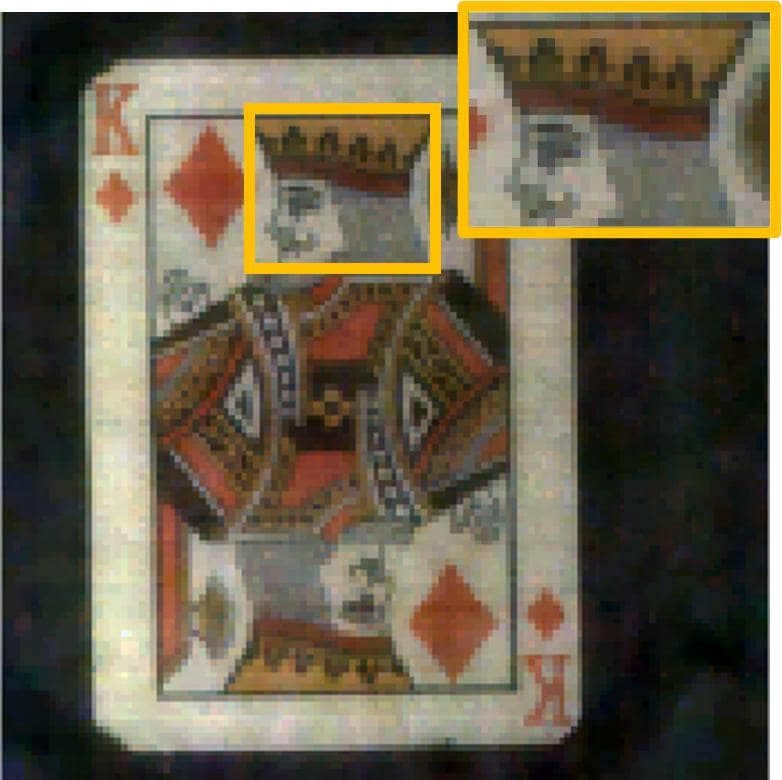} &
		\includegraphics[width=\figwidth,keepaspectratio]{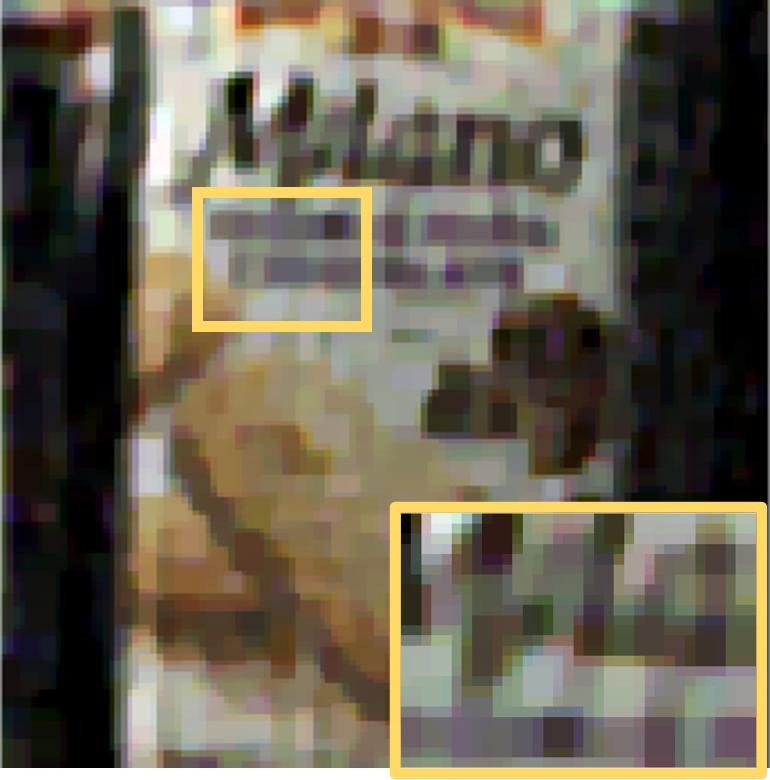} &
		\includegraphics[width=\figwidth,keepaspectratio]{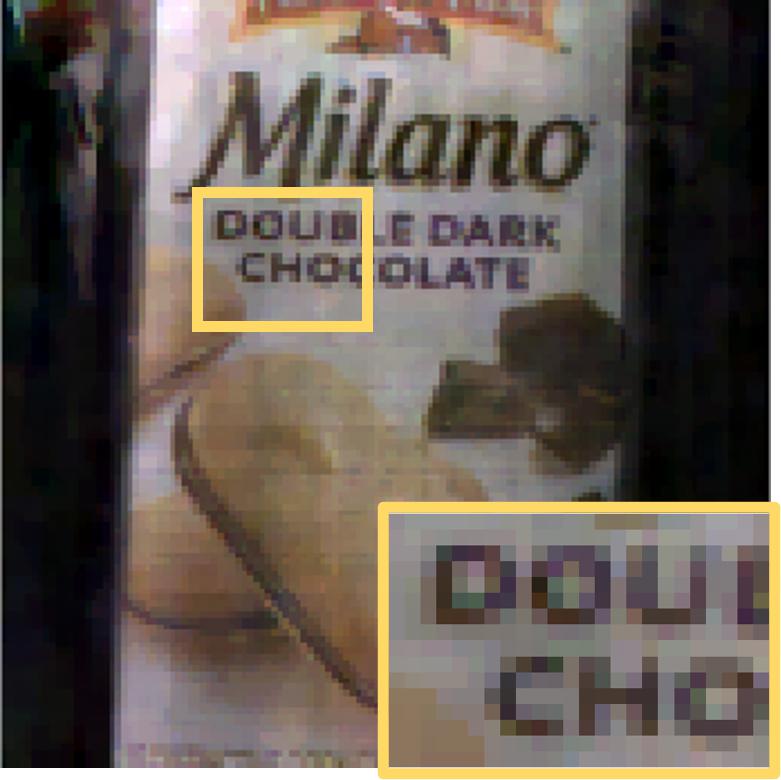} 
	\end{tabular}
	\caption{Reconstructing results using the proposed $\ell_2$-regularized least squares and TV-$\ell_1$ regularization on captured measurements. The TV-$\ell_1$ regularization is not visually not better than the proposed algorithm.
	}
	\label{fig:experiment_results_TV}
\end{figure*}

\clearpage

\section{Deep networks for refinement }
Deep learning-based methods have been widely used for image recovery and enhancement tasks.
We performed some additional experiments to compare the performance of four methods with coded and uniform illumination: least squares (LS) method proposed in the main paper, LS with a trained UNet that is used as a refinement network, LS with pretrained refinement network in FlatNet 
, and end-to-end trained FlatNet. 
The detailed description of the four methods are provided in the main text.
We present results on synthetic sensor measurements in Fig.~\ref{fig:simu_flatnet_unet} and on real  data captured using our prototype in Fig.~\ref{fig:exp_flatnet_unet}. The results with different numbers of illumination patterns using the end-to-end trained FlatNet is presented in Fig.~\ref{fig:exp_flatnet_number}. 

\renewcommand{\figwidth}{0.2\linewidth}
\begin{figure}[htb]
	\setlength\tabcolsep{1pt}
	\centering
	\footnotesize
	\begin{tabular}{cccc}
		(a) &
		(b) &
		(c) &
		(d)
		\\
		\rotatebox{90}{\parbox{2.3cm}{\centering uniforn}}
		\includegraphics[width=\figwidth,keepaspectratio]{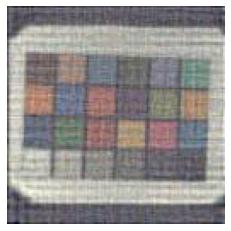} &
		\includegraphics[width=\figwidth,keepaspectratio]{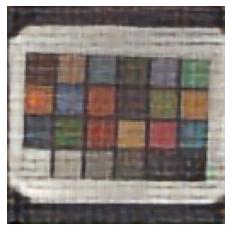}&
		\includegraphics[width=\figwidth,keepaspectratio]{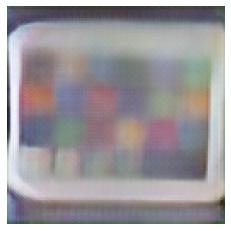} &
		\includegraphics[width=\figwidth,keepaspectratio]{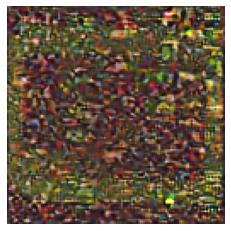} 
		\\
		\rotatebox{90}{\parbox{2.3cm}{\centering 49 Shifting dots}}
		\includegraphics[width=\figwidth,keepaspectratio]{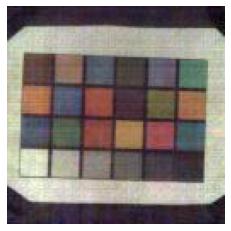} &
		\includegraphics[width=\figwidth,keepaspectratio]{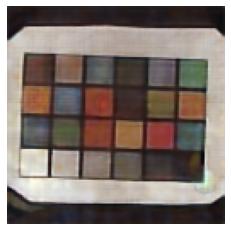}&
		\includegraphics[width=\figwidth,keepaspectratio]{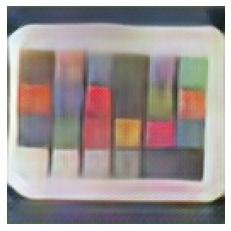} &
		\includegraphics[width=\figwidth,keepaspectratio]{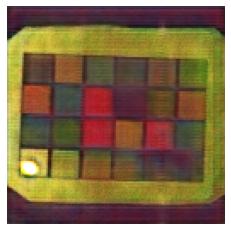} 
		\\
		\rotatebox{90}{\parbox{2.3cm}{\centering uniforn}}
		\includegraphics[width=\figwidth,keepaspectratio]{figures/img05_pattern01_num01.jpg} &
		\includegraphics[width=\figwidth,keepaspectratio]{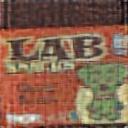}&
		\includegraphics[width=\figwidth,keepaspectratio]{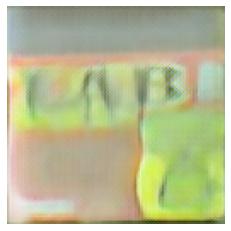} &
		\includegraphics[width=\figwidth,keepaspectratio]{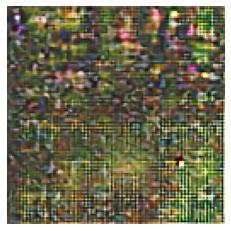} 
		\\
		\rotatebox{90}{\parbox{2.3cm}{\centering 49 Shifting dots}}
		\includegraphics[width=\figwidth,keepaspectratio]{figures/img05_pattern03_num49.jpg} &
		\includegraphics[width=\figwidth,keepaspectratio]{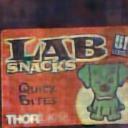}&
		\includegraphics[width=\figwidth,keepaspectratio]{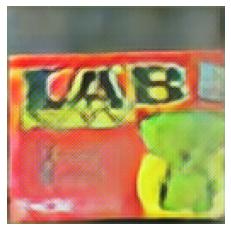} &
		\includegraphics[width=\figwidth,keepaspectratio]{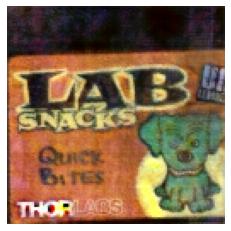} 
	\end{tabular}
	\caption{Reconstruction results for simulated measurements with 49 uniform and shifting dots illumination patterns. 
		Images in four columns show (a) LS solution, (b) LS solution with trained UNet refinement, (c) LS solution with pretrained FlatNet refinement, and (d) trained FlatNet that reconstructs image directly from measurements. 
	}
	\label{fig:exp_flatnet_unet}
\end{figure}

\renewcommand{\figwidth}{0.13\linewidth}
\begin{figure*}[htb]
	\setlength\tabcolsep{1pt}
	\centering
	\footnotesize
	\begin{tabular}{cccc}
		(a) &
		(b) &
		(c) &
		(d)
		\\
		\rotatebox{90}{\parbox{2.3cm}{\centering uniforn}}
		\includegraphics[width=\figwidth,keepaspectratio]{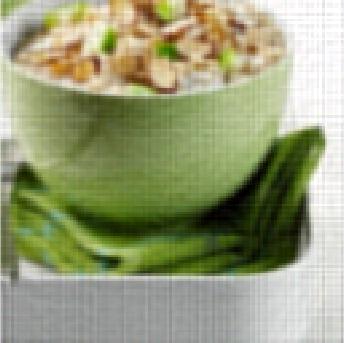} &
		\includegraphics[width=\figwidth,keepaspectratio]{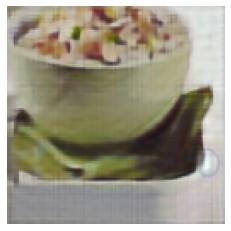} &
		\includegraphics[width=\figwidth,keepaspectratio]{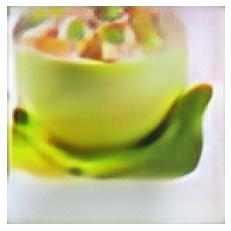} &
		\includegraphics[width=\figwidth,keepaspectratio]{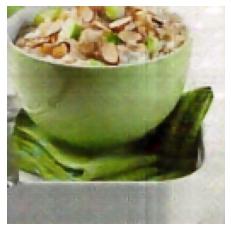} 
		\\
		SSIM:0.850 &
		0.833 &
		0.629 &
		0.909
		\\
		\rotatebox{90}{\parbox{2.3cm}{\centering 49 Shifting dots}}
		\includegraphics[width=\figwidth,keepaspectratio]{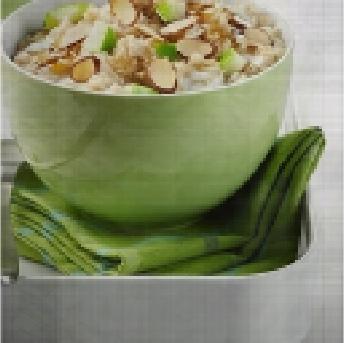} &
		\includegraphics[width=\figwidth,keepaspectratio]{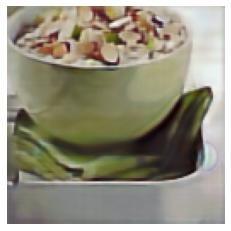} &
		\includegraphics[width=\figwidth,keepaspectratio]{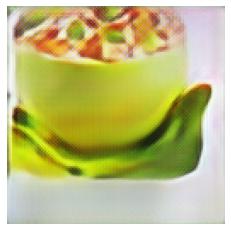} &
		\includegraphics[width=\figwidth,keepaspectratio]{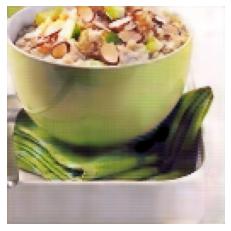} 
		\\
		SSIM:0.931 &
		0.891 &
		0.619 &
		0.931 
		\\
		\rotatebox{90}{\parbox{2.3cm}{\centering uniform}}
		\includegraphics[width=\figwidth,keepaspectratio]{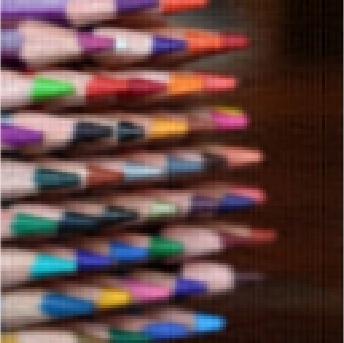} &
		\includegraphics[width=\figwidth,keepaspectratio]{figures/flatnet_finetuned_synthetic_illum01_01_img01.jpg}&
		\includegraphics[width=\figwidth,keepaspectratio]{figures/flatnet_pretrained_synthetic_illum01_01_img01.jpg} &
		\includegraphics[width=\figwidth,keepaspectratio]{figures/netProcessed_illum_01_01_img01.jpg} 
		\\
		SSIM:0.942 &
		0.851 &
		0.474 &
		0.950 
		\\
		\rotatebox{90}{\parbox{2.3cm}{\centering 49 Shifting dots}}
		\includegraphics[width=\figwidth,keepaspectratio]{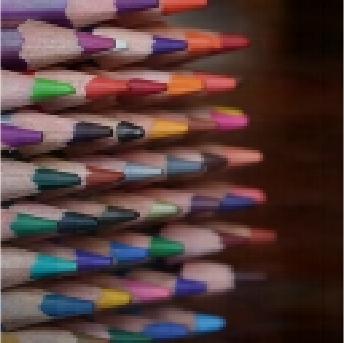} &
		\includegraphics[width=\figwidth,keepaspectratio]{figures/flatnet_finetuned_synthetic_illum03_49_img01.jpg}&
		\includegraphics[width=\figwidth,keepaspectratio]{figures/flatnet_pretrained_synthetic_illum03_49_img01.jpg} &
		\includegraphics[width=\figwidth,keepaspectratio]{figures/netProcessed_illum_03_49_img01.jpg} 
		\\
		SSIM:0.966 &
		0.898 &
		0.469 &
		0.936 
		\\
		\rotatebox{90}{\parbox{2.1cm}{\centering uniform}}
		\includegraphics[width=\figwidth,keepaspectratio]{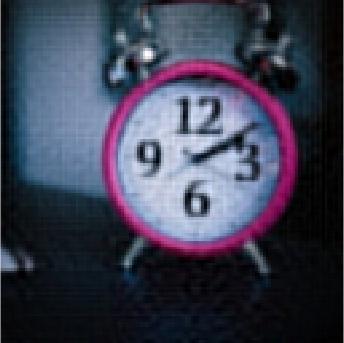} &
		\includegraphics[width=\figwidth,keepaspectratio]{figures/flatnet_finetuned_synthetic_illum01_01_img02.jpg} &
		\includegraphics[width=\figwidth,keepaspectratio]{figures/flatnet_pretrained_synthetic_illum01_01_img02.jpg} &
		\includegraphics[width=\figwidth,keepaspectratio]{figures/netProcessed_illum_01_01_img02.jpg} 
		\\
		SSIM: 0.803 &
		0.781 &		
		0.501 &
		0.903 
		\\
		\rotatebox{90}{\parbox{2.1cm}{\centering shifting dots}}
		\includegraphics[width=\figwidth,keepaspectratio]{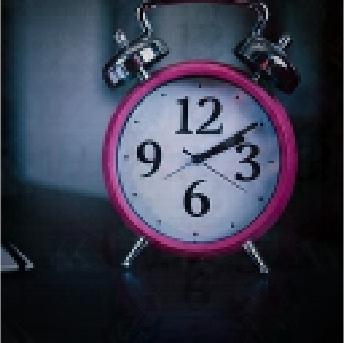} &
		\includegraphics[width=\figwidth,keepaspectratio]{figures/flatnet_finetuned_synthetic_illum03_49_img02.jpg} &
		\includegraphics[width=\figwidth,keepaspectratio]{figures/flatnet_pretrained_synthetic_illum03_49_img02.jpg} &
		\includegraphics[width=\figwidth,keepaspectratio]{figures/netProcessed_illum_03_49_img02.jpg} 
		\\
		SSIM: 0.936 &		
		0.831 &
		0.562 &
		0.904
		\\
		\rotatebox{90}{\parbox{2.1cm}{\centering uniform}}
		\includegraphics[width=\figwidth,keepaspectratio]{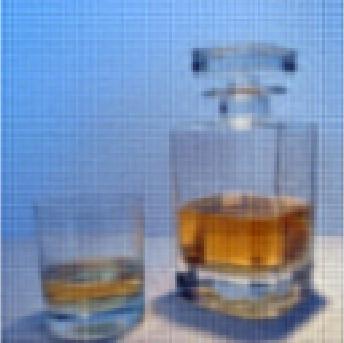} &
		\includegraphics[width=\figwidth,keepaspectratio]{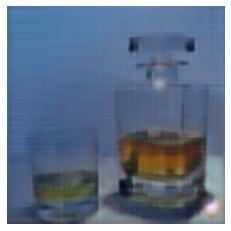}&
		\includegraphics[width=\figwidth,keepaspectratio]{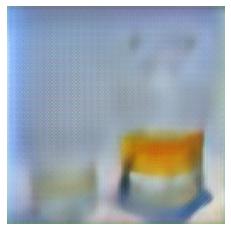} &
		\includegraphics[width=\figwidth,keepaspectratio]{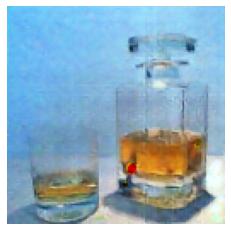} 
		\\
		SSIM: 0.945 &
		0.819 &
		0.712 &
		0.898 
		\\
		\rotatebox{90}{\parbox{2.1cm}{\centering shifting dots}}
		\includegraphics[width=\figwidth,keepaspectratio]{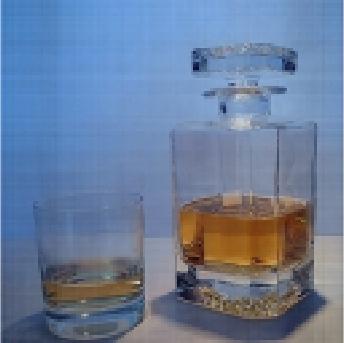} &
		\includegraphics[width=\figwidth,keepaspectratio]{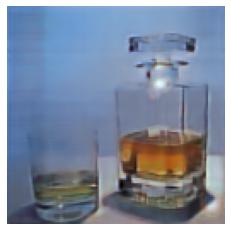}&
		\includegraphics[width=\figwidth,keepaspectratio]{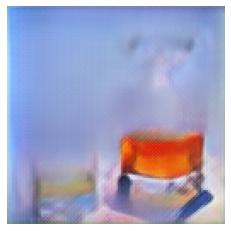} &
		\includegraphics[width=\figwidth,keepaspectratio]{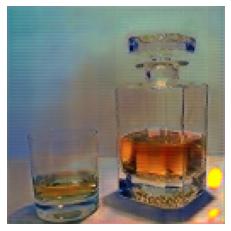} 
		\\
		SSIM: 0.980 &
		0.910 &
		0.708 &
		0.877
	\end{tabular}
	\caption{Reconstruction results for simulated measurements with 49 uniform and shifting dots illumination patterns. 
		Images in four columns show (a) LS solution, (b) LS solution with trained UNet refinement, (c) LS solution with pretrained FlatNet refinement, and (d) trained FlatNet that reconstructs image directly from measurements. For each image, we show the SSIM value underneath. 
	}
	\label{fig:simu_flatnet_unet}
\end{figure*}

\renewcommand{\figwidth}{0.2\linewidth}
\begin{figure*}[htb]
	\setlength\tabcolsep{1pt}
	\centering
	\footnotesize
	\begin{tabular}{ccccc}
		uniform &
		16 shifting dots &
		49 shifting dots &
		\\
		\rotatebox{90}{\parbox{2.4cm}{\centering Least-Squares}}
		\includegraphics[width=\figwidth,keepaspectratio]{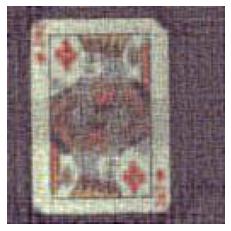} &
		\includegraphics[width=\figwidth,keepaspectratio]{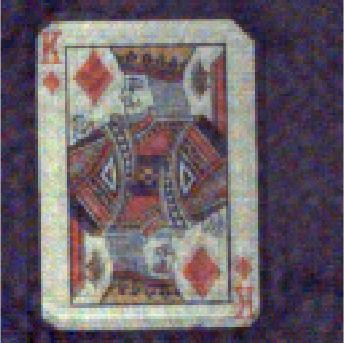} &
		\includegraphics[width=\figwidth,keepaspectratio]{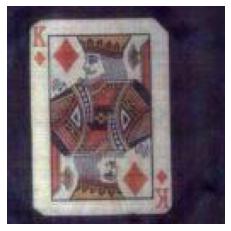}\\
		\rotatebox{90}{\parbox{2.4cm}{\centering FlatNet}}
		\includegraphics[width=\figwidth,keepaspectratio]{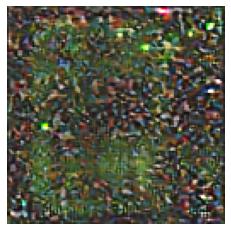} &
		\includegraphics[width=\figwidth,keepaspectratio]{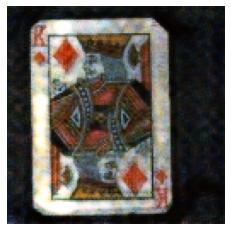} &
		\includegraphics[width=\figwidth,keepaspectratio]{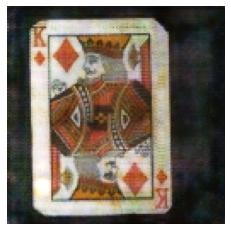}
	\end{tabular}
	\caption{Reconstruction results using uniform, 16 shifting dots, and 49 shifting dots illumination patterns. We compare LS solution and trained FlatNet. We observe that the FlatNet-based method fails to recover images with uniform illumination measurements of real hardware. 
	}
	\label{fig:exp_flatnet_number}
\end{figure*} 

% \clearpage
% \bibliographystyle{IEEEtran}
% \bibliography{TCI_bib}

\end{appendices}
		
\end{document}